\documentclass{aa}  
\usepackage[normalem]{ulem}
\usepackage{placeins}
\usepackage{color}
\usepackage{calc}
\usepackage{tensor}
\usepackage{bm}
\usepackage{newtxtext}
\usepackage{newtxmath}
\usepackage{dcolumn}
\usepackage[nolist,nohyperlinks]{acronym}
\usepackage{xspace}
\usepackage[abs]{overpic}
\usepackage{pict2e}
\usepackage{enumitem}
\usepackage[usenames,dvipsnames]{xcolor}
\usepackage[utf8]{inputenc}
\usepackage{acronym}
\usepackage{gensymb}
\usepackage{caption}
\usepackage{subcaption}
\usepackage{etoolbox}
\patchcmd{\equation}{\hfil}{\hfill}{}{}
\usepackage[normalem]{ulem}
\usepackage{longtable}
\usepackage{verbatim}
\usepackage{multirow}
\usepackage{amsmath,amssymb}
\usepackage{csquotes}
\usepackage[export]{adjustbox}
\usepackage{placeins}
\usepackage{soul}
\usepackage{fancyvrb} 
\usepackage{float}
\usepackage{dblfloatfix}

\usepackage{hhline}

\newcommand{\ID}[1]{{\color{blue}{#1}}} 

\usepackage[pdfborder={0 0 0},colorlinks=true,citecolor=NavyBlue]{hyperref}
\usepackage{cleveref}
\titlerunning{Binary neutron star populations across cosmic time}
\begin{document}

\title{Binary neutron star populations across cosmic time: 
the impact of binary stellar evolution uncertainties}

\author{Mathieu Venet \inst{1} \and Irina Dvorkin \inst{1,2}}

\institute{
  Institut d’Astrophysique de Paris, UMR 7095, CNRS and Sorbonne Université, 98 bis boulevard Arago, 75014 Paris, France 
\and
Institut Universitaire de France, Minist\`ere de l’Enseignement Sup\'erieur et de la Recherche, 1 rue Descartes, 75231 Paris Cedex F-05, France
  }

\abstract
{Direct observations of binary neutron star (BNS) systems remain scarce, but their number is expected to increase significantly in the coming years through both electromagnetic and gravitational-wave observations. Preparing robust tools to interpret these forthcoming datasets is therefore essential.}
{We investigate the impact of uncertainties in binary stellar evolution on the formation efficiency and cosmological merger rate of BNS systems. In particular, we aim to determine whether robust signatures of binary evolution can be identified across a broad range of metallicities.}
{We perform a systematic exploration of 72 binary evolution models using the population synthesis code \textsc{COSMIC}. We vary four key parameters controlling the onset and efficiency of common-envelope evolution, the natal kick prescription, and the treatment of kicks during the common envelope phase. For each model, we compute the BNS formation efficiency and cosmological merger rate using several prescriptions for the cosmic metallicity-redshift relation.}
{Despite the large uncertainties introduced by binary stellar evolution, we identify characteristic features in the metallicity dependence of the BNS formation efficiency that arise from specific evolutionary processes. These include the onset of first-giant-branch skipping at low metallicity and a non-monotonic evolution between $Z=10^{-3}$ and $7.5\times10^{-3}$. All models also predict a bimodal distribution of BNS progenitor masses associated with two distinct formation channels. The common envelope ejection efficiency and natal kick prescription remain the dominant sources of uncertainty, producing variations of several orders of magnitude in both the BNS formation efficiency and the cosmological merger rate, and strongly affecting the prominence of the metallicity-dependent features.}
{Finally, we show that different combinations of binary-evolution parameters can produce similar merger-rate histories, highlighting significant degeneracies among population-synthesis models. Current gravitational-wave observations can rule out only the least efficient formation scenarios. Breaking these degeneracies will require larger samples of BNS mergers expected from future third-generation gravitational-wave observatories, and improved observational constraints on the evolution of metallicity across cosmic time.}

\maketitle

\section{Introduction} \label{sec:Introduction}

Binary neutron star (BNS) systems are associated with a wide range of high-energy phenomena across various messengers. These include double neutron star systems (pulsar with a neutron star companion) detected through radio emission~\citep{2005yCat.7245....0M}, BNS mergers observed through gravitational waves~\citep{2017ApJ...848L..12A,2017PhRvL.119p1101A,2020ApJ...892L...3A}, and phenomena such as kilonovae \citep{Metzger_2019} and short gamma-ray bursts \citep{2014ARA&A..52...43B}. Furthermore, BNS mergers are believed to contribute significantly to the production of heavy elements through rapid neutron-capture nucleosynthesis (r-process), although their exact contribution remains an active area of research \citep{Cowan_2021}.

Despite their importance across several observational domains, the population of BNS systems remains poorly constrained by direct observations. To date, only slightly more than twenty Galactic double neutron stars have been accurately characterized \citep{Farrow_2019}. Likewise, the fifth Gravitational-Wave Transient Catalog (GWTC-5) of the LIGO-Virgo-KAGRA Collaboration currently contains only two confidently identified BNS merger events \citep{theligoscientificcollaboration2026gwtc50populationpropertiesmerging,2020ApJ...892L...3A}. However, the number of observed BNS systems is expected to increase substantially in the coming years. The Square Kilometre Array \citep[SKA;][]{lazio2009squarekilometrearray} is expected to detect several hundred double neutron star systems \citep[see][]{2018RSPTA.37670293S,2023MNRAS.526.2210S} out of the $\sim$2500 predicted to be currently present in the Milky Way \citep{2020MNRAS.494.1587C}. Furthermore, the LIGO-Virgo-KAGRA Collaboration \citep{2015CQGra..32g4001L,2015CQGra..32b4001A,2021PTEP.2021eA101A} is expected to greatly increase the number of detected BNS mergers as the detector network approaches its design sensitivity during the fifth observing run \citep{Abbott2020Prospects,Muccillo_2026}. Looking further ahead, third-generation gravitational-wave detectors such as the Einstein Telescope \citep{2010CQGra..27s4002P,2011CQGra..28i4013H} and  Cosmic Explorer \citep{2019BAAS...51g..35R,2021arXiv210909882E}, planned for the 2030s, are expected to detect up to $10^{5}$ BNS mergers per year \citep{2023JCAP...07..068B,Abac_2026}. Developing robust tools to analyse and interpret these forthcoming observations is therefore essential. Such tools are needed not only to prepare for the unprecedented volume of future data, but also to fully exploit current observations and assess the constraints that can already be placed on BNS formation and evolution. 

Population-synthesis codes \citep[e.g.][]{2002MNRAS.329..897H,2008ApJS..174..223B,2018MNRAS.474.2959G,2018MNRAS.481.1908K,2018MNRAS.479...75S,2018MNRAS.480.2011G,Breivik_2020,2019MNRAS.485..889S,2023MNRAS.524..426I,Fragos_2023,2025ApJS..280...43T,andrews2025posydonversion2population} provide a powerful framework for interpreting the limited observational sample of BNS systems while constructing self-consistent models of compact-object formation and binary stellar evolution. These codes follow the evolution of large populations of single and binary stars under a set of physical prescriptions, allowing synthetic populations to be directly compared with observations and thereby placing statistical constraints on the underlying evolutionary processes \citep[e.g.][]{2022MNRAS.514.1315B,2022MNRAS.509.1557C,deng2024formationdoubleneutronstars,2026ApJ...997...52C,2026PhRvD.113h3006T}. 

The canonical formation scenario of a BNS begins with a massive binary on the main sequence, with both stars typically having zero-age main-sequence masses above $\sim8\,M_\odot$ \citep{2017ApJ...846..170T}. As the stars evolve, they can undergo several phases of binary interaction before each experiences core collapse or electron capture supernovae explosions and, ultimately, two neutron stars are formed. A major challenge in forming BNS systems is that each supernova may impart a natal kick to the newly born neutron star owing to asymmetries in the explosion \citep[e.g.][]{2005MNRAS.360..974H}. These kicks frequently disrupt the binary, making surviving BNS systems intrinsically rare. Consequently, a successful BNS progenitor must remain gravitationally bound after both supernova explosions despite the associated mass loss and natal kicks. In addition to supernovae, binary interactions play a fundamental role in determining the final outcome of the system. Mass transfer may occur through stellar-wind accretion or Roche-lobe overflow (RLOF) \citep{Wang_2016}. Depending on the evolutionary stage of the donor star and the response of both the donor and its Roche lobe to mass loss, RLOF may proceed either stably or unstably \citep{1997A&A...327..620S,2015ApJ...812...40G}. In the latter case, the system may enter a common-envelope (CE) phase, during which the companion spirals into the envelope of the donor star~\citep{2002MNRAS.329..897H,2013A&ARv..21...59I}. The outcome of this poorly understood phase strongly affects the final orbital separation and therefore the likelihood of producing a merging BNS \citep[e.g.][]{2018MNRAS.480.2011G,2023ApJ...955..133G}. Population-synthesis codes generally model the onset and evolution of mass transfer and CE evolution using simplified prescriptions based on parameters such as the binary mass ratio, the envelope binding energy, and the efficiency with which orbital energy is converted into envelope ejection \citep{2002MNRAS.329..897H}. Because the formation of BNS systems results from the interplay between stellar evolution, binary interactions, and supernova physics, population-synthesis models remain indispensable for exploring their formation channels and predicting their observable properties. Their predictions can be confronted with radio pulsar surveys and gravitational-wave observations, providing increasingly stringent constraints on the physical processes governing binary evolution.

However, population-synthesis calculations rely on numerous analytical prescriptions and empirical models that attempt to describe complex processes in stellar and binary evolution. Several of these prescriptions remain affected by significant uncertainties, including models of single stellar evolution \citep{2000MNRAS.315..543H,Bressan2012PARSEC,Paxton2011MESA}, the treatment of CE evolution \citep{2002MNRAS.329..897H,2011MNRAS.411.2277D,Ricker_2018}, and supernova natal kick prescriptions \citep{2016MNRAS.456.4089B,2020ApJ...891..141G,Vigna_G_mez_2025,Disberg_2025}. Recent studies have even suggested that standard population-synthesis prescriptions may struggle to simultaneously reproduce the observed merger rates and properties of compact-object binaries \citep{2025A&A...698A.144S,2026arXiv260602725B}. Exploring the parameter space of population-synthesis models is therefore essential for assessing the robustness of theoretical predictions and identifying the physical processes that most strongly influence BNS formation.

In this work, we perform a systematic exploration of the parameter space of the population-synthesis code \textit{COSMIC}. Since \textit{COSMIC} contains more than fifty adjustable parameters and over one hundred alternative physical prescriptions, a complete exploration of the full parameter space is computationally prohibitive. We therefore focus on a reduced set of parameters selected according to both their current theoretical uncertainties and their expected impact on BNS formation. Specifically, we identify four key parameters related to CE evolution, including the conditions required for CE initiation, the efficiency of envelope ejection, the treatment of natal kicks occurring during CE phases and the kick prescription in itself. We explore these parameters across several metallicity bins in order to probe different star-formation environments and cosmic epochs.

Using this framework, we investigate how variations in these uncertain physical prescriptions affect BNS formation efficiencies and merger rates. Furthermore, we explore in detail their dependence on metallicity. Indeed, while the variation in BNS formation efficiencies across various metallicities is significantly smaller that in the much more studied case of binary black holes \citep{2021MNRAS.502.4877S,2024ApJ...976...23B,2025ApJ...979..209V}, several key features can be identified that are directly linked to specific single and binary stellar evolution processes.

This paper is structured as follows. In Section \ref{section:Methodology}, we present our main framework, including our population-synthesis simulation setup, the parameters explored in this work, and our methodology for calculating cosmological merger rates from the simulated BNS populations. In Section \ref{section:Results}, we describe the results obtained from our simulated populations. We first investigate the bimodal distribution of progenitor masses and the relevance of the explored parameters in shaping this feature in Section \ref{section:bimodal_distribution}. We then present the obtained formation efficiencies and discuss the impact of the explored parameters and assumptions in Section \ref{section : Formation efficiency}. Finally, we present the cosmological merger rates derived from our simulated populations and discuss their associated features in Section \ref{section:merger_rates_results}. We conclude in Section \ref{section:discussion}.

\section{Methodology} 
\label{section:Methodology}
\label{sec:methodo}
\subsection{Population synthesis simulation set-up } \label{section:pop synthesis set-up}

\subsubsection{Population synthesis code: COSMIC} \label{section:COSMIC}

In this work, we use the publicly available code \textbf{COSMIC}\footnote{\url{https://github.com/COSMIC-PopSynth/COSMIC}}~\citep[Compact Object Synthesis and Monte Carlo Investigation Code;][]{Breivik_2020}, a rapid binary population-synthesis framework with a specific focus on the formation and evolution of compact-object binaries.

\textit{COSMIC} implements stellar evolution using SSE  \citep{2000MNRAS.315..543H} and binary interactions using BSE \citep{2002MNRAS.329..897H}. Several modifications have been applied to the original BSE framework in order to incorporate more recent developments in binary evolution that are particularly relevant for compact-binary formation, such as metallicity-dependent stellar winds or updated prescriptions for black-hole natal kick strengths. A detailed description of these modifications is provided in \citet{Breivik_2020}.

    We use version 3.6 of \textit{COSMIC}, which introduces significant changes in the treatment of contact and merger conditions during stable mass transfer in binary systems by implementing a periastron collision check. In later versions of \textit{COSMIC} (from version 3.7.8 onwards), this check is deactivated until improved SSE models become available. We nevertheless adopt a slightly modified version of \textit{COSMIC} 3.6, as this prescription provides a more consistent treatment within the assumptions adopted in our study. This difference naturally explains why identical sets of binary-evolution parameters can lead to significantly different BNS populations depending on the \textit{COSMIC} version employed\footnote{The periastron collision check prevents binaries from undergoing stable mass transfer solutions when the stellar components would physically overlap at periastron. Without this criterion, systems in which both stellar radii expand rapidly may continue their evolution despite entering an unphysical configuration, without triggering any change in the adopted prescriptions. Including this additional condition significantly affects the predicted BNS formation efficiency: for specific regions of the initial parameter space, we find a decrease of more than one order of magnitude in the number of BNS systems produced from the same initial stellar population.}.

One of our  modifications consists in the implementation of a custom convergence criterion. Indeed, BNS populations exhibit particularly low formation efficiencies, making it essential to optimize the sampling procedure in order to avoid spending computational resources on already converged populations while still ensuring a robust characterization of rare-event tails. Since merging systems occupy only a small and highly localized region of the binary parameter space, merger rates are primarily determined by the tails of the underlying distributions, which motivates a dedicated convergence treatment of these regions. 
For this reason, we adopt a tail-convergence criterion specifically designed to assess the stabilization of the low-probability regions of the distribution. To this end, the initial population is divided into 100 subsets, each containing 1\% of the total number of systems, which are evolved sequentially. After each subset has been processed, the resulting BNS population is cumulatively added to those obtained from the previous subsets, thereby progressively reconstructing the complete simulated population. At each step, both the cumulative BNS population and the subset of systems merging within a Hubble time are binned into 500 mass bins. We then identify the bins whose occupation fraction remains below 10\% of the total population and evaluate the variation of their normalized bin probabilities between two consecutive cumulative populations. Convergence is considered to be achieved when this variation remains below a predefined threshold $T$ for all low-occupation bins. This criterion explicitly targets the convergence of the tails of the distribution, where statistical fluctuations are expected to be the largest and where standard global convergence metrics are generally less sensitive.
Convergence is assessed independently for both the complete population and the merging BNS population. It is declared only if all selected bins satisfy the convergence criterion and if the final population contains more than 1100 BNS systems in total and more than 80 merging BNS systems. These additional constraints ensure that convergence is driven by the stabilization of the distribution tails rather than by statistical noise in low-count regimes. We adopt thresholds of $T_{\rm{tot}} = 0.01$ for the complete population and $T_{\rm{merge}} = 0.25$ for the merging population, the latter being intentionally less stringent due to the intrinsically lower statistical support of merging systems. This asymmetry reflects the fact that the merger sample probes a significantly rarer region of parameter space, where relative fluctuations are naturally larger.

\subsubsection{Parameter space exploration} 

\label{section: parameter space}

In this work we explore four key parameters, expected to play a key role in BNS formation, across $20$ metallicity bins. The  latter span the range from $Z = 10^{-4}$ to $Z = Z_\odot$, thereby covering environments from present-day star formation to Population~II stellar populations \citep{2025A&A...704A..54M,2004ApJ...616L..87F}. The specific metallicity bins explored in this study are described in Table \ref{tab:model_parameters}.

For this stellar population and metallicity range, we adopt the \citet{2003PASP..115..763C} initial mass function (IMF) to generate the initial binary population. While the analysis presented in Section ~\ref{section:Results}  is primarily based on this IMF, we additionally explore an alternative IMF in order to estimate the order of magnitude of formation and merger rates expected for Population~II stars. Specifically, for the two lowest metallicity bins, $Z = 0.0001$ and $Z = 0.00013$, we rerun the same set of models assuming a top-heavy IMF with characteristic mass $m_{\rm cut} = 20\,M_\odot$ \citep{2003PASP..115..763C} following :

\begin{equation}
    \phi(m) \propto \left(\frac{m}{M_\odot}\right)^{-2.3} 
    \exp\left[-\left(\frac{m_{\rm cut}}{m}\right)^{1.6}\right]
    \label{eq:top-heavy IMF}
\end{equation}

Once implemented, this IMF is then normalized by \textit{COSMIC} using the fiducial lower and upper initial mass limits. This \textit{top-heavy} IMF is intended to probe an extreme scenario for the BNS formation and merger rates at such low metallicities. To complete our initial sample characterization, we draw the initial separation and eccentricity from \citet{2012Sci...337..444S}. 

We adopt a metallicity-dependent treatment of stellar winds for O/B stars and Wolf-Rayet stars following \citet{2001A&A...369..574V} and \citet{2005A&A...442..587V}, while including LBV-like mass loss for giants and non-degenerate stars beyond the Humphreys-Davidson limit (\textbf{windflag = 3} in \textsc{COSMIC}). The treatment of stellar winds is central to describing mass loss and the amount of material potentially accreted by the companion, particularly for stars in the $\sim 6$ to $\sim 30 \, M_\odot$ range \citep{2001A&A...369..574V,2005A&A...442..587V}. The adopted prescription is similar to other works such as \citet{Ablimit_2018,Broekgaarden_2022,Pellouin_2025}.

We explore four \textit{COSMIC} parameters across the full range of metallicities and IMFs considered in this work. \textbf{QCFLAG}, which selects the prescription used to determine the critical mass ratio $q = M_{\rm donor}/M_{\rm accretor} $ for the onset of unstable mass transfer and CE evolution during RLOF; the efficiency parameter of CE \textbf{ALPHA}; the natal kick prescription \textbf{KICKFLAG}; and a parameter that determines the treatment of a supernova during CE \textbf{CE\_kickflag}. We now describe in some detail each of these parameters.

The critical mass ratio for the onset of unstable mass plays a central role in BNS formation and merger rates, as CE evolution efficiently reduces the orbital separation and enables merger times shorter than a Hubble time.
We consider three prescriptions: \textit{QCFLAG = 1} from \citet{2002MNRAS.329..897H} with \citet{1987ApJ...318..794H} prescriptions for GB/AGB stars, \textit{QCFLAG = 3} from \citet{2014A&A...563A..83C} with  \citet{1987ApJ...318..794H} prescriptions for GB/AGB stars, and \textit{QCFLAG = 5} from \citet{2019MNRAS.490.3740N}, which span a representative range of mass-transfer stability assumptions for the stellar types relevant to BNS formation. In the case of \textit{QCFLAG = 1}, envelope-stripped stars outside the main sequence are assumed to become unstable to RLOF for critical mass ratios $q \lesssim 1$. Conversely, \textit{QCFLAG = 3} adopts significantly larger critical mass ratios ($q > 1$) for instability for the same stellar types. Finally, \textit{QCFLAG = 5} assumes that mass transfer is always stable for these stars, irrespective of the mass ratio. These differences directly control the occurrence of single or multiple CE phases and therefore have a strong impact on the predicted BNS formation and merger rates. 

We also explore six values of the \textbf{ALPHA} parameter, which characterizes the CE efficiency and quantifies how effectively orbital energy is converted into kinetic energy to eject the envelope during a CE phase as described in \citet{2002MNRAS.329..897H} and developped more in details in Appendix \ref{Appendix}. We consider the following values: $0.5,\,1.0,\,1.5,\,3.0,\,5.0,$ and $7.0$. These values correspond, respectively, to scenarios in which only a fraction of the available orbital energy contributes to envelope ejection, the full orbital energy budget is used, and additional energy reservoirs, such as internal or recombination energy, can participate in unbinding the envelope \citep{2008ASSL..352..233W,Li_2026}.
We extend our exploration to such high values of \textit{ALPHA} as several studies suggest that reproducing the observed BNS merger rates may require CE efficiencies significantly larger than unity, potentially reflecting the contribution of additional energy sources during envelope ejection or the limits of the $\alpha - \lambda$ formalism \citep[see][]{Chu_2021,2021MNRAS.502.4877S,deng2024formationdoubleneutronstars,Li_2026,2023MNRAS.526.2210S,Chattaraj_2026}.

We explore two parameters related to supernova natal kicks, which play a key role in BNS formation and remain subject to significant theoretical uncertainties. The primary kick-related parameter is \textbf{KICKFLAG}, which specifies the adopted natal kick prescription. We consider two alternative prescriptions.
For \textit{KICKFLAG = 1}, we adopt the standard \textit{COSMIC} prescription, in which natal kicks are drawn from a bimodal distribution. In this framework, neutron stars formed through Fe core-collapse supernovae (CCSN) receive kicks drawn from a Maxwellian distribution with dispersion $\sigma$, whereas neutron stars formed through electron-capture supernovae (ECSN) or ultra-stripped supernovae (USSN) are assigned reduced kicks as described in \citet{2005MNRAS.360..974H}. Throughout this work, we adopt the default \textit{COSMIC} values of $\sigma_{\rm CCSN}=265\,\mathrm{km\,s^{-1}}$ and $\sigma_{\rm ECSN}= \sigma_{\rm USSN}=20\,\mathrm{km\,s^{-1}}$.

However, \citet{Disberg_2025} argued that, due to an inconsistency in \citet{2005MNRAS.360..974H}, this Maxwellian fit, which peaks at $400 \ \mathrm{km.s^{-1}}$ for CCSNe and $30 \ \mathrm{km.s^{-1}}$ for ECSNe, is shifted towards high-kick values relative to observations (see Appendix~\ref{section:kickflag_1_distributions}). Although these natal kick amplitudes remain compatible with observational constraints, they correspond to a  scenario in which supernova kicks are systematically quite strong.

We also consider \textit{KICKFLAG = 2}, in which case natal kicks are initially drawn from a Maxwellian distribution with dispersion $\sigma$ and subsequently scaled by the ejecta mass and the remnant mass, following Eq.~(1) of \citet{2020ApJ...891..141G}, using their default parameters ($m_{\rm NS} = 1.2\ M_\odot$ and $m_{\rm ej} = 9\,M_\odot$). This prescription introduces a progressive reduction of natal kicks for systems experiencing significant mass loss prior to the supernova.

Finally, we explore the parameter \textbf{CE\_kickflag}, which determines which stellar masses and orbital separations are used when a supernova occurs during a CE phase and a natal kick is applied. We consider two options: \textit{CE\_kickflag = 1}, which uses pre-CE stellar masses and orbital separations, and \textit{CE\_kickflag = 2}, which uses post-CE values.
This parameter plays a key role in linking common-envelope evolution with natal kick strength. In particular, the \textit{CE\_kickflag = 2} prescription is more favorable to BNS formation, as it allows the orbital separation to be reduced immediately before the application of the natal kick. Moreover, when combined with \textit{KICKFLAG = 2}, the stellar mass at the time of the supernova directly impacts the kick magnitude, whereas for \textit{KICKFLAG = 1} the kick distribution is independent of mass. Taken together, these prescriptions allow us to bracket the impact of supernova kicks and common-envelope evolution by exploring both relatively favorable and unfavorable scenarios for BNS formation.

We explore these four parameters using a grid approach, leading to a total of 72 population models. Each model is simulated for our 20 + 2 metallicity bins, including the \citet{2003PASP..115..763C} prescription and an additional top-heavy IMF case.

\begin{table*}
\centering
\caption{Summary of the population-synthesis parameters explored and fixed assumptions used in this study.}
\label{tab:model_parameters}
\renewcommand{\arraystretch}{1.3}
\begin{tabular}{p{4.2cm} c p{8.5cm}}
\hline\hline
Parameter &
Number of values &
Values / description \\
\hline

\multicolumn{3}{l}{\textbf{Explored parameters}} \\
\hline

QCFLAG &
3 &
Flags: 1, 3, 5 \\

ALPHA &
6 &
Values: 0.5, 1.0, 1.5, 3.0, 5.0 7.0 \\

KICKFLAG &
2 &
Flags: 1, 2 \\

CE\_kickflag &
2 &
Flags: 1, 2 \\
\hline

\multicolumn{3}{l}{\textbf{IMF and metallicity sampling}} \\
\hline

Metallicity grid ($Z$) &
20 &
$Z = 10^{-4},\, 1.3\times10^{-4},\, 1.6\times10^{-4},\, 1.9\times10^{-4},\, 2.2\times10^{-4},\, 2.25\times10^{-4},\, 3.8\times10^{-4},\, 5\times10^{-4},\, 6.3\times10^{-4},\, 7.5\times10^{-4},\, 8.8\times10^{-4},\, 10^{-3},\, 2.5\times10^{-3},\, 3.8\times10^{-3},\, 5\times10^{-3},\, 6.3\times10^{-3},\, 7.5\times10^{-3},\, 8.8\times10^{-3},\, 10^{-2},\, 1.4\times10^{-2}$ \\

IMF model used to create the initial population &
2 &
Chabrier IMF \citep{2003PASP..115..763C} sampled over the full metallicity grid (20 models), and one top-heavy \citep{2003PASP..115..763C} IMF model at fixed metallicitiy $Z = 10^{-4}$ and $Z=1.3 \times 10^{-4}$ \\
\hline

\multicolumn{3}{l}{\textbf{Sampling and orbital prescriptions}} \\
\hline

$P_{\rm orb}$ distribution &
-- &
\texttt{sana12} \citep{2012Sci...337..444S}  \\

Eccentricity distribution &
-- &
\texttt{sana12}  \citep{2012Sci...337..444S} \\
\hline

\multicolumn{3}{l}{\textbf{Other fixed parameters}} \\
\hline

$Z_{\odot}$ &
-- &
0.014 \\
q
windflag &
-- &
Model 3  \mbox{\citep{2001A&A...369..574V,2005A&A...442..587V}} \\

\hline\hline

\end{tabular}
\end{table*}

\subsection{Cosmological merger rate calculations} \label{section : merger rate calculations}

In the following calculations, we assume $H_0 = 68 \ \mathrm{km\,s^{-1}\,Mpc^{-1}}$ for the Hubble constant, $\Omega_\Lambda = 0.69$ and $\Omega_m = 0.31$ for the dark energy and matter density parameters, respectively, following \citet{2016A&A...594A..13P}. We further adopt $R = 0.27$ as the fraction of stellar mass returned to the interstellar medium, $y = 0.019$ as the net metal yield, and 
$\rho_b = 2.77 \times 10^{11}\,\Omega_b\,h_0^2\,M_\odot\,\mathrm{Mpc}^{-3}$ as the present-day baryon density \ID{\citep{2016Natur.534..512B}}.

In order to compute the cosmological merger rate for each of our populations we use \textit{cosmoRate} \citep{2020ApJ...898..152S,2021MNRAS.502.4877S}, a numerical framework designed to compute the properties and merger-rate distributions of compact-object binaries across cosmic time\footnote{\url{https://gitlab.com/Filippo.santoliquido/cosmo_rate_public}}. 
It evaluates the contribution of star formation at different redshifts and metallicities through a double integral over redshift $z'$ and metallicity $Z$ to the merger rate $\mathcal{R}(z)$ :

\begin{equation}
\begin{split}
\mathcal{R}(z)
=&\,
\frac{d}{dt_{\rm lb}(z)}
\int_{z_{\max}}^{z}
\psi(z')\,
\frac{dt_{\rm lb}(z')}{dz'} \\
&\qquad \times
\int_{Z_{\min}(z')}^{Z_{\max}(z')}
\eta_{\rm merge}(Z)\,
\mathcal{F}(z',z,Z)\,dZ\,dz'
\end{split}
\label{eq:Merger rate}
\end{equation}
Here, $t_{\rm lb}(z)$ denotes the look-back time associated with redshift $z$, and $z_{\max}$ is the upper redshift limit considered. The function $\psi(z')$ corresponds to the cosmic star formation rate density, whereas $Z_{\min}(z')$ and $Z_{\max}(z')$ delimit the metallicity range of stars born at redshift $z'$. The quantity $\eta_{\rm merge}(Z)$ denotes the merger efficiency for a given metallicity $Z$. The function $\mathcal{F}(z',z,Z)$ represents the fraction of compact binaries formed at redshift $z'$ and metallicity $Z$ that merge at redshift $z$, normalized with respect to all binary and single star systems formed at the same metallicity.

In our models, we assume the star formation rate (SFR) described in \cite{2017ApJ...840...39M}, given by : 

\begin{equation}
    \mathrm{\psi}(z')=
    0.01 \
    \frac{(1+z)^{2.6}}
    {1+\left(\dfrac{1+z}{3.2}\right)^{6.2}}
    \ \rm{M}_\odot \ \rm{yr}^{-1} \ \rm{Mpc}^{-3}
    \label{eq: SFR madau}
\end{equation}

The merger efficiency of BNS is computed directly from our \textit{COSMIC} simulations at each metallicity bin and is defined as:

\begin{equation}
\eta_{\rm merge}(Z) = \frac{\mathcal{N}_{\mathrm{merge}}(Z)}{M_{\star}(Z)}
\label{eq: Merger efficiency}
\end{equation}
where $\mathcal{N}_{\mathrm{merge}}(Z)$ is the number of BNS systems formed in the simulation at metallicity $Z$ and merging within a Hubble time, and $M_{\star}(Z) = M_{\mathrm{single}} + M_{\mathrm{binary}}$ is the total stellar mass required to produce a statistically significant population of binaries.

To facilitate the analysis in Section \ref{section:Results}, we define the formation efficiency:

\begin{equation}
\eta(Z) = \frac{\mathcal{N}_{\mathrm{TOT}}(Z)}{M_{\star}(Z)}
\label{eq: formation efficiency}
\end{equation}

and the merger fraction:

\begin{equation}
\epsilon(Z) = \frac{\mathcal{N}_{\rm merge}(Z)}{\mathcal{N}_{\mathrm{TOT}}(Z)}
\label{eq: merger fraction}
\end{equation}
where $\mathcal{N}_{\mathrm{TOT}}(Z)$ denotes the total number of BNS systems formed at metallicity $Z$.

The quantity $M_{\star}(Z)$ depends on the adopted IMF and on the assumed binary fraction, taken here to be 50\%. The function $\mathcal{F}(z',z,Z)$ is expressed as:

\begin{equation}
\mathcal{F}(z',z,Z) =
\frac{\mathcal{N}(z,Z)}{\mathcal{N}_{\mathrm{TOT}}(Z)} \,
p(z'|Z)
\end{equation}

where $\mathcal{N}(z,Z)$ is the number of BNS progenitors formed at metallicity $Z$ that merge at redshift $z$. This quantity is obtained from the delay-time distribution extracted from our \textit{COSMIC} simulations, which constitutes a second key input to \textit{cosmoRate}.

The term $p(z'|Z)$ models the metallicity distribution of star formation and is assumed to follow a log-normal distribution:

\begin{equation}
p(z'|Z) =
\frac{1}{\sqrt{2\pi}\sigma}
\exp\!\left[
-\frac{\left(\log(Z/Z_\odot) - \log(\bar{Z}(z')/Z_\odot)\right)^2}{2\sigma^2}
\right]
\end{equation}

Here, $\bar{Z}(z')$ is the mean metallicity of star formation at redshift $z'$, and $\sigma$ is the intrinsic dispersion of the metallicity distribution in log-space. In this work, we adopt $\sigma = 0.2$.

The cosmological merger rate defined in Section~\ref{eq:Merger rate} is subject to several sources of uncertainty. Beyond the BNS merger efficiency and the delay-time distribution, both derived from \textit{COSMIC} simulations and dependent on population synthesis assumptions, several additional model ingredients remain uncertain. The main uncertainty factor that we decide to explore here is the mean metallicity model. We consider three prescriptions that are intended to span a representative range of uncertainties in merger rate predictions associated with metallicity evolution.

The first two are already implemented in \textit{cosmoRate}. The \textit{linear} prescription is based on observational fits to the evolution of stellar metallicity, combining measurements from the Sloan Digital Sky Survey (SDSS) with metallicity estimates of damped Ly$\alpha$ absorbers (DLAs) derived from absorption-line spectroscopy. It is given by

\begin{equation}
\log\left(\frac{\bar{Z}(z)}{Z_{\odot}}\right) = a + b z
\label{eq: linear met}
\end{equation}
with $a = \log(1.04)$ from \citet{2008MNRAS.383.1439G} and $b = -0.25$ from \citet{2018A&A...611A..76D}. The \textit{MandF2017} prescription is based on observational constraints on the evolution of the gas-phase metallicity of the galaxy population as a function of redshift. It is given by 

\begin{equation}
\log\left(\frac{\bar{Z}(z)}{Z_{\odot}}\right) = a - b \ z^{c}
\label{eq: Manf2017 met}
\end{equation}
with $a = 0.153$, $b = 0.074$, and $c = 1.34$ \citep{2017ApJ...840...39M}.

It should be noted that the parameters for both models originate from observational fits and are defined for a solar metallicity of $Z_{\odot} = 0.02$. In this work, we consistently rescale these relations to a solar metallicity of $Z_{\odot} = 0.014$.

Finally, we consider an alternative redshift–metallicity relation based on the analytical expression from \cite{2016Natur.534..512B}:

\begin{equation}
\bar{Z}(z) =
\frac{y(1 - R)}{\rho_b}
\int_{z}^{z_{\max}}
\frac{10^{0.5}\,\psi(z')}
{H_0 (1 + z') \sqrt{\Omega_m (1 + z')^3 + \Omega_\Lambda}}
\, dz'
\label{eq:Analytical met}
\end{equation}

All these prescriptions lead to different redshift–metallicity relations as discussed in Appendix \ref{sec:met-Z}, see Figure~\ref{fig:metallicity}.

\section{Results} \label{section:Results}
\FloatBarrier

\subsection{Bimodal progenitor mass distribution} 
\label{section:bimodal_distribution}

\begin{figure}
    \centering
    \includegraphics[width=1.0\linewidth]{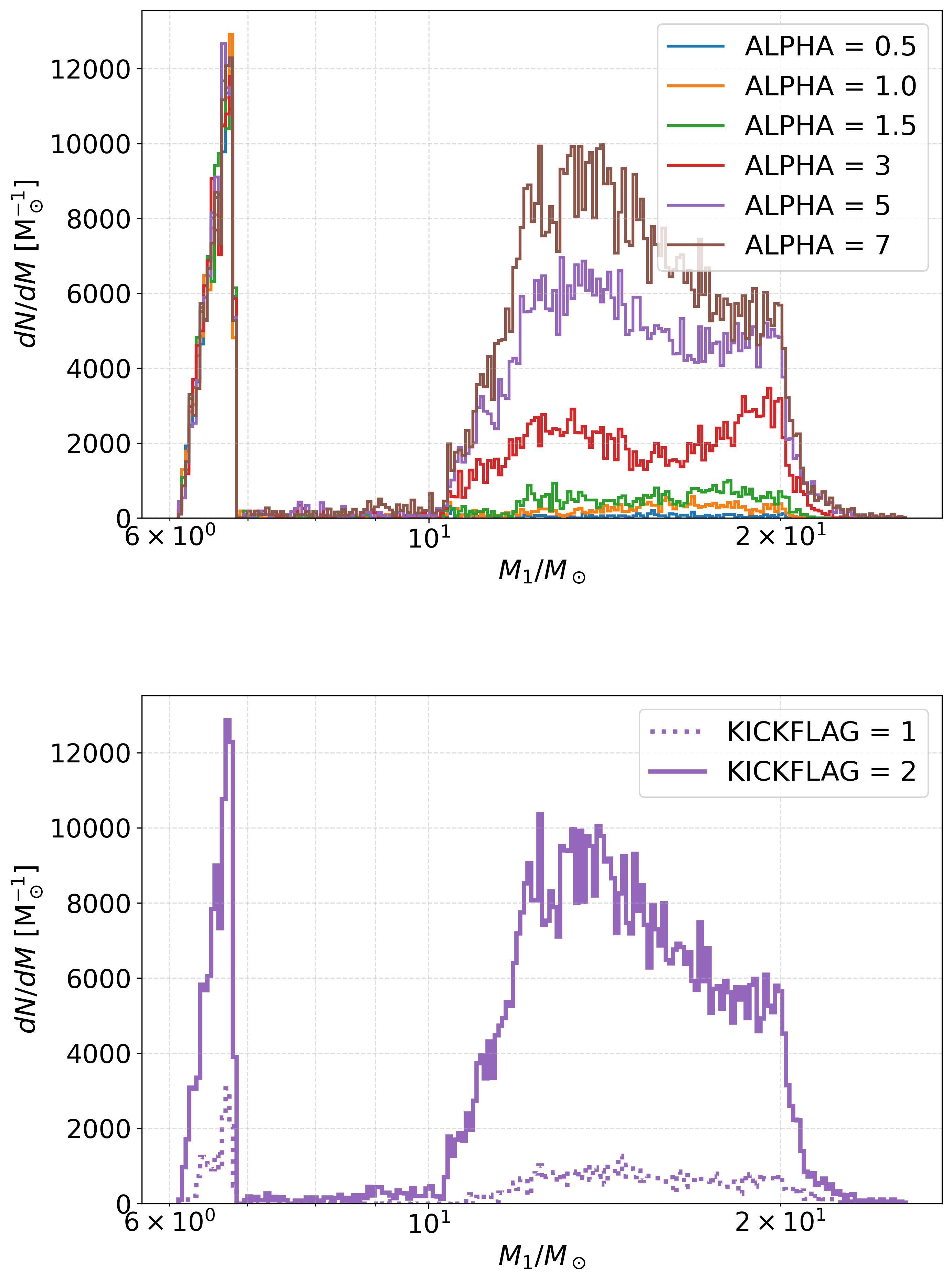}
    \caption{Histograms of the $dN/dM$ distribution as a function of the primary progenitor mass of BNS systems for $Z = 0.001$, \textbf{QCFLAG = 5}, \textbf{CE\_KICKFLAG = 2}, and a \citet{2003PASP..115..763C} IMF. The top panel shows the evolution of the distributions for different \textbf{ALPHA} values with a fixed \textbf{KICKFLAG = 2}. The bottom panel presents similar distributions, this time highlighting the impact of different \textbf{KICKFLAG} prescriptions for a fixed value of \textbf{ALPHA = 5}. Both panels exhibit a bimodal distribution in the progenitor masses of BNS systems and illustrate how this bimodality is affected by these two key parameters.}
    \label{fig:BNS_mass_distribution}
\end{figure}

\begin{figure}
    \centering
    \includegraphics[width=1.0\linewidth]{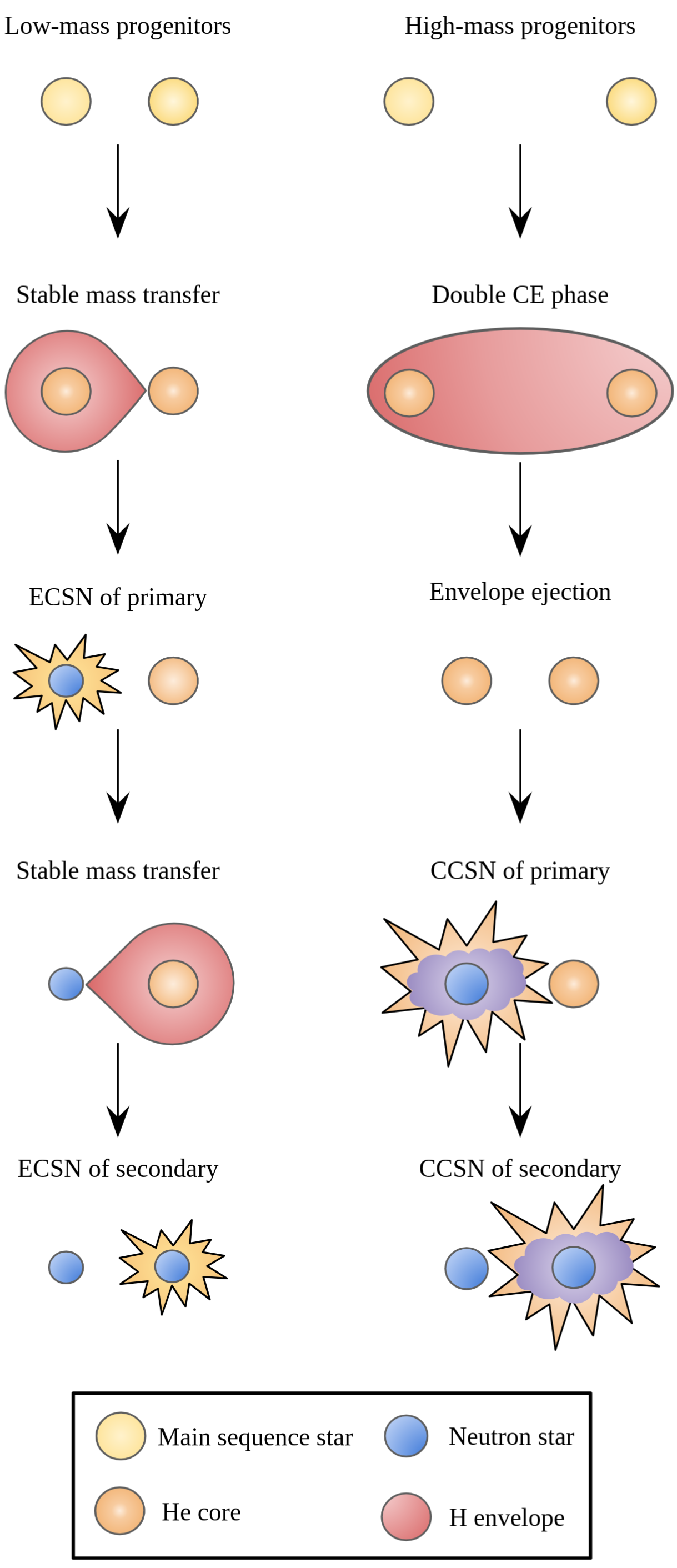}
    \caption{Schematic representation of the typical differences between the low-mass and high-mass BNS progenitor formation channels. Low-mass progenitors usually undergo one or multiple stable mass transfer episodes during their evolution and explode as ECSNe, which reduces the strength of the natal kick and helps to avoid binary disruption. On the contrary, high-mass progenitors typically undergo a CE phase to lose a significant fraction of their mass and enter the NS mass range. They explode as CCSNe, but the reduction in orbital separation due to the CE phase increases the probability of avoiding binary disruption.}
    \label{fig:formation_channels_schematic}
\end{figure}

The first main result of our parameter space exploration is the bimodal distribution of primary progenitor masses across all metallicities and model configurations exhibited by our BNS population, as shown in Figure~\ref{fig:BNS_mass_distribution}. Both the top and bottom panels exhibit a bimodal progenitor-mass distribution, with the two populations separated by a sparsely populated gap between approximately $6.8 \ M_\odot$ and $10\,M_\odot$. 
The bimodality reflects the existence of two dominant formation channels for BNS systems, associated with different physical processes such as distinct supernova types and the presence or absence of a CE phase. Low-mass progenitors usually undergo one or more stable mass transfer episodes during their evolution but do not experience a CE phase. In this case, their orbital separation typically remains relatively large prior to both supernova events. In this configuration the systems that survive binary disruption are the ones that usually undergo an ECSN, characterised by significantly smaller natal kicks (see Section \ref{section: parameter space}). On the contrary, high-mass progenitors often need to undergo a CE phase, or in some cases a double CE phase, to reduce their final pre-supernova mass. In the delayed supernova prescription of \citet{2012ApJ...749...91F}, the amount of fallback depends on both the CO-core mass and the final stellar mass. Therefore, envelope stripping through CE evolution can decrease the fallback contribution and allow some systems that would otherwise form black holes to produce neutron stars instead. These systems are typically above the ECSN mass range and experience CCSN events associated with stronger kicks. However, the strong reduction in orbital separation induced by the CE phase limits the disruptive effect of these kicks. An example of such typical formation pathways between low-mass and high-mass progenitors can be found in Figure ~\ref{fig:formation_channels_schematic}. 

This apparent desert of systems between the two progenitor mass populations can be explained as follows: in this intermediate regime, systems are too light to survive undergoing a CE phase at low separation (see Appendix \ref{Appendix}), yet too massive to experience ECSN events. They therefore retain relatively wide separations while undergoing CCSN events with strong kicks, which explains the low probability of such systems surviving as bound BNS systems. It must be noted, however, that in very specific circumstances some systems can overcome these unfavourable conditions if the natal kick is favourably oriented or if a CE phase occurs very late in the evolution, when only a fraction of the system’s mass is lost. Consequently, a limited number of systems may populate this otherwise underpopulated region, particularly in optimistic models combining weak natal kicks with a high common-envelope survival probability at very low metallicity. Nevertheless, even in these cases the progenitor-mass distribution remains clearly bimodal, although the intermediate region becomes partially populated.  Because the two components of this bimodal distribution are linked to different evolutionary processes, they do not respond in the same way to variations in metallicity or population synthesis parameters. As a result, the relative contribution and evolution of each progenitor population vary significantly throughout our parameter exploration.

In the following sections, we therefore distinguish between low-mass and high-mass primary progenitors, where the primary corresponds to the initially more massive star in the binary system. The transition mass separating these two populations depends on metallicity, since stellar winds influence both the minimum and maximum stellar masses able to produce BNS systems. Taking this dependence into account, we identify low-mass primary progenitors in the range of $\sim 6.5$ to $\sim 9\,M_\odot$, and high-mass primary progenitors in the range of $\sim 10$ to $\sim 30\,M_\odot$ as seen in Figure \ref{fig:BNS_mass_distribution}. \\

This bimodality is strongly affected by two specific parameters: \textbf{ALPHA} and \textbf{KICKFLAG}. Indeed, the relative contribution of low-mass and high-mass progenitors evolves significantly throughout the explored parameter space, as these two parameters strongly influence the dominant BNS formation pathways.

We first focus on the impact of the \textbf{ALPHA} parameter. As shown in the top panel of Figure~\ref{fig:BNS_mass_distribution}, varying \textbf{ALPHA} dramatically changes the relative contribution of low-mass and high-mass progenitors to the total BNS population. As discussed in Section \ref{section: parameter space}, larger values of \textbf{ALPHA} increase the efficiency with which binaries can eject their envelope during the CE phase, thereby enhancing the probability for systems to survive CE evolution without merging. This effect is particularly important for high-mass progenitors. These systems require an efficient mechanism to reduce both their stellar mass in order to produce a neutron star rather than a black hole, and orbital separation prior to the first supernova to increase the probability that the binary survives the strong natal kicks generally associated with massive progenitors. CE evolution is especially effective at simultaneously shrinking the orbit and ejecting the stellar envelope. Consequently, higher \textbf{ALPHA} values allow a larger fraction of high-mass systems to survive the CE phase and evolve into favourable BNS progenitors. As a result, increasing \textbf{ALPHA} significantly enhances the contribution of high-mass progenitors to the overall BNS population. In contrast, low-mass progenitors, which generally do not experience an early CE phase, remain comparatively unaffected by variations in this parameter. Both the increase in the high-mass progenitor population and the relative stability of the low-mass component can be clearly seen in the top panel of Figure~\ref{fig:BNS_mass_distribution}.

\textbf{KICKFLAG} also has a major impact on this bimodal distribution. As described in Section \ref{section: parameter space}, this parameter determines the natal-kick distribution from which supernova kicks are drawn. In this context, \textbf{KICKFLAG = 2} corresponds to an overall less disruptive kick prescription, thereby increasing the probability for binaries to survive both supernovae and eventually form BNS systems.
This effect influences both low-mass and high-mass progenitors, although its impact is significantly stronger for the high-mass population. Low-mass progenitors typically undergo at least one ECSN event. For both \textbf{KICKFLAG = 1} and \textbf{KICKFLAG = 2}, ECSN kicks remain substantially lower than CCSN kicks, allowing a higher fraction of these systems to survive without disruption. The more permissive kick prescription associated with \textbf{KICKFLAG = 2} mainly broadens the range of progenitors receiving relatively weak kicks beyond the narrow ECSN mass regime, thereby increasing the number of surviving low-mass systems.
The situation is markedly different for high-mass progenitors. Under \textbf{KICKFLAG = 1}, CCSN events produce a broad distribution of strong natal kicks, which disrupts a large fraction of binaries before they can form BNS systems. Reducing the kick amplitudes through the \textbf{KICKFLAG = 2} prescription therefore dramatically increases the survival probability of these systems. As a consequence, while both low-mass and high-mass progenitor populations increase for \textbf{KICKFLAG = 2}, the enhancement is considerably stronger for the high-mass channel, allowing it to become the dominant contribution to the total BNS population. This trend is clearly visible in the bottom panel of Figure~\ref{fig:BNS_mass_distribution}.

\FloatBarrier
\subsection{Formation efficiency} \label{section : Formation efficiency}

\begin{figure}
    \centering
    \includegraphics[width=\linewidth]{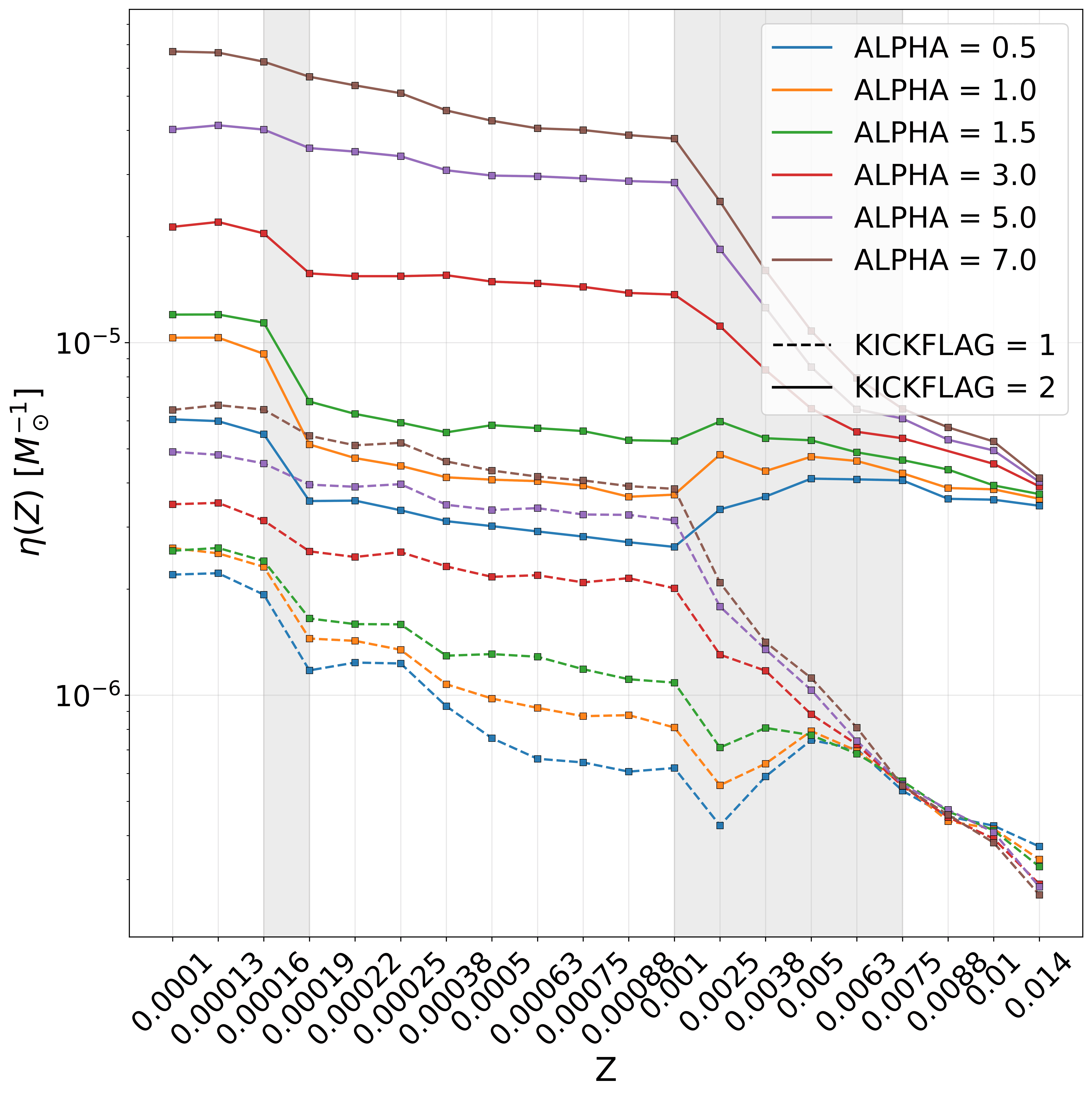}
    \caption{Formation efficiency $\eta(Z)$ as a function of metallicity $Z$ for the selected models with a \textit{Chabrier} IMF, \textit{QCFLAG} = 5, and \textit{CE\_kickflag} = 2. Solid and dashed lines correspond to \textbf{KICK\_FLAG = 2} and \textbf{KICK\_FLAG = 1}, respectively. Different colors indicate different values of the common-envelope efficiency parameter, with $\mathbf{ALPHA = 0.5, 1.0, 1.5, 3, 5, 7}$ shown in blue, orange, green, red, purple, and brown, respectively. The shaded regions highlight the two main metallicity-dependent features discussed in Section~\ref{section : Main parameter effects}. The $Z$ axis is not drawn to scale as each tick corresponds to one of the adopted metallicity bins in order to improve the readability of the figure.}
    \label{fig:formation_efficiency}
\end{figure}

\begin{figure}
    \centering
    \includegraphics[width=\linewidth]{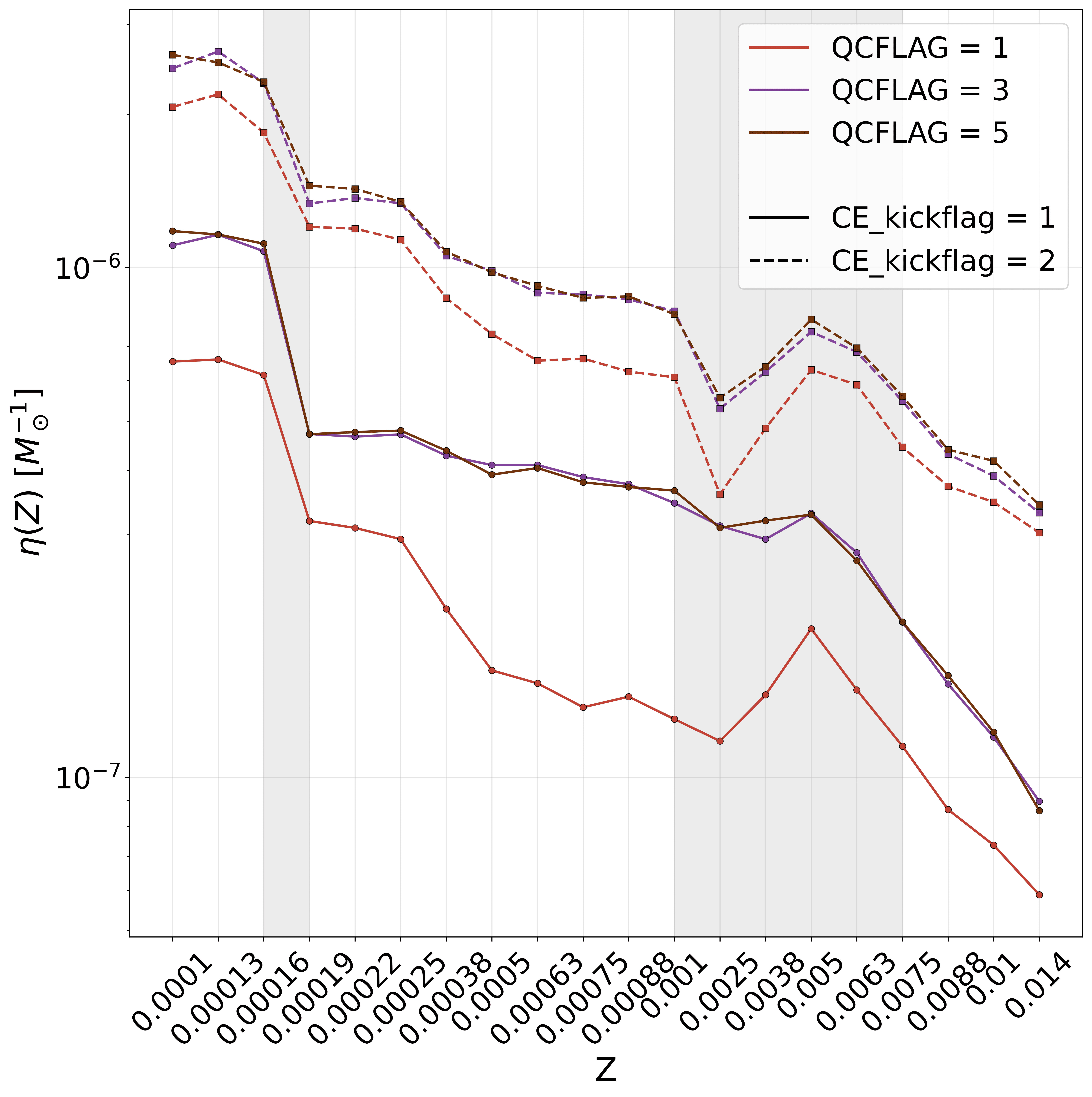}
    \caption{Formation efficiency $\eta(Z)$ as a function of metallicity $Z$ for the selected models with a \textit{Chabrier} IMF, \textit{ALPHA} = 1.0, and \textit{KICK\_FLAG} = 1. Solid and dashed lines correspond respectively to \textbf{CE\_KICKFLAG = 1} and \textbf{CE\_KICKFLAG = 2}. Different colors indicate different values of the mass transfer instability prescription with $\mathbf{QCFLAG = 1, 3, 5}$ shown respectively in red, purple and brown. The shaded regions represent the two main metallicity features discussed in Section~\ref{section : Main parameter effects}.  The $Z$ axis is not drawn to scale as each tick corresponds to one of the adopted metallicity bins in order to improve the readability of the figure.}
    \label{fig:formation_efficiency_qcflag}
\end{figure}

The formation efficiency, $\eta(Z)$, defined in equation~\ref{eq: formation efficiency}, is one of the main quantities that can be directly derived from our simulations. Figures~\ref{fig:formation_efficiency} and \ref{fig:formation_efficiency_qcflag} show $\eta(Z)$ as a function of metallicity for a subset of the models explored in our parameter space. We observe a significant spread in formation efficiencies both across metallicities and between models at fixed metallicity.
We generally observe a decreasing trend of $\eta(Z)$ with increasing metallicity. Stronger stellar winds at high metallicity lead to enhanced mass loss, requiring initially more massive progenitor systems in order to remain within the BNS progenitor mass range. This reduces the formation efficiency, as these progenitors move further away from the maximum of the \citet{2003PASP..115..763C} IMF, resulting in fewer systems being formed.
We note that while the variation of $\eta(Z)$ across metallicity is approximately one order of magnitude between the lowest and highest explored metallicities, the spread in formation efficiency between models at fixed metallicity is significantly larger, reaching two orders of magnitude. This indicates that uncertainties associated with binary evolution prescriptions dominate over the sole effect of metallicity.
Overall, we obtain values of $\eta(Z)$ ranging from a few $10^{-8}\,M_{\odot}^{-1}$ up to $\sim \ 7.10^{-5}\,M_{\odot}^{-1}$ for BNS formation, in agreement with the ranges reported by other population synthesis studies exploring broad parameter spaces \citep{2020ApJ...898..152S,Iorio_2023, 2023MNRAS.526.2210S,Grichener_2023}.

\subsubsection{Parameter study} \label{section : Main parameter effects}

In this section, we will describe how we can hierarchize and characterize the effects of the 4 main explored parameters on the BNSs formation efficiency $\eta(Z)$ and merger proportion $\epsilon(Z)$ as defined in equations \ref{eq: formation efficiency} and \ref{eq: merger fraction}.

The parameter with the strongest impact on the formation rate of the different BNS populations is the \textbf{KICKFLAG}. As discussed in Section \ref{sec:Introduction}, supernova natal kicks are one of the main mechanisms responsible for the disruption of binary systems and therefore for the rarity of BNS formation. The \textbf{KICKFLAG = 1} prescription, described in Section \ref{section: parameter space}, assumes a bimodal distribution of supernova kicks, CCSN and ECSN kicks are drawn from Maxwellian distributions with respectively $\sigma_{\rm CCSN}  \sim 265 \,\mathrm{km\,s^{-1}}$ and $\sigma_{\rm ECSN}  \sim 20 \,\mathrm{km\,s^{-1}}$. In this framework, the survival of a binary system after two successive supernovae requires very specific evolutionary pathways. Either the progenitor must evolve into the ECSN regime before explosion, allowing for a low-kick event, or the binary must become sufficiently tight during its prior evolution to increase its binding energy and thus improve its ability to survive high-velocity CCSN kicks. In both cases, these scenarios occupy a restricted region of the initial binary parameter space and rely on fine-tuned binary interactions. In contrast, the \textbf{KICKFLAG = 2} prescription introduces a smoother transition between CCSN and ECSN regimes, depending on the pre-supernova core mass and the ejecta mass. This formulation allows a significant fraction of CCSN events to receive lower natal kicks, thereby increasing the probability of binary survival. As a consequence, the formation rate of BNS populations is significantly higher for the \textbf{KICK\_FLAG = 2} models compared to \textbf{KICKFLAG = 1}, as shown in Fig.~\ref{fig:formation_efficiency}.\\

The second most impactful parameter is the \textbf{ALPHA} parameter. In our models, we observe the expected trend that the formation efficiency increases with increasing \textbf{ALPHA}, as shown in Fig.~\ref{fig:formation_efficiency}. This behaviour can be understood through two complementary effects. First, as discussed in Section \ref{section:bimodal_distribution}, high-mass progenitors require a CE phase in order to reduce both their fallback mass during CCSN and orbital separation and produce a neutron star rather than a black hole while surviving binary disruption by successive SN.  A higher \textbf{ALPHA} value therefore allows a larger fraction of systems to successfully survive a CE phase without merging, increasing the number of binaries that can reach these favorable configurations.
Systems that do not strictly require a CE phase to form BNSs (typically lower-mass progenitors that can directly evolve toward the ECSN regime) can also be affected in a less important way . A larger \textbf{ALPHA} extends the region of initial orbital parameter space in which a CE episode can occur without leading to a merger, thereby allowing specific systems with initially tighter separations to survive and evolve into BNSs rather than merging prematurely. Overall, \textbf{ALPHA} is therefore a key parameter, as it simultaneously affects the formation efficiency and the merger fraction (see annex \ref{fig:merger-proportion}) through its influence on both high- and low-mass progenitor channels in a complex manner.\\

The two remaining parameters have a lesser effect on the formation efficiency, as shown in Figure \ref{fig:formation_efficiency_qcflag}. We first examine the impact of the \textbf{CE\_KICKFLAG} parameter. Models with \textbf{CE\_KICKFLAG = 2} are systematically more efficient at forming BNS than their \textbf{CE\_KICKFLAG = 1} counterparts. This trend is expected, since, as defined in Section \ref{section: parameter space}, \textbf{CE\_KICKFLAG = 2} allows systems undergoing at least one supernova during a CE phase to account for their post-CE separation and mass when computing the kick velocity drawn from the natal kick distribution, as well as its effect on the binary orbit and potential disruption. Because both the separation and the component masses are reduced after the CE phase, the natal kick is less likely to disrupt the binary system. This is particularly relevant for \textbf{KICK\_FLAG = 2} models, in which the kick magnitude is drawn proportionally to the mass of the collapsing star and the ejected envelope, resulting in systematically lower kicks, and hence a lower probability of disruption. However, the significance of this parameter depends on several factors. For \textbf{CE\_KICKFLAG} to have a strong impact, the population must be dominated by formation channels in which a supernova occurs during a CE phase. Such channels require a very specific region of the initial parameter space and are not always the most efficient at forming BNS across all parameter combinations or metallicity bins. Furthermore, the effect of \textbf{CE\_KICKFLAG} is naturally reduced in \textbf{KICK\_FLAG = 2} models, where natal kicks are already intrinsically lower, thereby reducing the overall disruption pressure on the population.\\

Finally, the \textbf{QCFLAG} parameter has a notable impact on the population. As shown in Figure \ref{fig:formation_efficiency_qcflag}, while \textbf{QCFLAG = 3} and \textbf{QCFLAG = 5} models exhibit very similar formation efficiencies, \textbf{QCFLAG = 1} models are systematically less efficient across all metallicity bins. As described in Section \ref{section: parameter space}, the differences between these models lie in the mass ratio stability criterion applied when a RLOF occurs. The key distinction between the two groups is that \textbf{QCFLAG = 1} treats RLOF from naked helium stars as highly unstable, triggering a CE phase in most cases. In contrast, the other models adopt a significantly more permissive stability criterion, with \textbf{QCFLAG = 5} treating all such RLOF episodes as dynamically stable. These stripped stellar phases are critical to BNS formation, as they represent the evolutionary continuation of stars that have already undergone an earlier CE phase. If mass transfer from the naked helium star is again deemed unstable, the binary separation at that stage is already sufficiently reduced that a second CE phase will typically lead to a merger rather than a surviving tight binary. \textbf{QCFLAG = 1} models therefore suppress a significant fraction of BNS progenitors by channeling these systems into mergers during the second mass transfer episode. The small difference between \textbf{QCFLAG = 3} and \textbf{QCFLAG = 5} can be understood by noting that the stability threshold adopted in \textbf{QCFLAG = 3}, while finite, is sufficiently permissive that it is rarely reached by the progenitor systems that contribute to BNS formation, making it effectively equivalent to the always-stable prescription of \textbf{QCFLAG = 5}. As with the other parameters discussed above, the overall impact of \textbf{QCFLAG} depends on the fraction of BNS progenitors that undergo a second mass transfer episode following an initial CE phase, and is therefore sensitive to both the metallicity and the broader parameter configuration of the model.

\FloatBarrier
\subsubsection{Metallicity-dependent features}
\label{section : main features with metallicity}

\begin{figure}
    \centering
    \includegraphics[width=1.0\linewidth]{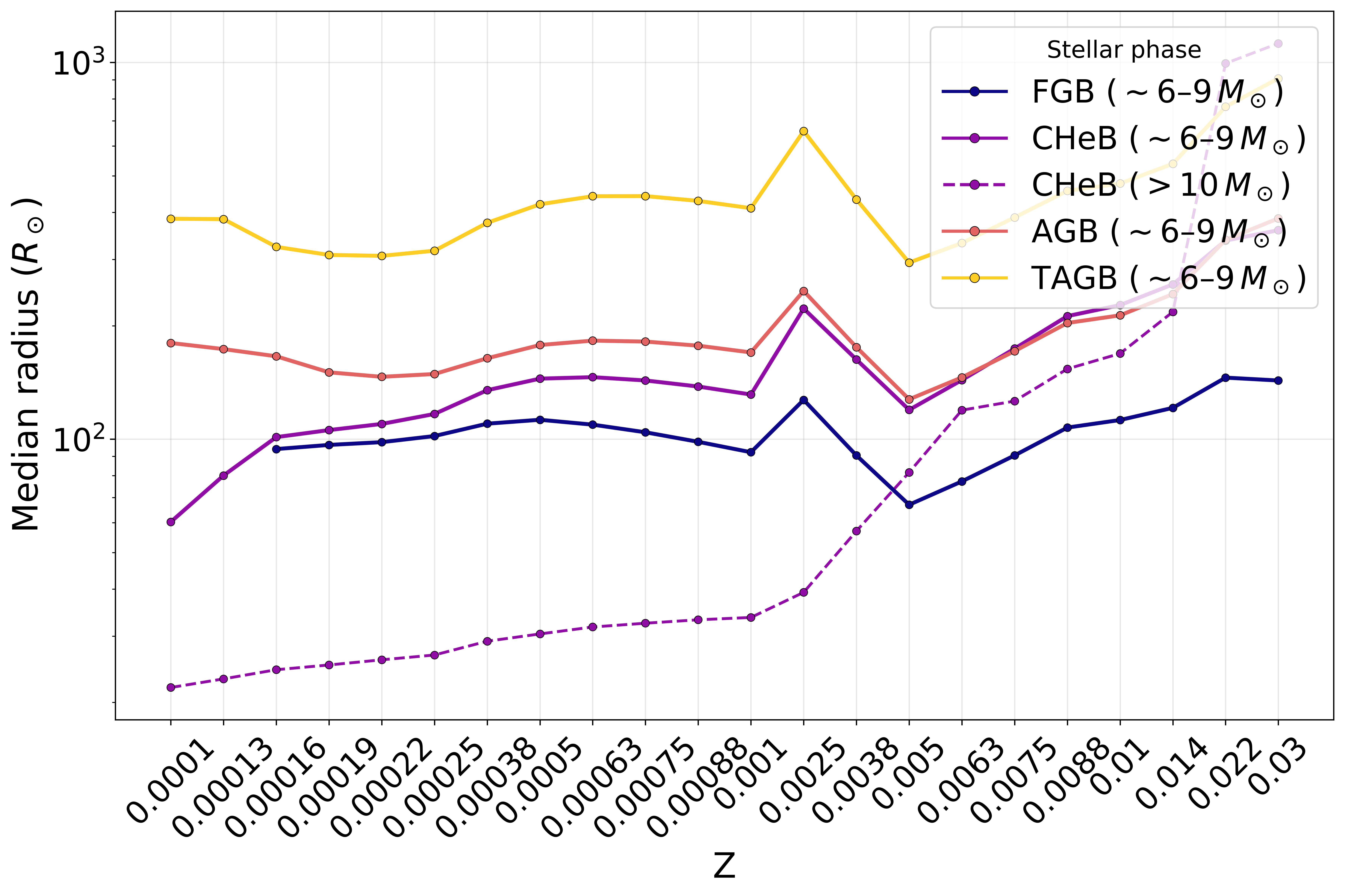}
    \caption{Mean stellar radii across different evolutionary phases and metallicity bins for the model \textit{IMF = Chabrier, QCFLAG = 5, CE\_KICKFLAG = 2, ALPHA = 7.0, KICKFLAG = 2}, shown separately for low- and high-mass progenitor stars. The mean radii during the First Giant Branch (FGB), Core Helium Burning (CHeB), Asymptotic Giant Branch (AGB), and Thermally Pulsing Asymptotic Giant Branch (TPAGB) phases are represented in dark blue, purple, pink, and yellow, respectively. Solid lines correspond to stars with masses between $6$ and $9\,M_\odot$, while dashed lines correspond to stars with masses above $10\,M_\odot$. For high-mass progenitors, only the CHeB phase is shown, as the other evolutionary phases are skipped in this mass range.}
    \label{fig:radii Z}
\end{figure}

\begin{figure*}[t]
    \centering
    \includegraphics[width=1.0\textwidth]{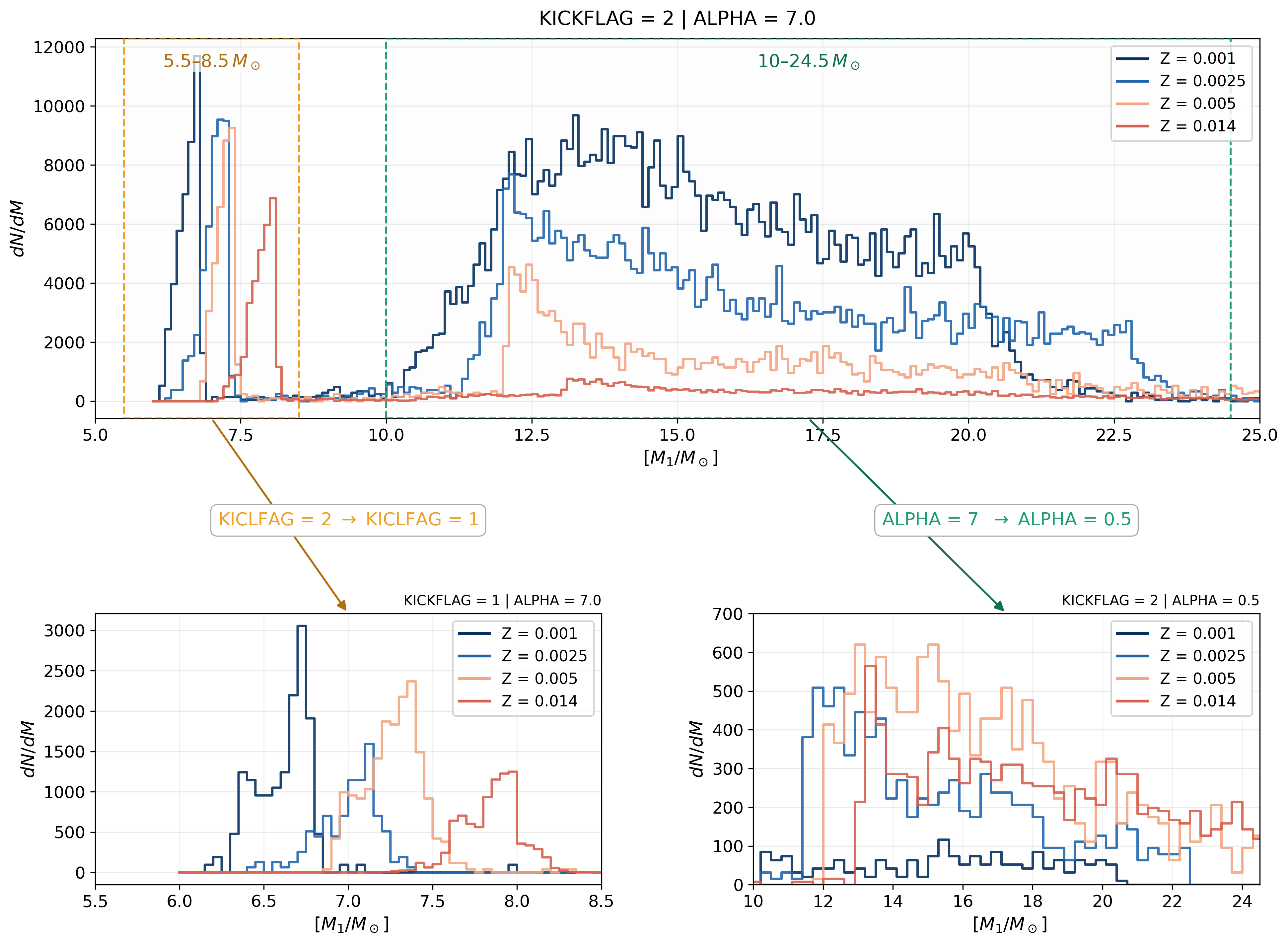}
    \caption{Histogrammes of the $dN/dM$ distribution as a function of the primary progenitor mass of BNS systems for $Z = 0.001$, $0.0025$, $0.005$, and $0.014$. All panels are extracted from models computed with the \citet{2003PASP..115..763C} IMF, \textbf{QCFLAG = 5}, and \textbf{CE\_KICKFLAG = 2}.
    The top panel shows the full histogram for a reference model with \textbf{KICKFLAG = 2} and \textbf{ALPHA = 7.0}. Two regions exhibiting distinct behaviours with respect to variations of these parameters are highlighted. The orange region corresponds to low-mass progenitors in the range $5.5$--$8.5 , M_\odot$. The bottom-right panel shows this same region for an otherwise identical model but with \textbf{KICKFLAG = 1}, highlighting the impact of this parameter on the low-mass progenitor distribution.
    The green region corresponds to high-mass progenitors in the range $10$--$24.5 , M_\odot$. The bottom-left panel shows the same region for an otherwise identical model but with \textbf{ALPHA = 0.5}, illustrating the effect of this parameter on the high-mass progenitor distribution.
    Overall, this figure highlights the bimodal progenitor-mass distribution obtained in our models, as well as its evolution with metallicity and with the two key parameters discussed in Section \ref{section : Main parameter effects}.}
    
    \label{fig:histogramms}
\end{figure*}

As discussed in Section~\ref{section : Main parameter effects}, the parameter space explored in this work produces a wide range of BNS formation efficiencies, reflecting the strong impact of the adopted stellar evolution prescriptions. Despite these differences, the metallicity dependence exhibits two characteristic features that are present across all models.

The first such feature is a sharp decrease in BNS formation efficiency between $Z = 0.00016$ and $Z = 0.00019$. This behaviour is again linked to the metallicity dependence of single-star evolution. High-mass progenitors systematically bypass the FGB, whereas low-mass progenitors generally evolve through this phase. This transition is associated with the onset of non-degenerate core helium ignition, which allows stars to skip the FGB entirely \citep{2000MNRAS.315..543H}. The threshold mass for this transition increases with metallicity.

In \textit{COSMIC}, this transition is modeled following \citet{2000MNRAS.315..543H} via the relation

\begin{equation}
    M_{\rm FGB} = \frac{13.048\,(Z/0.02)^{0.06}}
    {1 + 0.0012\,(0.02/Z)^{1.27}},
    \label{eq:MFGB}
\end{equation}

At $Z \simeq 0.0016$, the value of $M_{\rm FGB}$ is lower than the typical masses of the low-mass progenitors in our population (see Appendix~\ref{section:FGB skip}). Above this metallicity threshold, a large fraction of these progenitors therefore no longer bypasses the FGB phase. This stellar phase skip has important consequences for binary evolution. First, it suppresses the large radial expansion associated with the FGB, delaying the onset of mass transfer until later evolutionary stages, such as the Core Helium Burning (CHeB), Asymptotic Giant Branch (AGB), or Thermally Pulsing Asymptotic Giant Branch (TPAGB) phases. Second, it alters the stability of mass transfer. For all \textbf{QCFLAG} prescriptions considered in this work, the critical mass-ratio threshold for unstable mass transfer is significantly lower during the FGB phase than during these later evolutionary stages. Systems that would have entered a CE phase if mass transfer had occurred on the FGB can instead undergo stable mass transfer when the FGB is skipped. As a result, binaries with relatively short initial separations are more likely to avoid an early CE phase, which would otherwise lead to a merger before the formation of two neutron stars. Systems that skip the FGB can therefore continue their evolution and eventually form BNS systems, explaining the increase in formation efficiency observed below $Z \simeq 0.0016$. Above this metallicity, the BNS formation efficiency decreases because part of the initial parameter space leading to successful BNS formation is no longer accessible. In particular, low-mass binaries with relatively short initial separations and high mass ratios are more likely to initiate a CE phase during the FGB and subsequently merge before BNS formation.
Because this mechanism primarily affects low-mass progenitors, its impact depends on their relative contribution to the population. As discussed in Section~\ref{section:bimodal_distribution}, the fraction of high-mass progenitors increases with \textbf{ALPHA} which decreases the amplitude of this feature, consistent with the trends observed in Figs.~\ref{fig:formation_efficiency} and \ref{fig:formation_efficiency_qcflag}.\\

The second major feature is a non-monotonic variation in BNS formation efficiency between $Z = 0.001$ and the solar metallicity ($Z_\odot = 0.014$). This behaviour is visible across all model sets although with different amplitudes depending on the adopted prescriptions (see Figs.~\ref{fig:formation_efficiency} and \ref{fig:formation_efficiency_qcflag}). It is directly linked to the metallicity dependence of single-star evolution, in particular the evolution of stellar radii during advanced evolutionary phases. In this range, stellar radii can vary significantly during stages such as the FGB, CHeB, AGB, and TPAGB. The impact of these variations depends strongly on both stellar mass and evolutionary stage. High-mass progenitors follow a markedly different evolutionary path from lower-mass systems. In the \emph{SSE COSMIC} prescriptions, these stars bypass several late evolutionary phases such as the FGB and TPAGB, so their radius evolution is primarily governed by the CHeB phase. Figure~\ref{fig:radii Z} illustrates the metallicity dependence of stellar radii for different progenitor-mass ranges and evolutionary phases. For progenitors in the low-mass range, all late evolutionary phases exhibit a strong increase in radius at $Z = 0.0025$, followed by a decrease up to $Z = 0.005$, and then another increase toward solar metallicity. By contrast, high-mass progenitors show a smoother and more monotonic evolution, since only the CHeB phase contributes significantly to the evolution in radii. In this regime, the CHeB radius increases gradually with metallicity, with a much steeper trend between $Z = 0.001$ and $Z_\odot$.

We first examine the impact of the evolution of high-mass progenitor radii as a function of metallicity. High-mass progenitors typically require a CE phase during the CHeB stage in order to form a BNS system. If the CE is triggered too early, the system is more likely to merge during the CE phase or to lose too much mass to remain a viable progenitor. Since stellar radii increase with metallicity, the range of initial separations leading to a CE at the appropriate evolutionary stage shifts toward larger values. The efficiency of envelope ejection, parametrized by the \textbf{ALPHA} parameter in our models, is crucial for determining the fate of these systems. For high \textbf{ALPHA} values, systems with relatively small initial separations can still efficiently survive the CE phase. However, as metallicity increases and stellar radii expand, a progressively larger fraction of systems is removed from the viable region of parameter space, leading to a strong decrease in formation efficiency. This effect is further amplified by the initial separation distribution of \citet{2012Sci...337..444S}, which favours low initial separations. This behaviour is visible in the top panel of Figure~\ref{fig:histogramms}, where high \textbf{ALPHA} models rapidly lose systems with increasing metallicity for the high-mass progenitor channel.

In contrast, models with low \textbf{ALPHA} exhibit a different behaviour. In this regime, the increase in stellar radii with metallicity still shifts the initial separation to higher values. However, at low metallicity, this shift is initially beneficial because it moves the interacting systems away from the very shortest separations, where virtually no binary can survive a CE event with such a low \textbf{ALPHA}. As a result, an increasing fraction of systems survives the CE phase, leading to a rise in the BNS formation efficiency. As the metallicity increases further, the required initial separations continue to grow. However, the newly selected systems already belong to a region of the initial parameter space where survival through the CE is possible. Increasing the separation therefore no longer improves the survival probability, but instead shifts the progenitor population towards increasingly wider binaries. Since the initial orbital-separation distribution of \citet{2012Sci...337..444S} decreases towards large separations, progressively fewer binaries populate this region, resulting in a decline of the formation efficiency. While the increase in stellar radius shifts the range of initial separations leading to viable CE events towards larger values, its impact is determined by the competition between two effects. Larger initial separations allow a larger fraction of systems to undergo a CE phase without merging. This increase in survivability depends on \textbf{ALPHA}. However, this gain is partially counterbalanced by the lower number of systems available at these larger initial separations, as predicted by the initial separation distribution \citep{2012Sci...337..444S}. This competition naturally produces the turnover observed in the low-\textbf{ALPHA} models. This behaviour is illustrated in the bottom-right panel of Figure~\ref{fig:histogramms}, which shows the high-mass region of the $dN/dM$ distribution for a model identical to that shown in the top panel but with \textbf{ALPHA = 0.5}. In this case, the formation efficiency increases with metallicity up to $Z=0.005$, before decreasing at $Z=0.014$. As shown in Figure~\ref{fig:formation_efficiency}, the turnover shifts towards lower metallicities as \textbf{ALPHA} increases for \textbf{KICKFLAG = 2} models, and disappears entirely for sufficiently large values of \textbf{ALPHA}, as illustrated by the \textbf{ALPHA = 3}, \textbf{KICKFLAG = 2} model.

We now examine the impact of the evolution of low-mass progenitor radii as a function of metallicity. Most low-mass progenitors either undergo CE evolution only at relatively late evolutionary stages, after significant mass loss, or avoid CE episodes altogether. Therefore, the \textbf{ALPHA} parameter does not play an important role in the evolution of these systems. Instead, their fate is strongly affected by the adopted natal-kick prescription. At a given orbital separation, the sharp increase in stellar radius at $Z = 0.0025$ can trigger binary interactions at earlier evolutionary stages, either through stable mass transfer or through an early CE episode that is not always survivable. Such interactions may also drive the accretor out of the ECSN progenitor mass range, as a consequence of extended or enhanced mass-transfer phases. In the case of \textbf{KICKFLAG = 1}, low-mass progenitors must remain within a relatively narrow ECSN mass range to avoid binary disruption by strong CCSN kicks. In this configuration, the evolution of the population is therefore tightly coupled to the metallicity-driven evolution of stellar radii through its effect on the range of viable initial separations. This behaviour is visible in the bottom left panel of Figure~\ref{fig:histogramms}, where the number of low-mass progenitor systems decreases sharply between $Z = 0.001$ and $Z = 0.0025$, following the rapid increase in stellar radii shown in Figure~\ref{fig:radii Z}. The population then increases again at $Z = 0.005$, before decreasing once more at $Z = 0.014$, consistent with the metallicity dependence of stellar radii in the low-mass regime. 

In \textbf{KICKFLAG = 2} models, however, this effect is significantly mitigated. In this prescription, natal kicks depend on the progenitor mass (see Section~\ref{section: parameter space}) and remain relatively low over the relevant mass range. As a result, systems that move slightly outside the strict ECSN channel can still survive supernova events due to the reduced kick amplitudes. Moreover, if an early mass-transfer or CE phase occurs, the resulting binary may evolve toward a smaller pre-supernova orbital separation, provided that the adopted \textbf{ALPHA} value allows the system to survive the interaction. Since the natal kicks remain comparatively unchanged while the orbital separation decreases, the probability of surviving both supernova events can even increase. Consequently, the strong increase in stellar radius at $Z = 0.0025$ may, in some cases, enhance the formation of low-mass BNS progenitors in \textbf{KICKFLAG = 2} models. This behaviour is visible in the top panel of Figure~\ref{fig:histogramms}, where a smoother decrease in the number of systems is observed between $Z = 0.001$ and $Z = 0.0025$ compared to the corresponding \textbf{KICKFLAG = 1} model. The population remains approximately stable between $Z = 0.0025$ and $Z = 0.005$, highlighting the compensating effects discussed above, before decreasing again at $Z = 0.014$. To further demonstrate the distinct impact of \textbf{ALPHA} and \textbf{KICKFLAG} on the two progenitor-mass regimes, we provide in Appendix~\ref{section:Complementary histogramms} a complementary version of Figure~\ref{fig:histogramms}. While Figure~\ref{fig:histogramms} highlights the changes induced by variations of these two parameters in their respective dominant mass regimes, the complementary figure shows that these effects are not interchangeable: variations of \textbf{KICKFLAG} have a negligible impact on the high-mass progenitor population, while variations of \textbf{ALPHA} have little effect on the low-mass progenitor population. Together, these two figures demonstrate the distinct roles played by \textbf{ALPHA} and \textbf{KICKFLAG} in shaping the metallicity-dependent evolution of the low- and high-mass BNS progenitor populations. \\

By combining the metallicity dependence of both low-mass and high-mass progenitors discussed above, the relative dominance of these two formation channels described in Section \ref{section:bimodal_distribution}, and the overall decrease in formation efficiency with metallicity driven by stronger stellar winds discussed in Section \ref{section:bimodal_distribution}, we can explain the different evolutionary behaviours observed in Figures~\ref{fig:formation_efficiency} and \ref{fig:formation_efficiency_qcflag}. The metallicity evolution of each model is determined by two main factors: the dominant progenitor population (low-mass or high-mass) and the way this population responds to metallicity for a given parameter configuration. For example, the solid brown curve in Figure~\ref{fig:formation_efficiency} represents an extreme case dominated by high-mass progenitors with a large \textbf{ALPHA} value. As expected, the feature between $Z = 0.0016$ and $Z = 0.0019$ is almost absent, while the formation efficiency rapidly decreases between $Z = 0.001$ and $Z = 0.0075$, reflecting the strong metallicity dependence of the high-mass channel. Conversely, the blue dashed curve in Figure~\ref{fig:formation_efficiency} is dominated by low-mass progenitors in a \textbf{KICKFLAG = 1} configuration. In this case, a pronounced increase is observed between $Z = 0.0016$ and $Z = 0.0019$, followed by a more complex evolution between $Z = 0.001$ and $Z = 0.0075$, consistent with the metallicity dependence of late-stage stellar radii discussed previously. Intermediate models can simultaneously exhibit features associated with both progenitor populations, either because the relative contribution of each channel evolves with metallicity or because the rapid evolution of a subdominant population can still significantly affect the total formation efficiency. \\

It should be emphasized that both of the features discussed in this section strongly depend on the adopted SSE prescriptions and, in particular, on the interpolation methods used in the stellar models. As discussed in \citet{2022eas..conf.1076R} and \citet{2022MNRAS.516.5816X}, the evolution of stellar radii with mass and metallicity remains highly uncertain, making the second feature especially sensitive to the underlying interpolations. In particular, the abrupt radius evolution observed for low-mass progenitors around $Z \simeq 0.0025$ is suggested by \citet{2022eas..conf.1076R} to be an interpolation artefact that may not be physical. Nevertheless, stellar radii are still expected to increase with metallicity during the relevant evolutionary phases \citep{Choi2016}. Therefore, even if the sharp behaviour observed for low-mass progenitors were to disappear in more detailed stellar models, the qualitative trend identified for high-mass progenitors should remain similar. As a consequence, the overall dependence of the BNS formation efficiency on metallicity and \textbf{ALPHA} is expected to persist, although the slopes and transition metallicities may differ.

Similarly, the drop in formation efficiency at $Z\simeq 0.00016$ strongly depends on the adopted prescription for $M_{\rm FGB}$, which itself remains uncertain and is actively debated \citep{2000MNRAS.315..543H, Choi2016}. Therefore, the metallicity at which this drop occurs, as well as its amplitude and sharpness could vary significantly depending on the adopted stellar evolution model.

\subsubsection{IMF models} \label{section : top_heavy IMF}

\begin{figure}
    \centering
    \includegraphics[width=1.0\linewidth]{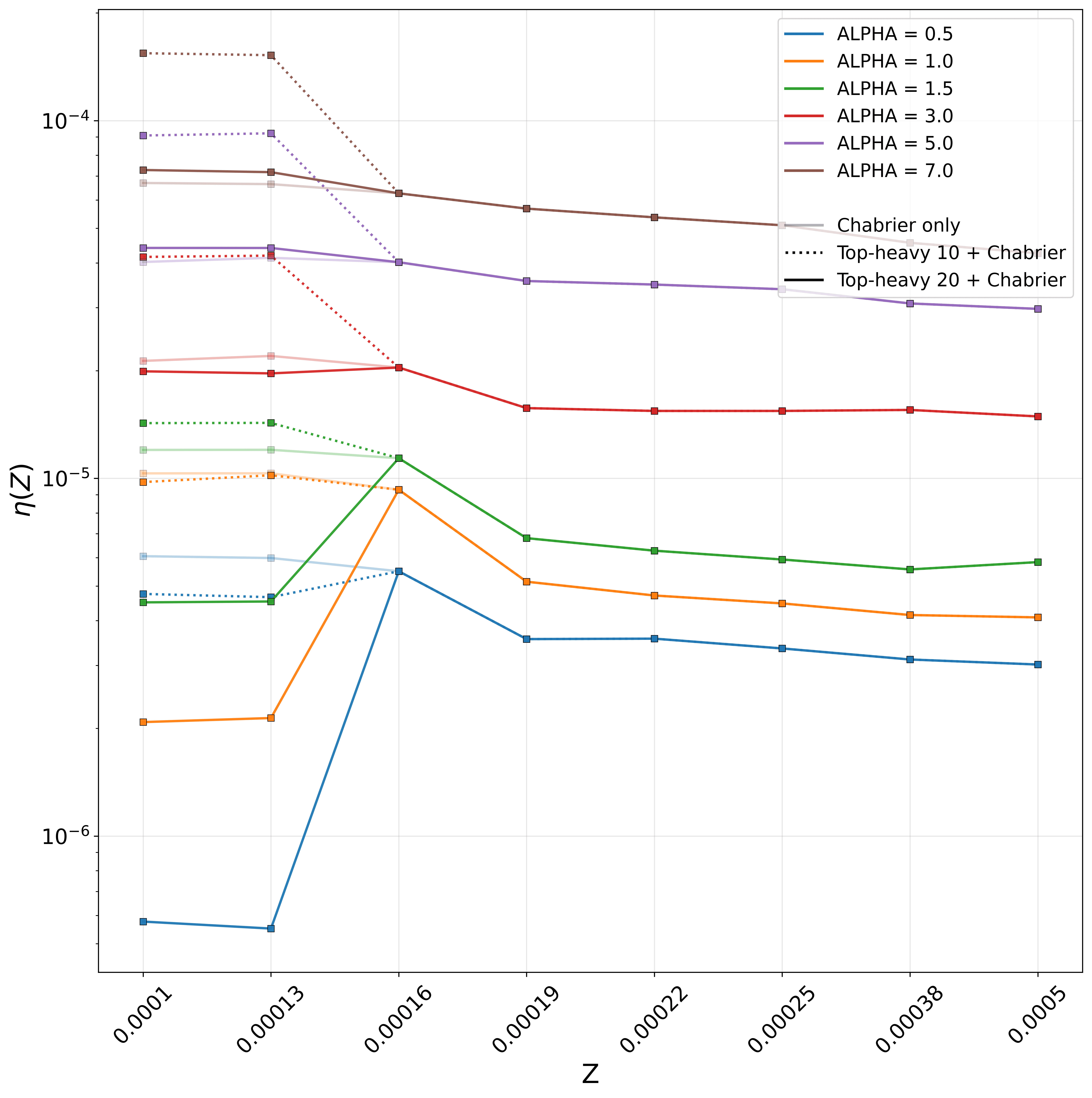}
    \caption{BNS formation efficiency as a function of metallicity for models with \textbf{QCFLAG = 5}, \textbf{CE\_KICKFLAG = 2}, and \textbf{KICKFLAG = 2}. Solid lines show the formation efficiency obtained assuming a \textit{Chabrier} IMF down to $Z = 0.00016$. For the two lowest metallicity bins, $Z = 0.00013$ and $Z = 0.0001$, the models instead adopt top-heavy IMFs with $m_{\mathrm{cut}} = 20,M_\odot$ and $m_{\mathrm{cut}} = 10,M_\odot$, respectively, labelled \textit{Top-heavy 20} and \textit{Top-heavy 10}. These models are represented by solid and dashed lines, respectively. The lower-opacity lines show the corresponding \textit{Chabrier} IMF models at the same metallicities for comparison. Different colors correspond to different values of \textbf{ALPHA}.}
    \label{fig:top_heavy comparison}
\end{figure}

As discussed in Section \ref{section: parameter space}, we simulated BNS populations using both a \textit{Chabrier} IMF \citep{2003PASP..115..763C} and a \textit{top-heavy} IMF with $m_{\rm cut} = 20 \ M_\odot$ for the two lowest metallicity bins, $Z = 0.00013$ and $Z = 0.0001$. The goal of this approach is to explore an extreme upper-limit scenario for the IMF at very low metallicity and to quantify its impact on the BNS formation efficiency.

Figure~\ref{fig:top_heavy comparison} compares the resulting formation efficiencies obtained with both IMFs. The solid lines correspond to the \textit{top-heavy} $m_{\rm cut} = 20 \ M_\odot$  IMF models, while the lower-opacity lines show the corresponding \textit{Chabrier} IMF models at the same metallicities for comparison. An interesting trend emerges from this comparison. Relative to the \textit{Chabrier} IMF, the \textit{top-heavy 20} IMF produces lower formation efficiencies for small values of \textbf{ALPHA}. This behaviour is expected, since low-\textbf{ALPHA} models strongly suppress the contribution of high-mass progenitors to the BNS population. As \textbf{ALPHA} increases, the relative efficiency of the \textit{top-heavy} IMF also increases.
This behaviour can be directly linked to the fraction of high-mass progenitors contributing to BNS formation in the corresponding \textit{Chabrier} IMF models (see Section \ref{section:bimodal_distribution}. Changing the IMF does not fundamentally modify the dominant BNS formation channels, but rather changes their relative weight within the total population. Consequently, models in which the formation efficiency is primarily driven by low-mass progenitors become less efficient when adopting an IMF that favors high-mass stars.

We also note that, even for the highest \textbf{ALPHA} values explored here, the \textit{top-heavy 20} IMF only marginally increases the BNS formation efficiency at very low metallicity. This suggests that the adopted IMF remains too strongly weighted toward progenitor masses that do not efficiently produce BNS systems.
To further investigate this point, we simulated, for a specific subset of models, the BNS population assuming a \textit{top-heavy} IMF with $m_{\mathrm{cut}} = 10\,M_\odot$. The resulting formation efficiencies are shown as dashed lines in Fig.~\ref{fig:top_heavy comparison}. For these models, we find that while the \textit{top-heavy 10} IMF consistently yields higher formation efficiencies than the \textit{top-heavy 20} case, it also exceeds the \citet{2003PASP..115..763C} IMF models already for lower \textbf{ALPHA} values (starting from \textbf{ALPHA} = 1.5), whereas the \textit{top-heavy 20} models only begin to show a similar trend at higher \textbf{ALPHA}. In the most optimistic cases, the \textit{top-heavy 10} IMF reaches formation efficiencies approximately a factor of two higher than the corresponding \textit{Chabrier} models.
We therefore find that, while the extreme \textit{top-heavy 20} IMF considered here can be detrimental to BNS formation, since its peak lies outside the progenitor mass range associated with the most efficient formation channels, a \textit{top-heavy} IMF peaking at lower characteristic masses can instead significantly enhance the formation efficiency under favorable conditions.

\FloatBarrier
\subsection{Cosmological merger rates} \label{section:merger_rates_results}

\subsubsection{Relative impact of stellar evolution parameters}

\begin{figure*}[t]
    \centering
    \includegraphics[width=0.97\textwidth]{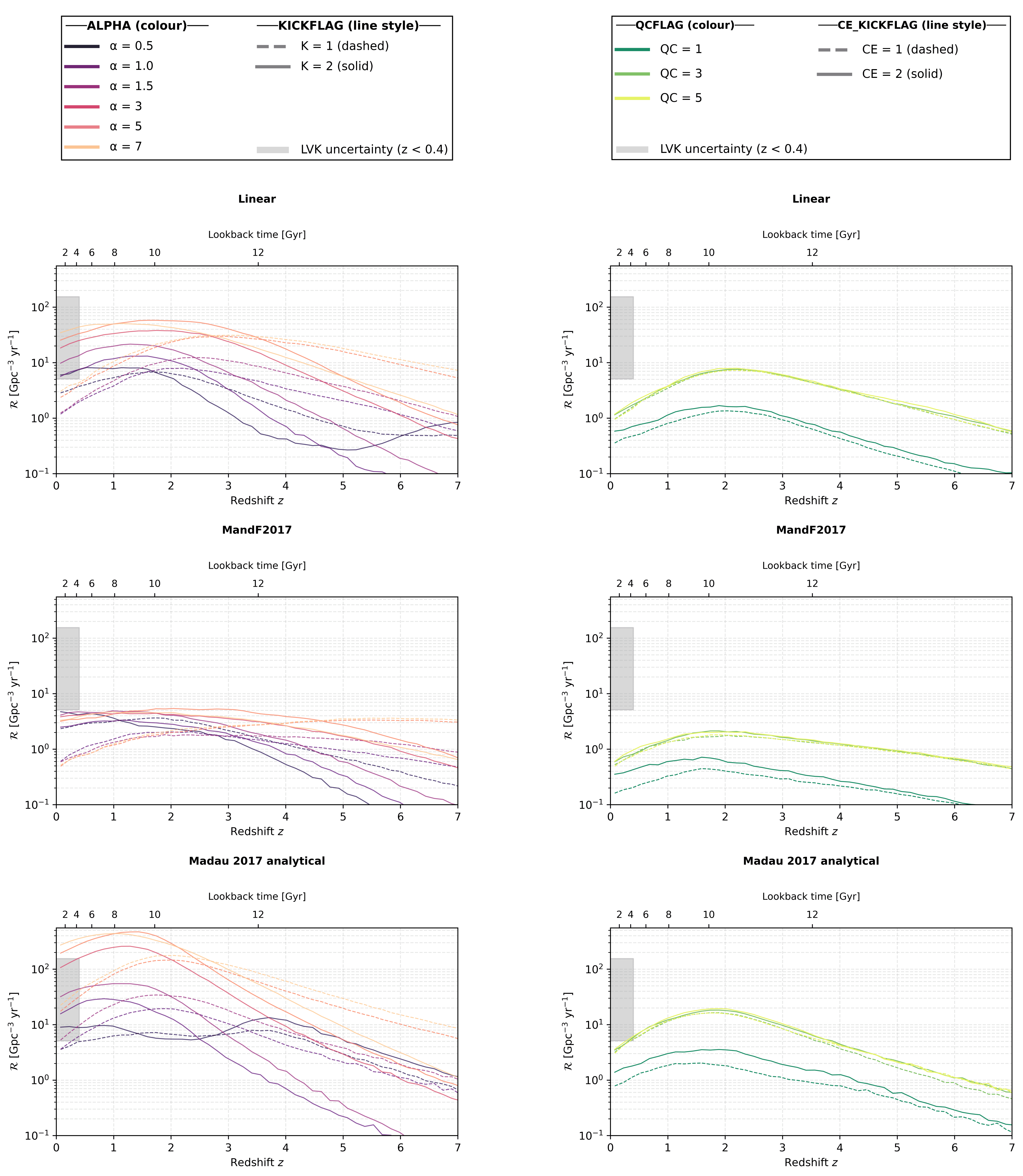}
    \caption{Cosmological merger rate as a function of redshift for three different methods described in Section~\ref{section : merger rate calculations}. For all three approaches, we adopt the SFR from \citet{2017ApJ...840...39M} and implement different prescriptions for the evolution of the mean metallicity. The \textbf{linear} and \textbf{MandF2017} models are already implemented in \textit{cosmo\_Rate} and are based on observational fits from \citet{2008MNRAS.383.1439G}, \citet{2018A&A...611A..76D}, and \citet{2017ApJ...840...39M}, respectively. The \textbf{Madau 2017 analytical} model relies on an analytical expression for $\bar{Z}(z)$ derived in \citet{2016Natur.534..512B}. The plots in the left column are obtained for the \citet{2003PASP..115..763C} IMF, with \textit{QCFLAG = 5} and \textit{CE\_KICKFLAG = 2}. For these models, the \textbf{ALPHA} values are represented by colors and the \textbf{KICKFLAG} by line styles. The plots in the right column are obtained for the \citet{2003PASP..115..763C} IMF, with \textit{ALPHA = 1} and \textit{KICKFLAG = 1}. The colors correspond to different \textbf{QCFLAG} values, while the line styles correspond to \textbf{CE\_KICKFLAG} variations. The grey shaded area represents the LVK uncertainty band for detected BNS mergers, as reported by \citet{theligoscientificcollaboration2026gwtc50populationpropertiesmerging}. Its extent along the redshift axis is shown for visual guidance only and does not represent the redshift range over which the constraint is measured.}
    \label{fig:merger_rates_special}
\end{figure*}

Figure~\ref{fig:merger_rates_special} shows the cosmological merger rates obtained for different parameter combinations, as well as for the three $\bar{Z}(z)$ prescriptions described in Section~\ref{section : merger rate calculations}. This figure explores the same combinations of models as those presented in Section~\ref{section : Formation efficiency} and shown in Figs.~\ref{fig:formation_efficiency} and \ref{fig:formation_efficiency_qcflag}. The left column illustrates the effects of the \textbf{KICKFLAG} and \textbf{ALPHA} parameters, while the right column explores the effects of the \textbf{QCFLAG} and \textbf{CE\_KICKFLAG} parameters. The complete set of models explored in this work is shown in the Appendix in Fig.~\ref{fig:merger rates}.

We find several features common to all three metallicity-redshift prescriptions. The most obvious one is the difference between the \textbf{KICKFLAG = 1} and \textbf{KICKFLAG = 2} models. As shown in Fig.~\ref{fig:merger_rates_special}, the latter models consistently yield higher merger rates at low redshifts (typically $z \lesssim 5 $) across all metallicity models. This is an expected result. As discussed in Section~\ref{section : Main parameter effects}, the lower natal kick prescription (\textbf{KICKFLAG = 2}) allows a larger fraction of systems to survive CCSN events, thereby increasing the formation efficiency of BNS. This effect is particularly important for high-mass progenitors, which are the dominant channel for CCSN events. These systems typically require CE evolution to sufficiently reduce the system's  mass and enter the NS mass range at the end of their evolution. In this context, CE events play a crucial role in shrinking orbital separations and enabling the formation of systems that merge within a Hubble time. Consequently, \textbf{KICKFLAG = 2} favours the production of more high-mass progenitors undergoing CE phases, leading to a higher number of rapidly merging binaries that dominate the cosmological merger rate.

\begin{figure}
    \centering
    \includegraphics[width=1.0\linewidth]{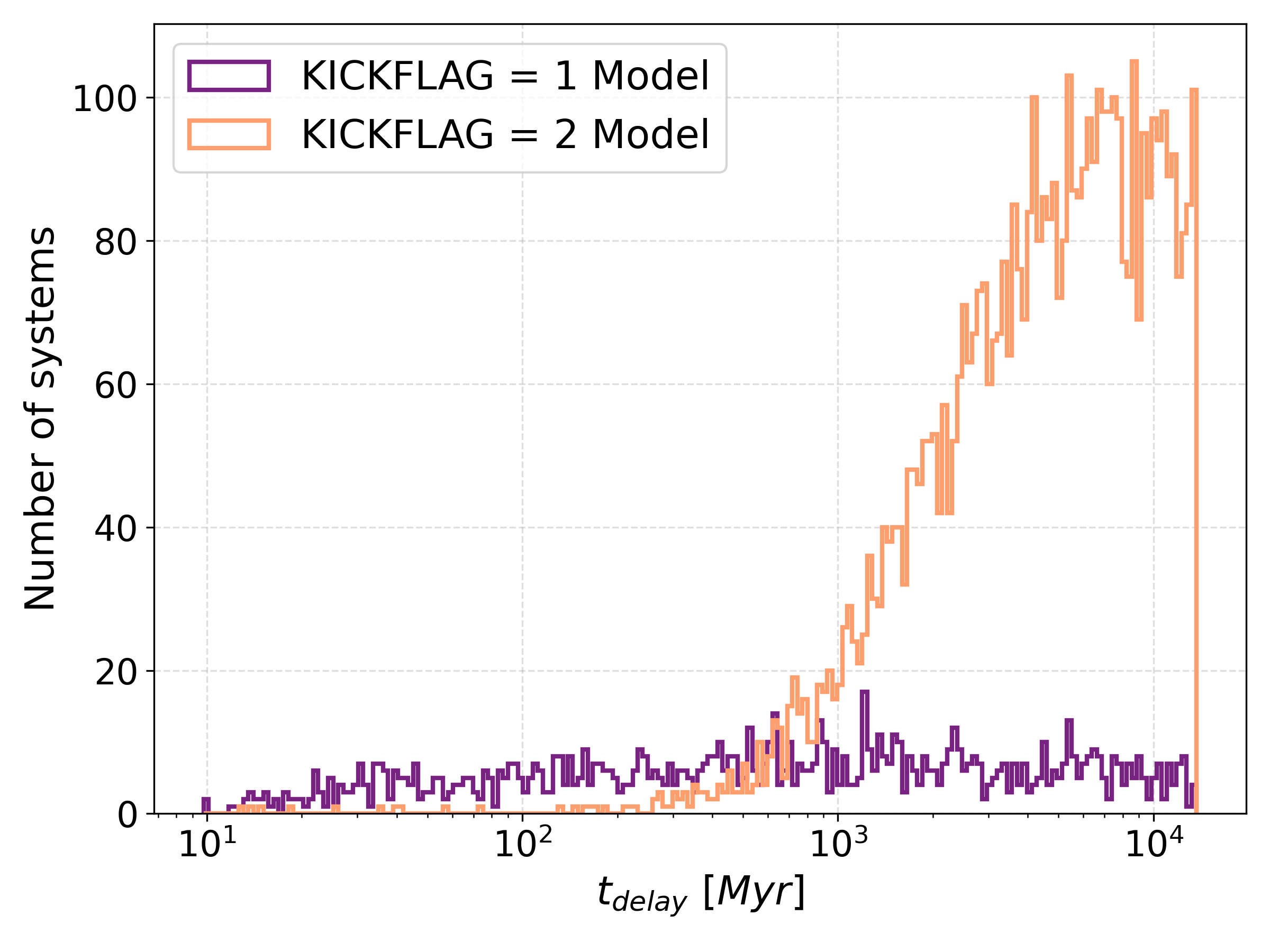}
\caption{Merger time-delay distributions of systems merging within a Hubble time for two representative models at $Z = 0.001$. Both models adopt \textit{ALPHA = 7}, \textit{QCFLAG = 5}, and \textit{CE\_KICKFLAG = 2}. The purple curve corresponds to \textit{KICKFLAG = 1}, while the orange curve corresponds to \textit{KICKFLAG = 2}. This comparison highlights that, although \textbf{KICKFLAG = 2} models produce a larger number of systems merging within a Hubble time, \textbf{KICKFLAG = 1} models yield a significant fraction of systems with very short merger time delays ($\lesssim 3 \times 10^2$ Myr), whereas in \textbf{KICKFLAG = 2} models the merger-time distribution is almost entirely shifted to values above this threshold.}
    \label{fig:merger_time_delays}
\end{figure}

We also note a second systematic difference between these model families. The \textbf{KICKFLAG = 1} models reach their merger rate maximum at systematically higher redshift than the \textbf{KICKFLAG = 2} models. This behaviour can also be understood in terms of the underlying physical assumptions. In the high-kick scenario (\textbf{KICKFLAG = 1}), the larger supernova kicks reduce the overall formation efficiency but preferentially select systems that experience strong orbital tightening in rare favourable configurations. In order for a binary to remain bound, the kick must be precisely oriented; otherwise, the system is disrupted due to the large imparted momentum. This introduces a survival bias, where only systems that benefit from a fortuitous kick geometry survive both supernova events. When such conditions are met, these strong kicks can significantly reduce the orbital separation more efficiently than in the lower-kick (\textbf{KICKFLAG = 2}) case. This behaviour can be seen in Fig.~\ref{fig:merger_time_delays}, which shows the merger time-delay distributions for two representative models. Although the orange curve, corresponding to a \textbf{KICKFLAG = 2} model, produces a larger number of systems merging within a Hubble time, the purple curve, corresponding to a \textbf{KICKFLAG = 1} model, exhibits a merger time-delay distribution that is significantly shifted towards shorter merger times.

The difference in merger delay-time distributions naturally explains the shift in the redshift of the peak merger rate. Both models exhibit a peak at a redshift close to the maximum of the \citet{2017ApJ...840...39M} SFR  at $z \sim 2$, with their offset reflecting their respective merger time-delay distributions. The \textbf{KICKFLAG = 1} models peak closer to this redshift due to their shorter typical delay times, whereas the \textbf{KICKFLAG = 2} models peak at lower redshift as a consequence of their longer merger delay-time distribution.\\

We observe another key feature associated with the \textbf{ALPHA} parameter. As described in Section~\ref{section : Main parameter effects}, increasing \textbf{ALPHA} leads to a larger fraction of high-mass progenitors, which naturally explains the tendency, across metallicity models, for higher \textbf{ALPHA} values to be associated with higher local merger rates. However, a straightforward interpretation of the behaviour of each \textbf{ALPHA}-dependent model family is non-trivial, since \textbf{ALPHA} simultaneously affects the formation and merger efficiencies, as well as the post-CE orbital separations.

For instance, for very low \textbf{ALPHA}, the formation efficiency can be significantly reduced. Nevertheless, systems that successfully survive the CE phase are left with much tighter orbits, resulting in substantially shorter merger time delays. This explains the second merger rate bump observed in Figure \ref{fig:merger rates} for \textbf{ALPHA = 0.5} at high redshift.\footnote{The cosmological merger rate can be enhanced if the resulting delay times align with epochs of high SFR. For example, in Fig.~\ref{fig:merger rates}, the model with \textbf{ALPHA = 0.5}, \textbf{KICKFLAG = 2}, \textbf{QCFLAG = 5}, and \textbf{CE\_KICKFLAG = 2}, under a \textit{Madau 2017 analytical} prescription, exhibits an interesting feature: a secondary peak in the merger rate at $z \sim 4$, which is comparable to or even exceeds its first peak at $z \sim 2$ associated with the SFR maximum. This behaviour can be explained by the enhanced formation efficiency at low metallicity as shown in Figure \ref{fig:formation_efficiency}, combined with the suppression of the FGB channel (Section~\ref{section : Main parameter effects}).In this regime, the low \textbf{ALPHA} value produces extremely compact post-CE systems, resulting in very short merger time delays that allow BNS systems to merge at a high redshift.}
For the other parameters, the trends follow more clearly the formation efficiency hierarchy described in Section~\ref{section : Main parameter effects}. In particular, \textbf{QCFLAG = 3} and \textbf{QCFLAG = 5} allow a larger fraction of systems to survive early CE phases, leading to more efficient merger production compared to \textbf{QCFLAG = 1} models. Similarly, \textbf{CE\_KICKFLAG = 2} improves the survival probability of systems undergoing SN events during or after CE phases, resulting in higher merger rates compared to \textbf{CE\_KICKFLAG = 1}.

\FloatBarrier

\subsubsection{Metallicity-redshift relations}

We now focus on the differences in the BNS merger-rate predictions obtained with the three metallicity-redshift relations introduced in Section~\ref{section : merger rate calculations}.

The cosmological merger rates associated with these metallicity prescriptions are primarily determined by two coupled effects: the metallicity dependence of the BNS formation efficiency and merger fraction, and the correspondence between the adopted metallicity-redshift relation and the chosen cosmic SFR history.

As discussed in Section \ref{section: parameter space}, our models span a metallicity range from $Z = Z_\odot = 0.014$ down to $Z = 10^{-4}$. Over this interval, we observe a significant increase in the BNS formation efficiency $\eta(Z)$ (eq. \ref{eq: formation efficiency}) as shown in Figures \ref{fig:formation_efficiency} and \ref{fig:formation_efficiency_qcflag}. In contrast, the merger fraction $\epsilon(Z)$ (eq. \ref{eq: merger fraction}) remains of the same order of magnitude across this metallicity range (see \ref{fig:merger-proportion}). Consequently, the metallicity-redshift prescriptions that maximize the cosmological merger rate are those for which the metallicities associated with the highest formation efficiencies are reached close to the peak of the cosmic SFR, which is $z\sim 2$ for the \citet{2017ApJ...840...39M} SFR adopted in this work.

\FloatBarrier

The three main trends observed in Fig.~\ref{fig:merger_rates_special} can therefore be understood from the metallicity evolution with redshift (see Appendix~\ref{sec:met-Z}). For the \textit{MandF2017} prescription, the metallicity of $\log_{10}(Z/Z_\odot) \sim -1$, which corresponds to high BNS formation efficiency (see Figure~\ref{fig:formation_efficiency}), is reached only at very high redshift, around $z \sim 8.8$ At such redshifts, the \citet{2017ApJ...840...39M} SFR has already declined by roughly one order of magnitude relative to its maximum. As a consequence, the resulting cosmological merger rates are the lowest among our three metallicity-redhisft prescriptions. In this case, the merger rate is largely dominated by progenitors formed at relatively high metallicity around $z \sim 2-3$.

In contrast, the \textit{Madau 2017 analytical} prescription reaches $\log_{10}(Z/Z_\odot) \sim -1$ at $z \sim 1.9$, and $\log_{10}(Z/Z_\odot) \sim -2$ at $z \sim 4$. This alignment between low-metallicity environments and the peak of the cosmic SFR allows progenitors with the highest formation efficiencies to form during the epoch of strongest star-formation activity. This naturally explains why this prescription produces both the highest overall cosmological merger rates and the largest merger-rate peaks. In this regime, the merger population is strongly dominated by low-metallicity progenitors.

The \textit{linear} prescription reaches $\log_{10}(Z/Z_\odot) \sim -1$ at $z \sim 4.5$ and $\log_{10}(Z/Z_\odot) \sim -2$ at $z \sim 8.5$. This intermediate behaviour leads to a merger population composed of both relatively high- and low-metallicity progenitors, resulting in cosmological merger rates that lie between those obtained for the \textit{MandF2017} and \textit{Madau 2017 analytical} prescriptions.

Finally, we compare the local merger rates obtained across our explored parameter space with the BNS merger-rate interval of $5.1$--$154.7 \ \mathrm{Gpc}^{-3} \ \mathrm{yr}^{-1}$ inferred by the LIGO-Virgo-KAGRA Collaboration \citep{theligoscientificcollaboration2026gwtc50populationpropertiesmerging}. We first observe that, for the \textit{MandF2017} prescription, none of our models falls within the LVK uncertainty band, as their predicted cosmological merger rates are systematically lower than the observational constraints. The models that come closest to the lower edge of the observational uncertainty range are characterized by \textbf{KICKFLAG = 2}, \textbf{QCFLAG} values in \textbf{[3,5]}, and low \textbf{ALPHA} values. This indicates that, within the parameter space explored in this study, the \textit{MandF2017} metallicity prescription tends to underpredict the observed BNS merger rate and is therefore marginally disfavoured compared to the other prescriptions considered.

For the \textit{Madau 2017 analytical} prescription, the explored models instead span nearly the entire LVK uncertainty range, with a large fraction of the parameter space remaining compatible with current observational constraints. In this configuration, models combining the lowest \textbf{ALPHA} values with \textbf{QCFLAG = 1} can already be disfavoured by current observations.
The most merger-rate-efficient models, characterised by \textbf{ALPHA = [5,7]}, \textbf{KICKFLAG = 2}, \textbf{QCFLAG = [3,5]}, and \textbf{CE\_KICKFLAG = 2}, actually exceed the LVK merger rates based on GWTC-5.0 reported by \citet{theligoscientificcollaboration2026gwtc50populationpropertiesmerging} and are therefore inconsistent with the observational data.
The remaining models cover a broad range of merger rates; however, the current observational uncertainties do not yet allow us to efficiently discriminate between the key parameters discussed in Section~\ref{section : Main parameter effects}, such as the different \textbf{KICKFLAG} prescriptions or \textbf{ALPHA} values below and above unity.

For the  \textit{linear} prescription, we find an intermediate behaviour. A significant number of explored models remain compatible with the LVK uncertainty interval, while not reaching its upper end. In this case, although several models remain observationally degenerate, the current constraints already allow partial discrimination between some model parameters. In particular, only models with \textbf{KICKFLAG = 2} are able to reproduce the observed local merger-rate estimates.

\section{Discussion}
\label{section:discussion}

In this work we have explored the parameter space of BNS formation models using the population synthesis code \emph{COSMIC}. We have specifically focused on four parameters that encode the onset of unstable mass transfer, the efficiency of CE ejection, natal kicks, and their treatment during the common envelope phase. This study revealed several noteworthy features.

First, all our populations exhibit a bimodal distribution of BNS progenitors, across all parameters combinations and metallicities. As discussed in Section \ref{section:bimodal_distribution}, this bimodality reflects the two main formation channels identified in this work. The first channel involves low-mass progenitors, generally avoiding early CE events while evolving toward the ECSN mass range, thereby limiting the disruptive effects of SN kicks. The second channel is associated with high-mass progenitors and relies on early CE phases to both reduce the orbital separation and strip the stars sufficiently to enter the NS mass range, increasing the probability of surviving the stronger CCSN kicks. These two broad categories can be related to the compact-object binary formation channels described in \citet{Iorio_2023}: channels II and 0 largely correspond to our low-mass progenitor pathways, whereas channels III and IV are associated with our high-mass progenitor channels. Channel I can be found in both regimes, although with characteristics specific to each progenitor population. We therefore argue that this bimodality is physically motivated. This feature may still be absent from specific parameter combinations that either suppress one progenitor population or make the other sufficiently dominant to mask its presence, as observed for the high \textit{ALPHA} models discussed in Section \ref{section:bimodal_distribution}.

A second important result is the large spread in formation efficiencies and merger rates obtained across our model grid, spanning several orders of magnitude between the most pessimistic and optimistic scenarios. This outcome is expected, as the explored parameters remain among the most uncertain aspects of binary evolution and are believed to play a fundamental role in BNS formation. The purpose of this study was therefore to characterize the effect of these uncertainties on direct observables, such as the cosmological BNS merger rate, and indirect quantities, such as the formation efficiencies discussed in Sections \ref{section : Formation efficiency} and \ref{section:merger_rates_results}. Despite this important variation, our results remain broadly consistent with previous population-synthesis studies. In particular, \citet{Santoliquido_2020}, using the \textit{MOBSE} \citep{2018MNRAS.480.2011G,2018MNRAS.474.2959G} population synthesis code, reported comparable merger efficiencies, although with somewhat higher merger rates. Similar conclusions were obtained by \citet{Iorio_2023} and \citet{2023MNRAS.526.2210S}, both using the \textit{Sevn} \citep{2019MNRAS.485..889S} population synthesis code, as well as by \citet{Grichener_2023} with \textit{COMPAS} \citep{2022ApJS..258...34R,2025ApJS..280...43T}. While these studies generally find formation and merger efficiencies comparable to ours, their predicted cosmological merger rates are often higher, in some cases by up to an order of magnitude relative to our most optimistic models. We attribute this discrepancy primarily to two factors. First, earlier LVK data releases favored higher BNS merger rates than current estimates, leading previous studies to preferentially adopt combinations of lower kicks and higher \textit{ALPHA} values. Second, for the \textit{linear} and \textit{MandF2017} metallicity-redshift relations, described in Section \ref{section : merger rate calculations} we applied a correction to account for our adopted solar metallicity value of $Z_\odot = 0.014$. This correction lowers the predicted cosmological merger rate, since the original relations favored metallicities exceeding our maximum solar metallicity at low redshift.

We have also identified the relative importance of the studied parameters for BNS formation. We find that the dominant parameters are the efficiency of common envelope ejection, parametrized by \textit{ALPHA} and SN natal kick prescription, encoded by the \textit{KICKFLAG} parameter in our models. This finding is in accordance with previous studies \citep[e.g.][]{Santoliquido_2020,Kruckow_2021,Vigna_G_mez_2025} that explored the importance of the CE phase and natal kicks for compact binary formation.

More specifically, we find that BNS formation efficiency systematically increases with increasing \textit{ALPHA}, as discussed in Section \ref{section : Main parameter effects}, although this trend appears to flatten at the highest values explored, suggesting a possible saturation regime in which further increases in \textit{ALPHA} would have a reduced impact on the efficiency. The cosmological merger rate shows a more complex dependence on \textit{ALPHA}, since this parameter simultaneously affects the number of formed BNS systems, their delay-time distribution, and their coupling with the adopted star formation history and metallicity–redshift relation.
At the same time, it remains one of the largest sources of uncertainty in binary evolution models, as emphasized in works such as \citet{Ricker_2018}, which highlight the strong sensitivity of compact binary merger predictions to CE physics. Other formalisms beyond the $\alpha-\lambda$ prescription have been proposed, such as the $\alpha-\gamma$ formalism \citep{nelemans2000reconstructingevolutiondoublehelium,Nelemans2005}. Instead of assuming that the envelope is ejected through the release of orbital energy during a spiral-in phase, this approach describes the orbital evolution through the loss of angular momentum associated with mass ejection.

For the \textit{KICKFLAG} prescriptions, we consistently find higher formation efficiencies and cosmological merger rates for the \textit{KICKFLAG = 2} prescription compared to \textit{KICKFLAG = 1}, across all model combinations and metallicity bins considered. In addition, the merger rate exhibits a shift in its redshift peak, indicating that natal kick assumptions not only regulate binary survival but also reshape the cosmic evolution of merger activity.

In addition to these well-established trends, we identify two characteristic features in the metallicity dependence of the BNS formation efficiency that are present across all simulated models and can be directly linked to specific stages of stellar evolution. The first is a decrease in the formation efficiency between $Z = 0.00016$ and $Z = 0.0019$, which is associated with the onset of the FGB skip for low-mass progenitors at $Z < 0.0019$. The second feature is a non-monotonic evolution of the formation efficiency between $Z = 0.001$ and $Z = 0.0075$. All models exhibit significant changes in the slope of the formation efficiency within this metallicity range. The amplitude and shape of this variation strongly depend on the dominant progenitor population (low- or high-mass progenitors) and on the combination of \textit{ALPHA} and \textit{KICKFLAG} parameters. As a result, the formation efficiency can display different behaviours, including increasing or decreasing trends and the presence of local maxima or minima. Both recurring features may be linked to potential observables in the cosmological merger rate and in the mass distribution of observed mergers. However, their interpretation remains strongly dependent on the adopted SSE method. The magnitude, slope, and even the existence of such features vary significantly between population-synthesis codes, reflecting the current lack of convergence in SSE prescriptions.

Finally, we find a significant degeneracy between models when considering the cosmological merger rate. Although each explored parameter has a measurable impact on this quantity, different combinations of prescriptions can produce remarkably similar merger-rate evolutions, both in terms of their redshift dependence and their local merger-rate density. As discussed in Section \ref{section:merger_rates_results}, our ability to break this degeneracy is currently limited by two major factors: the still modest BNS merger sample available from the latest LVK release \citep{theligoscientificcollaboration2026gwtc50populationpropertiesmerging} and the substantial uncertainties affecting the metallicity-redshift relation used in the cosmological merger rate calculations. Under these conditions, it is possible to disfavour some extreme parameter combinations, particularly those associated with very low BNS yields, such as \textit{KICKFLAG = 1} combined with low \textit{ALPHA} values. However, current BNS observational merger data do not yet allow a clear preference for a specific parameter value or prescription.

This limited discriminating power for key SSE and BSE parameters remains one of the main challenges for future population-synthesis studies. Recent investigations of BBH populations have already revealed tensions between population-synthesis predictions and LVK observations, as discussed in \citet{2026arXiv260602725B}. Although these discrepancies appear to depend on both the population-synthesis code considered and the specific combination of model parameters adopted \citep{broekgaarden2026lowerratesclaimsbinary}, such studies point to shortcomings in our current treatment of several critical evolutionary processes, including CE evolution and SN natal kicks. They also underscore the importance of systematically exploring the parameter space to better quantify the uncertainties arising from both the underlying evolutionary parameters and the specific assumptions and implementations of different population-synthesis codes. 

Further progress in this area will require a substantially larger sample of observed BNS mergers. This will be enabled by next-generation gravitational-wave observatories, such as the Einstein Telescope and Cosmic Explorer, which are expected to detect up to $\sim 10^5$ BNS mergers per year and measure the BNS merger rate out to redshifts of $z \sim 2-3$ \citep{2022ApJ...941..208I}. The resulting population will provide significantly tighter constraints on the properties of BNS populations and their formation channels \citep{2019ApJ...878L..13S,2026A&A...711A.211T}. In addition, improved observational constraints on the cosmic metallicity evolution, together with more realistic modelling of galaxy evolution and chemical enrichment \citep{2019MNRAS.490.3740N,2022MNRAS.514.1315B} will be essential for reducing the systematic uncertainties in predictions of the cosmological BNS merger rate.

Nevertheless, even with current observational data, part of this degeneracy may be alleviated by jointly considering multiple observational probes of the same underlying BNS population. For example, population-synthesis models can simultaneously predict the properties and abundance of Galactic double neutron stars, the contribution of BNS mergers to the r-process enrichment of the Milky Way, and the cosmological merger rate. Combining these complementary observables could provide significantly stronger constraints on the underlying evolutionary prescriptions than any individual probe alone. Exploring such a multi-messenger and multi-observable approach is the subject of ongoing work and will be investigated in future studies.

\begin{acknowledgements}

This project was provided with computing HPC and storage resources by GENCI at IDRIS thanks to grants AD010416184 and AD010416184R1 on the supercomputer Jean Zay’s CSL partition. This work made use of the Infinity cluster, hosted by the Institut d’Astrophysique de Paris, on which some of the simulations were run and post-processed. The authors warmly thank S. Rouberol for running it smoothly. The authors thank M. Zevin and K. Breivik for useful discussions and the entire COSMIC team for making the code publicly available. This study is part of the EVOLVE project (ANR-24-CE31-3262-01, PI: I. Dvorkin), funded by the Agence Nationale de la Recherche (ANR). 

\end{acknowledgements}

\FloatBarrier

\bibliographystyle{aa}
\bibliography{ref}

\begin{thebibliography}{101}
\expandafter\ifx\csname natexlab\endcsname\relax\def\natexlab#1{#1}\fi

\bibitem[{{Aasi} {et~al.}(2015){Aasi}, {Abbott}, {Abbott}, {Abbott}, {Abernathy}, {Ackley}, {Adams}, {Adams}, {Addesso}, {Adhikari}, {Adya}, {Affeldt}, {Aggarwal}, {Aguiar}, {Ain}, {Ajith}, {Alemic}, {Allen}, {Amariutei}, {Anderson}, {Anderson}, {Arai}, {Araya}, {Arceneaux}, {Areeda}, {Ashton}, {Ast}, {Aston}, {Aufmuth}, {Aulbert}, {Aylott}, {Babak}, {Baker}, {Ballmer}, {Barayoga}, {Barbet}, {Barclay}, {Barish}, {Barker}, {Barr}, {Barsotti}, {Bartlett}, {Barton}, {Bartos}, {Bassiri}, {Batch}, {Baune}, {Behnke}, {Bell}, {Bell}, {Benacquista}, {Bergman}, {Bergmann}, {Berry}, {Betzwieser}, {Bhagwat}, {Bhandare}, {Bilenko}, {Billingsley}, {Birch}, {Biscans}, {Biwer}, {Blackburn}, {Blackburn}, {Blair}, {Blair}, {Bock}, {Bodiya}, {Bojtos}, {Bond}, {Bork}, {Born}, {Bose}, {Brady}, {Braginsky}, {Brau}, {Bridges}, {Brinkmann}, {Brooks}, {Brown}, {Brown}, {Brown}, {Buchman}, {Buikema}, {Buonanno}, {Cadonati}, {Calder{\'o}n Bustillo}, {Camp}, {Cannon}, {Cao}, {Capano}, {Caride}, {Caudill}, {Cavagli{\`a}}, {Cepeda},
  {Chakraborty}, {Chalermsongsak}, {Chamberlin}, {Chao}, {Charlton}, {Chen}, {Cho}, {Cho}, {Chow}, {Christensen}, {Chu}, {Chung}, {Ciani}, {Clara}, {Clark}, {Collette}, {Cominsky}, {Constancio}, {Cook}, {Corbitt}, {Cornish}, {Corsi}, {Costa}, {Coughlin}, {Countryman}, {Couvares}, {Coward}, {Cowart}, {Coyne}, {Coyne}, {Craig}, {Creighton}, {Creighton}, {Cripe}, {Crowder}, {Cumming}, {Cunningham}, {Cutler}, {Dahl}, {Dal Canton}, {Damjanic}, {Danilishin}, {Danzmann}, {Dartez}, {Dave}, {Daveloza}, {Davies}, {Daw}, {DeBra}, {Del Pozzo}, {Denker}, {Dent}, {Dergachev}, {DeRosa}, {DeSalvo}, {Dhurandhar}, {D́{\i}az}, {Di Palma}, {Dojcinoski}, {Dominguez}, {Donovan}, {Dooley}, {Doravari}, {Douglas}, {Downes}, {Driggers}, {Du}, {Dwyer}, {Eberle}, {Edo}, {Edwards}, {Edwards}, {Effler}, {Eggenstein}, {Ehrens}, {Eichholz}, {Eikenberry}, {Essick}, {Etzel}, {Evans}, {Evans}, {Factourovich}, {Fairhurst}, {Fan}, {Fang}, {Farr}, {Farr}, {Favata}, {Fays}, {Fehrmann}, {Fejer}, {Feldbaum}, {Ferreira}, {Fisher}, {Frei}, {Freise},
  {Frey}, {Fricke}, {Fritschel}, {Frolov}, {Fuentes-Tapia}, {Fulda}, {Fyffe}, \& {Gair}}]{2015CQGra..32g4001L}
{Aasi}, J., {Abbott}, B.~P., {Abbott}, R., {et~al.} 2015, Classical and Quantum Gravity, 32, 074001

\bibitem[{Abac {et~al.}(2026)Abac, Abramo, Albanesi, Albertini, Agapito, Agathos, Albertus, Andersson, Andrade, Andreoni, Angeloni, Antonelli, Antoniadis, Antonini, Sedda, Celeste~Artale, Ascenzi, Auclair, Bachetti, Badger, Banerjee, Barba-González, Barta, Bartolo, Bauswein, Begnoni, Beirnaert, Bejger, Belgacem, Bellomo, Bernard, Bernardini, Bernuzzi, Berry, Berti, Bertone, Bettoni, Bezares, Bhagwat, Bisero, Bizouard, Blanco-Pillado, Blasi, Bonino, Borghese, Borghi, Borhanian, Bortolas, Botticella, Branchesi, Breschi, Brito, Brocato, Broekgaarden, Bulik, Buonanno, Burgio, Burrows, Calcagni, Canevarolo, Cappellaro, Capurri, Carbone, Casadio, Cayuso, Cerdá-Durán, Char, Chaty, Chiarusi, Chruslinska, Cireddu, Cole, Colombo, Colpi, Compère, Contaldi, Corman, Crescimbeni, Cristallo, Cuoco, Cusin, Canton, Dálya, D’Avanzo, Davari, De~Luca, De~Renzis, Della~Valle, Del~Pozzo, De~Santi, De~Santis, Dietrich, Dimastrogiovanni, Domenech, Doneva, Drago, Dupletsa, Duval, Dvorkin, Elias-Rosa, Fairhurst, Fantina,
  Fasiello, Fays, Fender, Fischer, Foucart, Fragos, Foffa, Franciolini, Fumagalli, Gair, Gamba, Garcia-Bellido, García-Quirós, Gergely, Ghirlanda, Ghosh, Giacomazzo, Gittins, Giudice, Goncharov, Gonzalez, Goriély, Graziani, Greco, Gualtieri, Guidi, Gupta, Haney, Hannam, Harms, Harutyunyan, Haskell, Haungs, Hazra, Hemming, Heng, Hinderer, van~der Horst, Hu, Husa, Iacovelli, Illuminati, Inguglia, Villalba, Janquart, Janssens, Jenkins, Jones, Kacskovics, Klessen, Kokkotas, Kuan, Kumar, Kuroyanagi, Laghi, Lamberts, Lambiase, Larrouturou, Leaci, Lenzi, Levan, Li, Li, Liang, Limongi, Liu, Llanes-Estrada, Loffredo, Long, Lope-Oter, Lukes-Gerakopoulos, Maggio, Maggiore, Mancarella, Mapelli, Marchant, Margiotta, Mariotti, Marriott-Best, Marsat, Martínez-Pinedo, Maselli, Mastrogiovanni, Matos, Melandri, Mendes, de~Souza, Mentasti, Mezcua, Mösta, Mondal, Moresco, Mukherjee, Muttoni, Nagar, Narola, Nava, Moreno, Nelemans, Nielsen, Nissanke, Obergaulinger, Oertel, Oganesyan, Onori, Pacilio, Pagliaroli, Palomba, Pang,
  Pani, Papalini, Patricelli, Patruno, Pedrotti, Perego, Pérez-García, Périgois, Perna, Péroux, Perret, Perrodin, Pesci, Pfeiffer, Piccinni, Pieroni, Piranomonte, Pompili, Porter, Porto, Pound, Powell, Puech, Pratten, Puecher, Pujolas, Quartin, Raduta, Ramos-Buades, Rase, Razzano, Rea, Regimbau, Renzini, Rettegno, Ricciardone, Riotto, Romero-Rodriguez, Ronchini, Rosinska, Rossi, Roy, Rubiera-Garcia, Rubio, Ruiz-Lapuente, Sagun, Sakellariadou, Salafia, Samajdar, Sanchis-Gual, Sanna, Santoliquido, Sathyaprakash, Schmidt, Schmidt, Schneider, Schneider, Sedrakian, Servant, Sevrin, Shao, Silva, Simakachorn, Smartt, Sotiriou, Spera, Stamerra, Steer, Steinhoff, Stergioulas, Sturani, Suárez, Suresh, Swain, Tagliazucchi, Tamanini, Tasinato, Tauris, Tissino, Tomaselli, Toonen, Torres-Forné, Turski, Ugolini, Vagenas, Dall’Armi, Valenti, Valiante, Broeck, van~de Meent, van Son, Vanvlasselaer, Vaglio, Varma, Veitch, Vaskonen, Vergani, Wils, Witek, Wong, Yazadjiev, Yim, Acernese, Ahn, Allocca, Amato,
  Andrés-Carcasona, Avallone, Bachlechner, Baer, Bagnasco, Balbi, Barone, Benedetti, Benning, Bini, Salcedo, Bozza, Bruno, Butz, Califano, Calloni, Carapella, Cardini, Subrahmanya, Chiadini, Chiummo, Cianetti, Ciani, Coccia, Contu, Cornelissen, Cozzumbo, Croney, Crosta, D’Agostino, Danilishin, D’Antonio, Bolle, Degallaix, Laurentis, della Monica, Marco, de~Martino, Rosa, Salvo, Simone, Detavernier, Diaferia, Cesare, Fiore, Giovanni, Pace, Docherty, D’Urso, Mecherfi, Errico, Fabrizi, Fafone, Fanti, Fittipaldi, Fiumara, Freise, Funk, Gaedtke, Garufi, Gerberding, Giangrandi, Giunchi, Graham, Granata, Granata, Green, Haughian, Heisenberg, Hennig, Hild, Hoang, Holland, Iannone, Isleif, Joppe, Kim, Kim, Kim, Korb, Korobko, Kranzhoff, Kuhlbusch, Lacaille, Lartaux-Vollard, Lavezzi, Laycock, Lee, Lee, Lee, Losurdo, Lucchesi, Lück, Macquet, Majorana, Mangano, Martelli, Martin, Martinez, Masoni, Massaro, Melini, Mercurio, Mereni, Miller, Mirasola, Mitchell, Molinari, Montani, Mow-Lowry, Murgia, Murray, Muscas,
  Naticchioni, Nela, Nery, Niggemann, Nippe, Novak, Numic, Olivieri, Orsini, Park, Pascucci, Perreca, Piergiovanni, Pierro, Pinard, Pinto, Punturo, Puppo, Quochi, Rading, Rapagnani, Ricci, Rodrigues, Romano, Rozza, Saffarieh, Santucci, Schramm, Schwab, Sequino, Neto, Silenzi, Sintes, Sopuerta, Spencer, Stahl, Steinlechner, Steinlechner, Szabó, Thümmler, Tofani, Torniamenti, Travaglini, Trozzo, Paola~Vaccaro, Valentini, Ván, van Dongen, van Heijningen, van Ranst, Vardaro, Verdier, Vernieri, Wagner, Woehler, Wolf, Zavattini, Zink, \& Zmija}]{Abac_2026}
Abac, A., Abramo, R., Albanesi, S., {et~al.} 2026, Journal of Cosmology and Astroparticle Physics, 2026, 081

\bibitem[{{Abbott} {et~al.}(2020{\natexlab{a}}){Abbott}, {Abbott}, {Abbott}, {Abraham}, {Acernese}, {Ackley}, {Adams}, {Adhikari}, {Adya}, {Affeldt}, {Agathos}, {Agatsuma}, {Aggarwal}, {Aguiar}, {Aiello}, {Ain}, {Ajith}, {Allen}, {Allocca}, {Aloy}, {Altin}, {Amato}, {Anand}, {Ananyeva}, {Anderson}, {Anderson}, {Angelova}, {Antier}, {Appert}, {Arai}, {Araya}, {Areeda}, {Ar{\`e}ne}, {Arnaud}, {Aronson}, {Arun}, {Ascenzi}, {Ashton}, {Aston}, {Astone}, {Aubin}, {Aufmuth}, {AultONeal}, {Austin}, {Avendano}, {Avila-Alvarez}, {Babak}, {Bacon}, {Badaracco}, {Bader}, {Bae}, {Baird}, {Baker}, {Baldaccini}, {Ballardin}, {Ballmer}, {Bals}, {Banagiri}, {Barayoga}, {Barbieri}, {Barclay}, {Barish}, {Barker}, {Barkett}, {Barnum}, {Barone}, {Barr}, {Barsotti}, {Barsuglia}, {Barta}, {Bartlett}, {Bartos}, {Bassiri}, {Basti}, {Bawaj}, {Bayley}, {Baylor}, {Bazzan}, {B{\'e}csy}, {Bejger}, {Belahcene}, {Bell}, {Beniwal}, {Benjamin}, {Berger}, {Bergmann}, {Bernuzzi}, {Berry}, {Bersanetti}, {Bertolini}, {Betzwieser}, {Bhandare},
  {Bidler}, {Biggs}, {Bilenko}, {Bilgili}, {Billingsley}, {Birney}, {Birnholtz}, {Biscans}, {Bischi}, {Biscoveanu}, {Bisht}, {Bitossi}, {Bizouard}, {Blackburn}, {Blackman}, {Blair}, {Blair}, {Blair}, {Bloemen}, {Bobba}, {Bode}, {Boer}, {Boetzel}, {Bogaert}, {Bondu}, {Bonnand}, {Booker}, {Boom}, {Bork}, {Boschi}, {Bose}, {Bossilkov}, {Bosveld}, {Bouffanais}, {Bozzi}, {Bradaschia}, {Brady}, {Bramley}, {Branchesi}, {Brau}, {Breschi}, {Briant}, {Briggs}, {Brighenti}, {Brillet}, {Brinkmann}, {Brockill}, {Brooks}, {Brooks}, {Brown}, {Brunett}, {Buikema}, {Bulik}, {Bulten}, {Buonanno}, {Buskulic}, {Buy}, {Byer}, {Cabero}, {Cadonati}, {Cagnoli}, {Cahillane}, {Calder{\'o}n Bustillo}, {Callister}, {Calloni}, {Camp}, {Campbell}, {Canepa}, {Cannon}, {Cao}, {Cao}, {Carapella}, {Carbognani}, {Caride}, {Carney}, {Carullo}, {Casanueva Diaz}, {Casentini}, {Caudill}, {Cavagli{\`a}}, {Cavalier}, {Cavalieri}, {Cella}, {Cerd{\'a}-Dur{\'a}n}, {Cesarini}, {Chaibi}, {Chakravarti}, {Chamberlin}, {Chan}, {Chao}, {Charlton}, {Chase},
  {Chassande-Mottin}, {Chatterjee}, {Chaturvedi}, {Chatziioannou}, {Cheeseboro}, {Chen}, {Chen}, {Chen}, {Cheng}, {Cheong}, {Chia}, {Chiadini}, {Chincarini}, {Chiummo}, {Cho}, \& {Cho}}]{2020ApJ...892L...3A}
{Abbott}, B.~P., {Abbott}, R., {Abbott}, T.~D., {et~al.} 2020{\natexlab{a}}, \apjl, 892, L3

\bibitem[{{Abbott} {et~al.}(2017{\natexlab{a}}){Abbott}, {Abbott}, {Abbott}, {Acernese}, {Ackley}, {Adams}, {Adams}, {Addesso}, {Adhikari}, {Adya}, {Affeldt}, {Afrough}, {Agarwal}, {Agathos}, {Agatsuma}, {Aggarwal}, {Aguiar}, {Aiello}, {Ain}, {Ajith}, {Allen}, {Allen}, {Allocca}, {Altin}, {Amato}, {Ananyeva}, {Anderson}, {Anderson}, {Angelova}, {Antier}, {Appert}, {Arai}, {Araya}, {Areeda}, {Arnaud}, {Arun}, {Ascenzi}, {Ashton}, {Ast}, {Aston}, {Astone}, {Atallah}, {Aufmuth}, {Aulbert}, {AultONeal}, {Austin}, {Avila-Alvarez}, {Babak}, {Bacon}, {Bader}, {Bae}, {Baker}, {Baldaccini}, {Ballardin}, {Ballmer}, {Banagiri}, {Barayoga}, {Barclay}, {Barish}, {Barker}, {Barkett}, {Barone}, {Barr}, {Barsotti}, {Barsuglia}, {Barta}, {Barthelmy}, {Bartlett}, {Bartos}, {Bassiri}, {Basti}, {Batch}, {Bawaj}, {Bayley}, {Bazzan}, {B{\'e}csy}, {Beer}, {Bejger}, {Belahcene}, {Bell}, {Berger}, {Bergmann}, {Bero}, {Berry}, {Bersanetti}, {Bertolini}, {Betzwieser}, {Bhagwat}, {Bhandare}, {Bilenko}, {Billingsley}, {Billman}, {Birch},
  {Birney}, {Birnholtz}, {Biscans}, {Biscoveanu}, {Bisht}, {Bitossi}, {Biwer}, {Bizouard}, {Blackburn}, {Blackman}, {Blair}, {Blair}, {Blair}, {Bloemen}, {Bock}, {Bode}, {Boer}, {Bogaert}, {Bohe}, {Bondu}, {Bonilla}, {Bonnand}, {Boom}, {Bork}, {Boschi}, {Bose}, {Bossie}, {Bouffanais}, {Bozzi}, {Bradaschia}, {Brady}, {Branchesi}, {Brau}, {Briant}, {Brillet}, {Brinkmann}, {Brisson}, {Brockill}, {Broida}, {Brooks}, {Brown}, {Brown}, {Brunett}, {Buchanan}, {Buikema}, {Bulik}, {Bulten}, {Buonanno}, {Buskulic}, {Buy}, {Byer}, {Cabero}, {Cadonati}, {Cagnoli}, {Cahillane}, {Calder{\'o}n Bustillo}, {Callister}, {Calloni}, {Camp}, {Canepa}, {Canizares}, {Cannon}, {Cao}, {Cao}, {Capano}, {Capocasa}, {Carbognani}, {Caride}, {Carney}, {Casanueva Diaz}, {Casentini}, {Caudill}, {Cavagli{\`a}}, {Cavalier}, {Cavalieri}, {Cella}, {Cepeda}, {Cerd{\'a}-Dur{\'a}n}, {Cerretani}, {Cesarini}, {Chamberlin}, {Chan}, {Chao}, {Charlton}, {Chase}, {Chassande-Mottin}, {Chatterjee}, {Chatziioannou}, {Cheeseboro}, {Chen}, {Chen}, {Chen},
  {Cheng}, {Chia}, {Chincarini}, {Chiummo}, {Chmiel}, {Cho}, {Cho}, {Chow}, {Christensen}, {Chu}, {Chua}, {Chua}, {Chung}, {Chung}, \& {Ciani}}]{2017ApJ...848L..12A}
{Abbott}, B.~P., {Abbott}, R., {Abbott}, T.~D., {et~al.} 2017{\natexlab{a}}, \apjl, 848, L12

\bibitem[{{Abbott} {et~al.}(2017{\natexlab{b}}){Abbott}, {Abbott}, {Abbott}, {Acernese}, {Ackley}, {Adams}, {Adams}, {Addesso}, {Adhikari}, {Adya}, {Affeldt}, {Afrough}, {Agarwal}, {Agathos}, {Agatsuma}, {Aggarwal}, {Aguiar}, {Aiello}, {Ain}, {Ajith}, {Allen}, {Allen}, {Allocca}, {Altin}, {Amato}, {Ananyeva}, {Anderson}, {Anderson}, {Angelova}, {Antier}, {Appert}, {Arai}, {Araya}, {Areeda}, {Arnaud}, {Arun}, {Ascenzi}, {Ashton}, {Ast}, {Aston}, {Astone}, {Atallah}, {Aufmuth}, {Aulbert}, {AultONeal}, {Austin}, {Avila-Alvarez}, {Babak}, {Bacon}, {Bader}, {Bae}, {Bailes}, {Baker}, {Baldaccini}, {Ballardin}, {Ballmer}, {Banagiri}, {Barayoga}, {Barclay}, {Barish}, {Barker}, {Barkett}, {Barone}, {Barr}, {Barsotti}, {Barsuglia}, {Barta}, {Barthelmy}, {Bartlett}, {Bartos}, {Bassiri}, {Basti}, {Batch}, {Bawaj}, {Bayley}, {Bazzan}, {B{\'e}csy}, {Beer}, {Bejger}, {Belahcene}, {Bell}, {Berger}, {Bergmann}, {Bernuzzi}, {Bero}, {Berry}, {Bersanetti}, {Bertolini}, {Betzwieser}, {Bhagwat}, {Bhandare}, {Bilenko},
  {Billingsley}, {Billman}, {Birch}, {Birney}, {Birnholtz}, {Biscans}, {Biscoveanu}, {Bisht}, {Bitossi}, {Biwer}, {Bizouard}, {Blackburn}, {Blackman}, {Blair}, {Blair}, {Blair}, {Bloemen}, {Bock}, {Bode}, {Boer}, {Bogaert}, {Bohe}, {Bondu}, {Bonilla}, {Bonnand}, {Boom}, {Bork}, {Boschi}, {Bose}, {Bossie}, {Bouffanais}, {Bozzi}, {Bradaschia}, {Brady}, {Branchesi}, {Brau}, {Briant}, {Brillet}, {Brinkmann}, {Brisson}, {Brockill}, {Broida}, {Brooks}, {Brown}, {Brown}, {Brunett}, {Buchanan}, {Buikema}, {Bulik}, {Bulten}, {Buonanno}, {Buskulic}, {Buy}, {Byer}, {Cabero}, {Cadonati}, {Cagnoli}, {Cahillane}, {Calder{\'o}n Bustillo}, {Callister}, {Calloni}, {Camp}, {Canepa}, {Canizares}, {Cannon}, {Cao}, {Cao}, {Capano}, {Capocasa}, {Carbognani}, {Caride}, {Carney}, {Carullo}, {Casanueva Diaz}, {Casentini}, {Caudill}, {Cavagli{\`a}}, {Cavalier}, {Cavalieri}, {Cella}, {Cepeda}, {Cerd{\'a}-Dur{\'a}n}, {Cerretani}, {Cesarini}, {Chamberlin}, {Chan}, {Chao}, {Charlton}, {Chase}, {Chassande-Mottin}, {Chatterjee},
  {Chatziioannou}, {Cheeseboro}, {Chen}, {Chen}, {Chen}, {Cheng}, {Chia}, {Chincarini}, {Chiummo}, {Chmiel}, {Cho}, {Cho}, {Chow}, {Christensen}, {Chu}, {Chua}, \& {Chua}}]{2017PhRvL.119p1101A}
{Abbott}, B.~P., {Abbott}, R., {Abbott}, T.~D., {et~al.} 2017{\natexlab{b}}, \prl, 119, 161101

\bibitem[{{Abbott} {et~al.}(2020{\natexlab{b}})}]{Abbott2020Prospects}
{Abbott}, B.~P. {et~al.} 2020{\natexlab{b}}, Living Reviews in Relativity, 23, 3

\bibitem[{Ablimit \& Maeda(2018)}]{Ablimit_2018}
Ablimit, I. \& Maeda, K. 2018, The Astrophysical Journal, 866, 151

\bibitem[{{Acernese} {et~al.}(2015){Acernese}, {Agathos}, {Agatsuma}, {Aisa}, {Allemandou}, {Allocca}, {Amarni}, {Astone}, {Balestri}, {Ballardin}, {Barone}, {Baronick}, {Barsuglia}, {Basti}, {Basti}, {Bauer}, {Bavigadda}, {Bejger}, {Beker}, {Belczynski}, {Bersanetti}, {Bertolini}, {Bitossi}, {Bizouard}, {Bloemen}, {Blom}, {Boer}, {Bogaert}, {Bondi}, {Bondu}, {Bonelli}, {Bonnand}, {Boschi}, {Bosi}, {Bouedo}, {Bradaschia}, {Branchesi}, {Briant}, {Brillet}, {Brisson}, {Bulik}, {Bulten}, {Buskulic}, {Buy}, {Cagnoli}, {Calloni}, {Campeggi}, {Canuel}, {Carbognani}, {Cavalier}, {Cavalieri}, {Cella}, {Cesarini}, {Mottin}, {Chincarini}, {Chiummo}, {Chua}, {Cleva}, {Coccia}, {Cohadon}, {Colla}, {Colombini}, {Conte}, {Coulon}, {Cuoco}, {Dalmaz}, {D'Antonio}, {Dattilo}, {Davier}, {Day}, {Debreczeni}, {Degallaix}, {Del{\'e}glise}, {Pozzo}, {Dereli}, {Rosa}, {Fiore}, {Lieto}, {Virgilio}, {Doets}, {Dolique}, {Drago}, {Ducrot}, {Endr{\H{o}}czi}, {Fafone}, {Farinon}, {Ferrante}, {Ferrini}, {Fidecaro}, {Fiori}, {Flaminio},
  {Fournier}, {Franco}, {Frasca}, {Frasconi}, {Gammaitoni}, {Garufi}, {Gaspard}, {Gatto}, {Gemme}, {Gendre}, {Genin}, {Gennai}, {Ghosh}, {Giacobone}, {Giazotto}, {Gouaty}, {Granata}, {Greco}, {Groot}, {Guidi}, {Harms}, {Heidmann}, {Heitmann}, {Hello}, {Hemming}, {Hennes}, {Hofman}, {Jaranowski}, {Jonker}, {Kasprzack}, {K{\'e}f{\'e}lian}, {Kowalska}, {Kraan}, {Kr{\'o}lak}, {Kutynia}, {Lazzaro}, {Leonardi}, {Leroy}, {Letendre}, {Li}, {Lieunard}, {Lorenzini}, {Loriette}, {Losurdo}, {Magazz{\`u}}, {Majorana}, {Maksimovic}, {Malvezzi}, {Man}, {Mangano}, {Mantovani}, {Marchesoni}, {Marion}, {Marque}, {Martelli}, {Martellini}, {Masserot}, {Meacher}, {Meidam}, {Mezzani}, {Michel}, {Milano}, {Minenkov}, {Moggi}, {Mohan}, {Montani}, {Morgado}, {Mours}, {Mul}, {Nagy}, {Nardecchia}, {Naticchioni}, {Nelemans}, {Neri}, {Neri}, {Nocera}, {Pacaud}, {Palomba}, {Paoletti}, {Paoli}, {Pasqualetti}, {Passaquieti}, {Passuello}, {Perciballi}, {Petit}, {Pichot}, {Piergiovanni}, {Pillant}, {Piluso}, {Pinard}, {Poggiani}, {Prijatelj},
  {Prodi}, {Punturo}, {Puppo}, {Rabeling}, {R{\'a}cz}, {Rapagnani}, {Razzano}, {Re}, {Regimbau}, {Ricci}, {Robinet}, {Rocchi}, {Rolland}, {Romano}, {Rosi{\'n}ska}, {Ruggi}, \& {Saracco}}]{2015CQGra..32b4001A}
{Acernese}, F., {Agathos}, M., {Agatsuma}, K., {et~al.} 2015, Classical and Quantum Gravity, 32, 024001

\bibitem[{{Akutsu} {et~al.}(2021){Akutsu}, {Ando}, {Arai}, {Arai}, {Araki}, {Araya}, {Aritomi}, {Aso}, {Bae}, {Bae}, {Baiotti}, {Bajpai}, {Barton}, {Cannon}, {Capocasa}, {Chan}, {Chen}, {Chen}, {Chen}, {Chu}, {Chu}, {Eguchi}, {Enomoto}, {Flaminio}, {Fujii}, {Fukunaga}, {Fukushima}, {Ge}, {Hagiwara}, {Haino}, {Hasegawa}, {Hayakawa}, {Hayama}, {Himemoto}, {Hiranuma}, {Hirata}, {Hirose}, {Hong}, {Hsieh}, {Huang}, {Huang}, {Huang}, {Ikenoue}, {Imam}, {Inayoshi}, {Inoue}, {Ioka}, {Itoh}, {Izumi}, {Jung}, {Jung}, {Kajita}, {Kamiizumi}, {Kanda}, {Kang}, {Kawaguchi}, {Kawai}, {Kawasaki}, {Kim}, {Kim}, {Kim}, {Kim}, {Kimura}, {Kita}, {Kitazawa}, {Kojima}, {Kokeyama}, {Komori}, {Kong}, {Kotake}, {Kozakai}, {Kozu}, {Kumar}, {Kume}, {Kuo}, {Kuo}, {Kuroyanagi}, {Kusayanagi}, {Kwak}, {Lee}, {Lee}, {Lee}, {Leonardi}, {Lin}, {Lin}, {Lin}, {Liu}, {Luo}, {Marchio}, {Michimura}, {Mio}, {Miyakawa}, {Miyamoto}, {Miyazaki}, {Miyo}, {Miyoki}, {Morisaki}, {Moriwaki}, {Nagano}, {Nagano}, {Nakamura}, {Nakano}, {Nakano}, {Nakashima},
  {Narikawa}, {Negishi}, {Ni}, {Nishizawa}, {Obuchi}, {Ogaki}, {Oh}, {Oh}, {Ohashi}, {Ohishi}, {Ohkawa}, {Okutomi}, {Oohara}, {Ooi}, {Oshino}, {Pan}, {Pang}, {Park}, {Arellano}, {Pinto}, {Sago}, {Saito}, {Saito}, {Sakai}, {Sakai}, {Sakuno}, {Sato}, {Sato}, {Sawada}, {Sekiguchi}, {Sekiguchi}, {Shibagaki}, {Shimizu}, {Shimoda}, {Shimode}, {Shinkai}, {Shishido}, {Shoda}, {Somiya}, {Son}, {Sotani}, {Sugimoto}, {Suzuki}, {Suzuki}, {Tagoshi}, {Takahashi}, {Takahashi}, {Takamori}, {Takano}, {Takeda}, {Takeda}, {Tanaka}, {Tanaka}, {Tanaka}, {Tanaka}, {Tanaka}, {Tanioka}, {Tapia San Martin}, {Telada}, {Tomaru}, {Tomigami}, {Tomura}, {Travasso}, {Trozzo}, {Tsang}, {Tsubono}, {Tsuchida}, {Tsuzuki}, {Tuyenbayev}, {Uchikata}, {Uchiyama}, {Ueda}, {Uehara}, {Ueno}, {Ueshima}, {Uraguchi}, {Ushiba}, {van Putten}, {Vocca}, {Wang}, {Wu}, {Wu}, {Wu}, {Xu}, {Yamada}, {Yamamoto}, {Yamamoto}, {Yamamoto}, {Yokogawa}, {Yokoyama}, {Yokozawa}, {Yoshioka}, {Yuzurihara}, {Zeidler}, {Zhao}, \& {Zhu}}]{2021PTEP.2021eA101A}
{Akutsu}, T., {Ando}, M., {Arai}, K., {et~al.} 2021, Progress of Theoretical and Experimental Physics, 2021, 05A101

\bibitem[{{Andrews} {et~al.}(2025){Andrews}, {Bavera}, {Briel}, {Chattaraj}, {Dotter}, {Fragos}, {Gallegos-Garcia}, {Gossage}, {Kalogera}, {Kasdagli}, {Katsaggelos}, {Kimball}, {Kovlakas}, {Kruckow}, {Liotine}, {Misra}, {Rocha}, {Souropanis}, {Srivastava}, {Sun}, {Teng}, {Xing}, {Zapartas}, \& {Zevin}}]{andrews2025posydonversion2population}
{Andrews}, J.~J., {Bavera}, S.~S., {Briel}, M., {et~al.} 2025, \apjs, 281, 3

\bibitem[{{Belczynski} {et~al.}(2016){Belczynski}, {Holz}, {Bulik}, \& {O'Shaughnessy}}]{2016Natur.534..512B}
{Belczynski}, K., {Holz}, D.~E., {Bulik}, T., \& {O'Shaughnessy}, R. 2016, \nat, 534, 512

\bibitem[{{Belczynski} {et~al.}(2008){Belczynski}, {Kalogera}, {Rasio}, {Taam}, {Zezas}, {Bulik}, {Maccarone}, \& {Ivanova}}]{2008ApJS..174..223B}
{Belczynski}, K., {Kalogera}, V., {Rasio}, F.~A., {et~al.} 2008, \apjs, 174, 223

\bibitem[{{Beniamini} \& {Piran}(2016)}]{2016MNRAS.456.4089B}
{Beniamini}, P. \& {Piran}, T. 2016, \mnras, 456, 4089

\bibitem[{{Berger}(2014)}]{2014ARA&A..52...43B}
{Berger}, E. 2014, \araa, 52, 43

\bibitem[{{Boco} {et~al.}(2026){Boco}, {Bosi}, {Sgalletta}, {Romagnolo}, \& {Mapelli}}]{2026arXiv260602725B}
{Boco}, L., {Bosi}, M., {Sgalletta}, C., {Romagnolo}, A., \& {Mapelli}, M. 2026, arXiv e-prints, arXiv:2606.02725

\bibitem[{{Boesky} {et~al.}(2024){Boesky}, {Broekgaarden}, \& {Berger}}]{2024ApJ...976...23B}
{Boesky}, A.~P., {Broekgaarden}, F.~S., \& {Berger}, E. 2024, \apj, 976, 23

\bibitem[{{Branchesi} {et~al.}(2023){Branchesi}, {Maggiore}, {Alonso}, {Badger}, {Banerjee}, {Beirnaert}, {Belgacem}, {Bhagwat}, {Boileau}, {Borhanian}, {Brown}, {Leong Chan}, {Cusin}, {Danilishin}, {Degallaix}, {De Luca}, {Dhani}, {Dietrich}, {Dupletsa}, {Foffa}, {Franciolini}, {Freise}, {Gemme}, {Goncharov}, {Ghosh}, {Gulminelli}, {Gupta}, {Kumar Gupta}, {Harms}, {Hazra}, {Hild}, {Hinderer}, {Siong Heng}, {Iacovelli}, {Janquart}, {Janssens}, {Jenkins}, {Kalaghatgi}, {Koroveshi}, {Li}, {Li}, {Loffredo}, {Maggio}, {Mancarella}, {Mapelli}, {Martinovic}, {Maselli}, {Meyers}, {Miller}, {Mondal}, {Muttoni}, {Narola}, {Oertel}, {Oganesyan}, {Pacilio}, {Palomba}, {Pani}, {Pasqualetti}, {Perego}, {P{\'e}rigois}, {Pieroni}, {Piccinni}, {Puecher}, {Puppo}, {Ricciardone}, {Riotto}, {Ronchini}, {Sakellariadou}, {Samajdar}, {Santoliquido}, {Sathyaprakash}, {Steinlechner}, {Steinlechner}, {Utina}, {Van Den Broeck}, \& {Zhang}}]{2023JCAP...07..068B}
{Branchesi}, M., {Maggiore}, M., {Alonso}, D., {et~al.} 2023, \jcap, 2023, 068

\bibitem[{Breivik {et~al.}(2020)Breivik, Coughlin, Zevin, Rodriguez, Kremer, Ye, Andrews, Kurkowski, Digman, Larson, \& Rasio}]{Breivik_2020}
Breivik, K., Coughlin, S., Zevin, M., {et~al.} 2020, The Astrophysical Journal, 898, 71

\bibitem[{{Bressan} {et~al.}(2012){Bressan}, {Marigo}, {Girardi}, {Salasnich}, {Dal Cero}, {Rubele}, \& {Nanni}}]{Bressan2012PARSEC}
{Bressan}, A., {Marigo}, P., {Girardi}, L., {et~al.} 2012, MNRAS, 427, 127

\bibitem[{{Briel} {et~al.}(2022){Briel}, {Eldridge}, {Stanway}, {Stevance}, \& {Chrimes}}]{2022MNRAS.514.1315B}
{Briel}, M.~M., {Eldridge}, J.~J., {Stanway}, E.~R., {Stevance}, H.~F., \& {Chrimes}, A.~A. 2022, \mnras, 514, 1315

\bibitem[{{Broekgaarden}(2026)}]{broekgaarden2026lowerratesclaimsbinary}
{Broekgaarden}, F.~S. 2026, arXiv e-prints, arXiv:2606.28515

\bibitem[{Broekgaarden {et~al.}(2022)Broekgaarden, Berger, Stevenson, Justham, Mandel, Chruślińska, van Son, Wagg, Vigna-Gómez, de Mink, Chattopadhyay, \& Neijssel}]{Broekgaarden_2022}
Broekgaarden, F.~S., Berger, E., Stevenson, S., {et~al.} 2022, Monthly Notices of the Royal Astronomical Society, 516, 5737–5761

\bibitem[{{Chabrier}(2003)}]{2003PASP..115..763C}
{Chabrier}, G. 2003, \pasp, 115, 763

\bibitem[{{Chattaraj} {et~al.}(2026){Chattaraj}, {Andrews}, {Bavera}, {Briel}, {Chattopadhyay}, {Fragos}, {Gossage}, {Kalogera}, {Kovlakas}, {Kruckow}, {Liotine}, {Rocha}, {Srivastava}, {Sun}, {Teng}, {Xing}, \& {Zapartas}}]{2026ApJ...997...52C}
{Chattaraj}, A., {Andrews}, J.~J., {Bavera}, S.~S., {et~al.} 2026, \apj, 997, 52

\bibitem[{Chattaraj {et~al.}(2026)Chattaraj, Andrews, Bavera, Briel, Chattopadhyay, Fragos, Gossage, Kalogera, Kovlakas, Kruckow, Liotine, Rocha, Srivastava, Sun, Teng, Xing, \& Zapartas}]{Chattaraj_2026}
Chattaraj, A., Andrews, J.~J., Bavera, S.~S., {et~al.} 2026, The Astrophysical Journal, 997, 52

\bibitem[{{Chattopadhyay} {et~al.}(2020){Chattopadhyay}, {Stevenson}, {Hurley}, {Rossi}, \& {Flynn}}]{2020MNRAS.494.1587C}
{Chattopadhyay}, D., {Stevenson}, S., {Hurley}, J.~R., {Rossi}, L.~J., \& {Flynn}, C. 2020, \mnras, 494, 1587

\bibitem[{{Choi} {et~al.}(2016){Choi}, {Dotter}, {Conroy}, {Cantiello}, {Paxton}, \& {Johnson}}]{Choi2016}
{Choi}, J., {Dotter}, A., {Conroy}, C., {et~al.} 2016, \apj, 823, 102

\bibitem[{Chu {et~al.}(2021)Chu, Yu, \& Lu}]{Chu_2021}
Chu, Q., Yu, S., \& Lu, Y. 2021, Monthly Notices of the Royal Astronomical Society, 509, 1557–1586

\bibitem[{{Chu} {et~al.}(2022){Chu}, {Yu}, \& {Lu}}]{2022MNRAS.509.1557C}
{Chu}, Q., {Yu}, S., \& {Lu}, Y. 2022, \mnras, 509, 1557

\bibitem[{{Claeys} {et~al.}(2014){Claeys}, {Pols}, {Izzard}, {Vink}, \& {Verbunt}}]{2014A&A...563A..83C}
{Claeys}, J.~S.~W., {Pols}, O.~R., {Izzard}, R.~G., {Vink}, J., \& {Verbunt}, F.~W.~M. 2014, \aap, 563, A83

\bibitem[{Cowan {et~al.}(2021)Cowan, Sneden, Lawler, Aprahamian, Wiescher, Langanke, Martínez-Pinedo, \& Thielemann}]{Cowan_2021}
Cowan, J.~J., Sneden, C., Lawler, J.~E., {et~al.} 2021, Reviews of Modern Physics, 93

\bibitem[{{De Cia} {et~al.}(2018){De Cia}, {Ledoux}, {Petitjean}, \& {Savaglio}}]{2018A&A...611A..76D}
{De Cia}, A., {Ledoux}, C., {Petitjean}, P., \& {Savaglio}, S. 2018, \aap, 611, A76

\bibitem[{{De Marco} {et~al.}(2011){De Marco}, {Passy}, {Moe}, {Herwig}, {Mac Low}, \& {Paxton}}]{2011MNRAS.411.2277D}
{De Marco}, O., {Passy}, J.-C., {Moe}, M., {et~al.} 2011, \mnras, 411, 2277

\bibitem[{{Deng} {et~al.}(2024){Deng}, {Li}, {Shao}, \& {Xu}}]{deng2024formationdoubleneutronstars}
{Deng}, Z.-L., {Li}, X.-D., {Shao}, Y., \& {Xu}, K. 2024, \apj, 963, 80

\bibitem[{{Dewdney} {et~al.}(2009){Dewdney}, {Hall}, {Schilizzi}, \& {Lazio}}]{lazio2009squarekilometrearray}
{Dewdney}, P.~E., {Hall}, P.~J., {Schilizzi}, R.~T., \& {Lazio}, T.~J.~L.~W. 2009, IEEE Proceedings, 97, 1482

\bibitem[{Disberg \& Mandel(2025)}]{Disberg_2025}
Disberg, P. \& Mandel, I. 2025, The Astrophysical Journal Letters, 989, L8

\bibitem[{{Evans} {et~al.}(2021){Evans}, {Adhikari}, {Afle}, {Ballmer}, {Biscoveanu}, {Borhanian}, {Brown}, {Chen}, {Eisenstein}, {Gruson}, {Gupta}, {Hall}, {Huxford}, {Kamai}, {Kashyap}, {Kissel}, {Kuns}, {Landry}, {Lenon}, {Lovelace}, {McCuller}, {Ng}, {Nitz}, {Read}, {Sathyaprakash}, {Shoemaker}, {Slagmolen}, {Smith}, {Srivastava}, {Sun}, {Vitale}, \& {Weiss}}]{2021arXiv210909882E}
{Evans}, M., {Adhikari}, R.~X., {Afle}, C., {et~al.} 2021, arXiv e-prints, arXiv:2109.09882

\bibitem[{{Fang} \& {Cen}(2004)}]{2004ApJ...616L..87F}
{Fang}, T. \& {Cen}, R. 2004, \apjl, 616, L87

\bibitem[{Farrow {et~al.}(2019)Farrow, Zhu, \& Thrane}]{Farrow_2019}
Farrow, N., Zhu, X.-J., \& Thrane, E. 2019, The Astrophysical Journal, 876, 18

\bibitem[{Fragos {et~al.}(2023)Fragos, J.~Andrews, Bavera, Berry, Coughlin, Dotter, Giri, Kalogera, Katsaggelos, Kovlakas, Lalvani, Misra, Srivastava, Qin, Rocha, Román-Garza, Serra, Stahle, Sun, Teng, Trajcevski, Tran, Xing, Zapartas, \& Zevin}]{Fragos_2023}
Fragos, T., J.~Andrews, J., Bavera, S.~S., {et~al.} 2023, The Astrophysical Journal Supplement Series, 264, 45

\bibitem[{{Fryer} {et~al.}(2012){Fryer}, {Belczynski}, {Wiktorowicz}, {Dominik}, {Kalogera}, \& {Holz}}]{2012ApJ...749...91F}
{Fryer}, C.~L., {Belczynski}, K., {Wiktorowicz}, G., {et~al.} 2012, \apj, 749, 91

\bibitem[{{Gallazzi} {et~al.}(2008){Gallazzi}, {Brinchmann}, {Charlot}, \& {White}}]{2008MNRAS.383.1439G}
{Gallazzi}, A., {Brinchmann}, J., {Charlot}, S., \& {White}, S. D.~M. 2008, \mnras, 383, 1439

\bibitem[{{Gallegos-Garcia} {et~al.}(2023){Gallegos-Garcia}, {Berry}, \& {Kalogera}}]{2023ApJ...955..133G}
{Gallegos-Garcia}, M., {Berry}, C. P.~L., \& {Kalogera}, V. 2023, \apj, 955, 133

\bibitem[{{Ge} {et~al.}(2015){Ge}, {Webbink}, {Chen}, \& {Han}}]{2015ApJ...812...40G}
{Ge}, H., {Webbink}, R.~F., {Chen}, X., \& {Han}, Z. 2015, \apj, 812, 40

\bibitem[{{Giacobbo} \& {Mapelli}(2018)}]{2018MNRAS.480.2011G}
{Giacobbo}, N. \& {Mapelli}, M. 2018, \mnras, 480, 2011

\bibitem[{{Giacobbo} \& {Mapelli}(2020)}]{2020ApJ...891..141G}
{Giacobbo}, N. \& {Mapelli}, M. 2020, \apj, 891, 141

\bibitem[{{Giacobbo} {et~al.}(2018){Giacobbo}, {Mapelli}, \& {Spera}}]{2018MNRAS.474.2959G}
{Giacobbo}, N., {Mapelli}, M., \& {Spera}, M. 2018, \mnras, 474, 2959

\bibitem[{Grichener(2023)}]{Grichener_2023}
Grichener, A. 2023, Monthly Notices of the Royal Astronomical Society, 523, 221–232

\bibitem[{{Hild} {et~al.}(2011){Hild}, {Abernathy}, {Acernese}, {Amaro-Seoane}, {Andersson}, {Arun}, {Barone}, {Barr}, {Barsuglia}, {Beker}, {Beveridge}, {Birindelli}, {Bose}, {Bosi}, {Braccini}, {Bradaschia}, {Bulik}, {Calloni}, {Cella}, {Chassande Mottin}, {Chelkowski}, {Chincarini}, {Clark}, {Coccia}, {Colacino}, {Colas}, {Cumming}, {Cunningham}, {Cuoco}, {Danilishin}, {Danzmann}, {De Salvo}, {Dent}, {De Rosa}, {Di Fiore}, {Di Virgilio}, {Doets}, {Fafone}, {Falferi}, {Flaminio}, {Franc}, {Frasconi}, {Freise}, {Friedrich}, {Fulda}, {Gair}, {Gemme}, {Genin}, {Gennai}, {Giazotto}, {Glampedakis}, {Gr{\"a}f}, {Granata}, {Grote}, {Guidi}, {Gurkovsky}, {Hammond}, {Hannam}, {Harms}, {Heinert}, {Hendry}, {Heng}, {Hennes}, {Hough}, {Husa}, {Huttner}, {Jones}, {Khalili}, {Kokeyama}, {Kokkotas}, {Krishnan}, {Li}, {Lorenzini}, {L{\"u}ck}, {Majorana}, {Mandel}, {Mandic}, {Mantovani}, {Martin}, {Michel}, {Minenkov}, {Morgado}, {Mosca}, {Mours}, {M{\"u}ller─Ebhardt}, {Murray}, {Nawrodt}, {Nelson}, {Oshaughnessy}, {Ott},
  {Palomba}, {Paoli}, {Parguez}, {Pasqualetti}, {Passaquieti}, {Passuello}, {Pinard}, {Plastino}, {Poggiani}, {Popolizio}, {Prato}, {Punturo}, {Puppo}, {Rabeling}, {Rapagnani}, {Read}, {Regimbau}, {Rehbein}, {Reid}, {Ricci}, {Richard}, {Rocchi}, {Rowan}, {R{\"u}diger}, {Santamar{\'\i}a}, {Sassolas}, {Sathyaprakash}, {Schnabel}, {Schwarz}, {Seidel}, {Sintes}, {Somiya}, {Speirits}, {Strain}, {Strigin}, {Sutton}, {Tarabrin}, {Th{\"u}ring}, {van den Brand}, {van Veggel}, {van den Broeck}, {Vecchio}, {Veitch}, {Vetrano}, {Vicere}, {Vyatchanin}, {Willke}, {Woan}, \& {Yamamoto}}]{2011CQGra..28i4013H}
{Hild}, S., {Abernathy}, M., {Acernese}, F., {et~al.} 2011, Classical and Quantum Gravity, 28, 094013

\bibitem[{{Hjellming} \& {Webbink}(1987)}]{1987ApJ...318..794H}
{Hjellming}, M.~S. \& {Webbink}, R.~F. 1987, \apj, 318, 794

\bibitem[{{Hobbs} {et~al.}(2005){Hobbs}, {Lorimer}, {Lyne}, \& {Kramer}}]{2005MNRAS.360..974H}
{Hobbs}, G., {Lorimer}, D.~R., {Lyne}, A.~G., \& {Kramer}, M. 2005, \mnras, 360, 974

\bibitem[{{Hurley} {et~al.}(2000){Hurley}, {Pols}, \& {Tout}}]{2000MNRAS.315..543H}
{Hurley}, J.~R., {Pols}, O.~R., \& {Tout}, C.~A. 2000, \mnras, 315, 543

\bibitem[{{Hurley} {et~al.}(2002){Hurley}, {Tout}, \& {Pols}}]{2002MNRAS.329..897H}
{Hurley}, J.~R., {Tout}, C.~A., \& {Pols}, O.~R. 2002, \mnras, 329, 897

\bibitem[{{Iacovelli} {et~al.}(2022){Iacovelli}, {Mancarella}, {Foffa}, \& {Maggiore}}]{2022ApJ...941..208I}
{Iacovelli}, F., {Mancarella}, M., {Foffa}, S., \& {Maggiore}, M. 2022, \apj, 941, 208

\bibitem[{{Igoshev} {et~al.}(2021){Igoshev}, {Chruslinska}, {Dorozsmai}, \& {Toonen}}]{2021MNRAS.508.3345I}
{Igoshev}, A.~P., {Chruslinska}, M., {Dorozsmai}, A., \& {Toonen}, S. 2021, \mnras, 508, 3345

\bibitem[{{Iorio} {et~al.}(2023){Iorio}, {Mapelli}, {Costa}, {Spera}, {Escobar}, {Sgalletta}, {Trani}, {Korb}, {Santoliquido}, {Dall'Amico}, {Gaspari}, \& {Bressan}}]{2023MNRAS.524..426I}
{Iorio}, G., {Mapelli}, M., {Costa}, G., {et~al.} 2023, \mnras, 524, 426

\bibitem[{Iorio {et~al.}(2023)Iorio, Mapelli, Costa, Spera, Escobar, Sgalletta, Trani, Korb, Santoliquido, Dall’Amico, Gaspari, \& Bressan}]{Iorio_2023}
Iorio, G., Mapelli, M., Costa, G., {et~al.} 2023, Monthly Notices of the Royal Astronomical Society, 524, 426–470

\bibitem[{{Ivanova} {et~al.}(2013){Ivanova}, {Justham}, {Chen}, {De Marco}, {Fryer}, {Gaburov}, {Ge}, {Glebbeek}, {Han}, {Li}, {Lu}, {Marsh}, {Podsiadlowski}, {Potter}, {Soker}, {Taam}, {Tauris}, {van den Heuvel}, \& {Webbink}}]{2013A&ARv..21...59I}
{Ivanova}, N., {Justham}, S., {Chen}, X., {et~al.} 2013, \aapr, 21, 59

\bibitem[{Kruckow {et~al.}(2021)Kruckow, Neunteufel, Di~Stefano, Gao, \& Kobayashi}]{Kruckow_2021}
Kruckow, M.~U., Neunteufel, P.~G., Di~Stefano, R., Gao, Y., \& Kobayashi, C. 2021, The Astrophysical Journal, 920, 86

\bibitem[{{Kruckow} {et~al.}(2018){Kruckow}, {Tauris}, {Langer}, {Kramer}, \& {Izzard}}]{2018MNRAS.481.1908K}
{Kruckow}, M.~U., {Tauris}, T.~M., {Langer}, N., {Kramer}, M., \& {Izzard}, R.~G. 2018, \mnras, 481, 1908

\bibitem[{Li {et~al.}(2026)Li, Wei, Jia, Chen, Ge, Chen, Zhang, Chen, \& Han}]{Li_2026}
Li, Z., Wei, D., Jia, S., {et~al.} 2026, The Astrophysical Journal, 1004, 31

\bibitem[{{Madau} \& {Fragos}(2017)}]{2017ApJ...840...39M}
{Madau}, P. \& {Fragos}, T. 2017, \apj, 840, 39

\bibitem[{{Manchester} {et~al.}(2005){Manchester}, {Hobbs}, {Teoh}, \& {Hobbs}}]{2005yCat.7245....0M}
{Manchester}, R.~N., {Hobbs}, G.~B., {Teoh}, A., \& {Hobbs}, M. 2005, {VizieR Online Data Catalog: ATNF Pulsar Catalog (Manchester+, 2005)}, VizieR On-line Data Catalog: VII/245. Originally published in: 2005AJ....129.1993M

\bibitem[{{Mestichelli} {et~al.}(2025){Mestichelli}, {Mapelli}, {Santoliquido}, {Arca Sedda}, {Branchesi}, {Paiella}, {Costa}, {Iorio}, {Mould}, {Lipatova}, {Liu}, \& {Klessen}}]{2025A&A...704A..54M}
{Mestichelli}, B., {Mapelli}, M., {Santoliquido}, F., {et~al.} 2025, \aap, 704, A54

\bibitem[{Metzger(2019)}]{Metzger_2019}
Metzger, B.~D. 2019, Living Reviews in Relativity, 23

\bibitem[{Muccillo {et~al.}(2026)Muccillo, Patricelli, \& Razzano}]{Muccillo_2026}
Muccillo, L., Patricelli, B., \& Razzano, M. 2026, Journal of Physics: Conference Series, 3177, 012079

\bibitem[{{Neijssel} {et~al.}(2019){Neijssel}, {Vigna-G{\'o}mez}, {Stevenson}, {Barrett}, {Gaebel}, {Broekgaarden}, {de Mink}, {Sz{\'e}csi}, {Vinciguerra}, \& {Mandel}}]{2019MNRAS.490.3740N}
{Neijssel}, C.~J., {Vigna-G{\'o}mez}, A., {Stevenson}, S., {et~al.} 2019, \mnras, 490, 3740

\bibitem[{{Nelemans} \& {Tout}(2005)}]{Nelemans2005}
{Nelemans}, G. \& {Tout}, C.~A. 2005, \mnras, 356, 753

\bibitem[{{Nelemans} {et~al.}(2000){Nelemans}, {Verbunt}, {Yungelson}, \& {Portegies Zwart}}]{nelemans2000reconstructingevolutiondoublehelium}
{Nelemans}, G., {Verbunt}, F., {Yungelson}, L.~R., \& {Portegies Zwart}, S.~F. 2000, \aap, 360, 1011

\bibitem[{{Paxton} {et~al.}(2011){Paxton}, {Bildsten}, {Dotter}, {Herwig}, {Lesaffre}, \& {Timmes}}]{Paxton2011MESA}
{Paxton}, B., {Bildsten}, L., {Dotter}, A., {et~al.} 2011, ApJS, 192, 3

\bibitem[{Pellouin {et~al.}(2025)Pellouin, Dvorkin, \& Lehoucq}]{Pellouin_2025}
Pellouin, C., Dvorkin, I., \& Lehoucq, L. 2025, Astronomy \& Astrophysics, 693, A283

\bibitem[{{Planck Collaboration} {et~al.}(2016){Planck Collaboration}, {Ade}, {Aghanim}, {Arnaud}, {Ashdown}, {Aumont}, {Baccigalupi}, {Banday}, {Barreiro}, {Bartlett}, {Bartolo}, {Battaner}, {Battye}, {Benabed}, {Beno{\^\i}t}, {Benoit-L{\'e}vy}, {Bernard}, {Bersanelli}, {Bielewicz}, {Bock}, {Bonaldi}, {Bonavera}, {Bond}, {Borrill}, {Bouchet}, {Boulanger}, {Bucher}, {Burigana}, {Butler}, {Calabrese}, {Cardoso}, {Catalano}, {Challinor}, {Chamballu}, {Chary}, {Chiang}, {Chluba}, {Christensen}, {Church}, {Clements}, {Colombi}, {Colombo}, {Combet}, {Coulais}, {Crill}, {Curto}, {Cuttaia}, {Danese}, {Davies}, {Davis}, {de Bernardis}, {de Rosa}, {de Zotti}, {Delabrouille}, {D{\'e}sert}, {Di Valentino}, {Dickinson}, {Diego}, {Dolag}, {Dole}, {Donzelli}, {Dor{\'e}}, {Douspis}, {Ducout}, {Dunkley}, {Dupac}, {Efstathiou}, {Elsner}, {En{\ss}lin}, {Eriksen}, {Farhang}, {Fergusson}, {Finelli}, {Forni}, {Frailis}, {Fraisse}, {Franceschi}, {Frejsel}, {Galeotta}, {Galli}, {Ganga}, {Gauthier}, {Gerbino}, {Ghosh}, {Giard},
  {Giraud-H{\'e}raud}, {Giusarma}, {Gjerl{\o}w}, {Gonz{\'a}lez-Nuevo}, {G{\'o}rski}, {Gratton}, {Gregorio}, {Gruppuso}, {Gudmundsson}, {Hamann}, {Hansen}, {Hanson}, {Harrison}, {Helou}, {Henrot-Versill{\'e}}, {Hern{\'a}ndez-Monteagudo}, {Herranz}, {Hildebrandt}, {Hivon}, {Hobson}, {Holmes}, {Hornstrup}, {Hovest}, {Huang}, {Huffenberger}, {Hurier}, {Jaffe}, {Jaffe}, {Jones}, {Juvela}, {Keih{\"a}nen}, {Keskitalo}, {Kisner}, {Kneissl}, {Knoche}, {Knox}, {Kunz}, {Kurki-Suonio}, {Lagache}, {L{\"a}hteenm{\"a}ki}, {Lamarre}, {Lasenby}, {Lattanzi}, {Lawrence}, {Leahy}, {Leonardi}, {Lesgourgues}, {Levrier}, {Lewis}, {Liguori}, {Lilje}, {Linden-V{\o}rnle}, {L{\'o}pez-Caniego}, {Lubin}, {Mac{\'\i}as-P{\'e}rez}, {Maggio}, {Maino}, {Mandolesi}, {Mangilli}, {Marchini}, {Maris}, {Martin}, {Martinelli}, {Mart{\'\i}nez-Gonz{\'a}lez}, {Masi}, {Matarrese}, {McGehee}, {Meinhold}, {Melchiorri}, {Melin}, {Mendes}, {Mennella}, {Migliaccio}, {Millea}, {Mitra}, {Miville-Desch{\^e}nes}, {Moneti}, {Montier}, {Morgante}, {Mortlock},
  {Moss}, {Munshi}, {Murphy}, {Naselsky}, {Nati}, {Natoli}, {Netterfield}, {N{\o}rgaard-Nielsen}, {Noviello}, {Novikov}, {Novikov}, {Oxborrow}, {Paci}, {Pagano}, {Pajot}, {Paladini}, {Paoletti}, {Partridge}, {Pasian}, {Patanchon}, {Pearson}, {Perdereau}, {Perotto}, {Perrotta}, {Pettorino}, {Piacentini}, {Piat}, {Pierpaoli}, {Pietrobon}, {Plaszczynski}, {Pointecouteau}, {Polenta}, {Popa}, {Pratt}, \& {Pr{\'e}zeau}}]{2016A&A...594A..13P}
{Planck Collaboration}, {Ade}, P.~A.~R., {Aghanim}, N., {et~al.} 2016, \aap, 594, A13

\bibitem[{{Punturo} {et~al.}(2010){Punturo}, {Abernathy}, {Acernese}, {Allen}, {Andersson}, {Arun}, {Barone}, {Barr}, {Barsuglia}, {Beker}, {Beveridge}, {Birindelli}, {Bose}, {Bosi}, {Braccini}, {Bradaschia}, {Bulik}, {Calloni}, {Cella}, {Chassande Mottin}, {Chelkowski}, {Chincarini}, {Clark}, {Coccia}, {Colacino}, {Colas}, {Cumming}, {Cunningham}, {Cuoco}, {Danilishin}, {Danzmann}, {De Luca}, {De Salvo}, {Dent}, {De Rosa}, {Di Fiore}, {Di Virgilio}, {Doets}, {Fafone}, {Falferi}, {Flaminio}, {Franc}, {Frasconi}, {Freise}, {Fulda}, {Gair}, {Gemme}, {Gennai}, {Giazotto}, {Glampedakis}, {Granata}, {Grote}, {Guidi}, {Hammond}, {Hannam}, {Harms}, {Heinert}, {Hendry}, {Heng}, {Hennes}, {Hild}, {Hough}, {Husa}, {Huttner}, {Jones}, {Khalili}, {Kokeyama}, {Kokkotas}, {Krishnan}, {Lorenzini}, {L{\"u}ck}, {Majorana}, {Mandel}, {Mandic}, {Martin}, {Michel}, {Minenkov}, {Morgado}, {Mosca}, {Mours}, {M{\"u}ller─Ebhardt}, {Murray}, {Nawrodt}, {Nelson}, {Oshaughnessy}, {Ott}, {Palomba}, {Paoli}, {Parguez}, {Pasqualetti},
  {Passaquieti}, {Passuello}, {Pinard}, {Poggiani}, {Popolizio}, {Prato}, {Puppo}, {Rabeling}, {Rapagnani}, {Read}, {Regimbau}, {Rehbein}, {Reid}, {Rezzolla}, {Ricci}, {Richard}, {Rocchi}, {Rowan}, {R{\"u}diger}, {Sassolas}, {Sathyaprakash}, {Schnabel}, {Schwarz}, {Seidel}, {Sintes}, {Somiya}, {Speirits}, {Strain}, {Strigin}, {Sutton}, {Tarabrin}, {Th{\"u}ring}, {van den Brand}, {van Leewen}, {van Veggel}, {van den Broeck}, {Vecchio}, {Veitch}, {Vetrano}, {Vicere}, {Vyatchanin}, {Willke}, {Woan}, {Wolfango}, \& {Yamamoto}}]{2010CQGra..27s4002P}
{Punturo}, M., {Abernathy}, M., {Acernese}, F., {et~al.} 2010, Classical and Quantum Gravity, 27, 194002

\bibitem[{{Reitze} {et~al.}(2019){Reitze}, {Adhikari}, {Ballmer}, {Barish}, {Barsotti}, {Billingsley}, {Brown}, {Chen}, {Coyne}, {Eisenstein}, {Evans}, {Fritschel}, {Hall}, {Lazzarini}, {Lovelace}, {Read}, {Sathyaprakash}, {Shoemaker}, {Smith}, {Torrie}, {Vitale}, {Weiss}, {Wipf}, \& {Zucker}}]{2019BAAS...51g..35R}
{Reitze}, D., {Adhikari}, R.~X., {Ballmer}, S., {et~al.} 2019, in Bulletin of the American Astronomical Society, Vol.~51, 35

\bibitem[{Ricker {et~al.}(2018)Ricker, Timmes, Taam, \& Webbink}]{Ricker_2018}
Ricker, P.~M., Timmes, F.~X., Taam, R.~E., \& Webbink, R.~F. 2018, Proceedings of the International Astronomical Union, 14, 449–454

\bibitem[{{Riley} {et~al.}(2022){Riley}, {Agrawal}, {Barrett}, {Boyett}, {Broekgaarden}, {Chattopadhyay}, {Gaebel}, {Gittins}, {Hirai}, {Howitt}, {Justham}, {Khandelwal}, {Kummer}, {Lau}, {Mandel}, {de Mink}, {Neijssel}, {Riley}, {van Son}, {Stevenson}, {Vigna-G{\'o}mez}, {Vinciguerra}, {Wagg}, {Willcox}, \& {Team Compas}}]{2022ApJS..258...34R}
{Riley}, J., {Agrawal}, P., {Barrett}, J.~W., {et~al.} 2022, \apjs, 258, 34

\bibitem[{{Romagnolo}(2022)}]{2022eas..conf.1076R}
{Romagnolo}, A. 2022, in EAS2022, European Astronomical Society Annual Meeting, 1076

\bibitem[{{Safarzadeh} {et~al.}(2019){Safarzadeh}, {Berger}, {Ng}, {Chen}, {Vitale}, {Whittle}, \& {Scannapieco}}]{2019ApJ...878L..13S}
{Safarzadeh}, M., {Berger}, E., {Ng}, K. K.~Y., {et~al.} 2019, \apjl, 878, L13

\bibitem[{{Sana} {et~al.}(2012){Sana}, {de Mink}, {de Koter}, {Langer}, {Evans}, {Gieles}, {Gosset}, {Izzard}, {Le Bouquin}, \& {Schneider}}]{2012Sci...337..444S}
{Sana}, H., {de Mink}, S.~E., {de Koter}, A., {et~al.} 2012, Science, 337, 444

\bibitem[{Santoliquido {et~al.}(2020)Santoliquido, Mapelli, Bouffanais, Giacobbo, Di~Carlo, Rastello, Artale, \& Ballone}]{Santoliquido_2020}
Santoliquido, F., Mapelli, M., Bouffanais, Y., {et~al.} 2020, The Astrophysical Journal, 898, 152

\bibitem[{{Santoliquido} {et~al.}(2020){Santoliquido}, {Mapelli}, {Bouffanais}, {Giacobbo}, {Di Carlo}, {Rastello}, {Artale}, \& {Ballone}}]{2020ApJ...898..152S}
{Santoliquido}, F., {Mapelli}, M., {Bouffanais}, Y., {et~al.} 2020, \apj, 898, 152

\bibitem[{{Santoliquido} {et~al.}(2021){Santoliquido}, {Mapelli}, {Giacobbo}, {Bouffanais}, \& {Artale}}]{2021MNRAS.502.4877S}
{Santoliquido}, F., {Mapelli}, M., {Giacobbo}, N., {Bouffanais}, Y., \& {Artale}, M.~C. 2021, \mnras, 502, 4877

\bibitem[{{Sgalletta} {et~al.}(2023){Sgalletta}, {Iorio}, {Mapelli}, {Artale}, {Boco}, {Chattopadhyay}, {Lapi}, {Possenti}, {Rinaldi}, \& {Spera}}]{2023MNRAS.526.2210S}
{Sgalletta}, C., {Iorio}, G., {Mapelli}, M., {et~al.} 2023, \mnras, 526, 2210

\bibitem[{{Sgalletta} {et~al.}(2025){Sgalletta}, {Mapelli}, {Boco}, {Santoliquido}, {Artale}, {Iorio}, {Lapi}, \& {Spera}}]{2025A&A...698A.144S}
{Sgalletta}, C., {Mapelli}, M., {Boco}, L., {et~al.} 2025, \aap, 698, A144

\bibitem[{{Soberman} {et~al.}(1997){Soberman}, {Phinney}, \& {van den Heuvel}}]{1997A&A...327..620S}
{Soberman}, G.~E., {Phinney}, E.~S., \& {van den Heuvel}, E.~P.~J. 1997, \aap, 327, 620

\bibitem[{{Spera} {et~al.}(2019){Spera}, {Mapelli}, {Giacobbo}, {Trani}, {Bressan}, \& {Costa}}]{2019MNRAS.485..889S}
{Spera}, M., {Mapelli}, M., {Giacobbo}, N., {et~al.} 2019, \mnras, 485, 889

\bibitem[{{Stanway} \& {Eldridge}(2018)}]{2018MNRAS.479...75S}
{Stanway}, E.~R. \& {Eldridge}, J.~J. 2018, \mnras, 479, 75

\bibitem[{{Stappers} {et~al.}(2018){Stappers}, {Keane}, {Kramer}, {Possenti}, \& {Stairs}}]{2018RSPTA.37670293S}
{Stappers}, B.~W., {Keane}, E.~F., {Kramer}, M., {Possenti}, A., \& {Stairs}, I.~H. 2018, Philosophical Transactions of the Royal Society of London Series A, 376, 20170293

\bibitem[{{Tauris} {et~al.}(2017){Tauris}, {Kramer}, {Freire}, {Wex}, {Janka}, {Langer}, {Podsiadlowski}, {Bozzo}, {Chaty}, {Kruckow}, {van den Heuvel}, {Antoniadis}, {Breton}, \& {Champion}}]{2017ApJ...846..170T}
{Tauris}, T.~M., {Kramer}, M., {Freire}, P.~C.~C., {et~al.} 2017, \apj, 846, 170

\bibitem[{{Team Compas} {et~al.}(2025){Team Compas}, {Mandel}, {Riley}, {Boesky}, {Br{\v{c}}ek}, {Hirai}, {Kapil}, {Lau}, {Merritt}, {Rodr{\'\i}guez-Segovia}, {Romero-Shaw}, {Song}, {Stevenson}, {Vajpeyi}, {van Son}, {Vigna-G{\'o}mez}, \& {Willcox}}]{2025ApJS..280...43T}
{Team Compas}, {Mandel}, I., {Riley}, J., {et~al.} 2025, \apjs, 280, 43

\bibitem[{{The LIGO Scientific Collaboration} {et~al.}(2026){The LIGO Scientific Collaboration}, {the Virgo Collaboration}, {the KAGRA Collaboration}, {Abac}, {Abe}, {Abouelfettouh}, {Acernese}, {Ackley}, {Adam}, {Adhicary}, {Adhikari}, {Adhikari}, {Adkins}, {Afroz}, {Agapito}, {Agarwal}, {Agathos}, {Aggarwal}, {Aggarwal}, {Aguiar}, {Ahrend}, {Aiello}, {Ain}, {Ajith}, {Akutsu}, {Albers}, {Ali}, {Al-Kershi}, {Allene}, {Allocca}, {Al-Shammari}, {Alvarez}, {Alvarez-Lopez}, {Amar}, {Amarasinghe}, {Amato}, {Amicucci}, {Amra}, {Anand}, {Anand}, {Ananyeva}, {Anderson}, {Anderson}, {Andia}, {Ando}, {Andrade-Oliveira}, {Andr{\'e}s-Carcasona}, {Andrey}, {Andri{\'c}}, {Anglin}, {Anna}, {Antelis}, {Antier}, {Antonini}, {Aoki}, {Aoumi}, {Appavuravther}, {Appelt}, {Appert}, {Apple}, {Arai}, {Araya}, {Araya}, {Arca Sedda}, {Arciprete}, {Areeda}, {Aritomi}, {Armato}, {Armstrong}, {Arnaud}, {Arogeti}, {Aronson}, {Ashton}, {Aso}, {Asprea}, {Assiduo}, {Assis de Souza Melo}, {Aston}, {Astone}, {Aswathi}, {Attadio}, {Aubin},
  {AultONeal}, {Avallone}, {Avdeev}, {Avila}, {Babak}, {Badger}, {Bae}, {Bagnasco}, {Baimukhametova}, {Baiotti}, {Baka}, {Baker}, {Baker}, {Balbi}, {Baldi}, {Baldicchi}, {Ball}, {Ballardin}, {Ballelli}, {Ballmer}, {Banagiri}, {Banerjee}, {Bankar}, {Baptiste}, {Baral}, {Baratti}, {Barayoga}, {Baric}, {Barish}, {Barker}, {Barman}, {Barone}, {Barr}, {Barrios}, {Barsotti}, {Barsuglia}, {Barta}, {Barton}, {Bartos}, {Basalaev}, {Bassiri}, {Basti}, {Bawaj}, {Bayley}, {Baylor}, {Baynard}, {Bazzan}, {Bedakihale}, {Beirnaert}, {Bejger}, {Bell}, {Bellani}, {Bellie}, {Beltran-Martinez}, {Benedetti}, {Benoit}, {Bentara}, {Ben Yaala}, {Bera}, {Bergamin}, {Berger}, {Beroiz}, {Berry}, {Berry}, {Bersanetti}, {Bertheas}, {Bertolini}, {Betzwieser}, {Beveridge}, {Bevins}, {Bezerra-Sobrinho}, {Bhandare}, {Bhatt}, {Bhattacharjee}, {Bhattacharjee}, {Bhattacharyya}, {Bhaumik}, {Biancalana}, {Bianchi}, {Bilenko}, {Bilicki}, {Billingsley}, {Binetti}, {Bini}, {Biot}, {Birnholtz}, {Biscoveanu}, {Bisht}, {Bitossi}, {Bizouard}, {Blaber},
  {Blackburn}, {Blagg}, {Blair}, {Blair}, {Bloch}, {Bode}, {Boettner}, {Bogdan}, {Boileau}, {Boldrini}, {Bolingbroke}, {Bonavena}, {Bonhomme}, {Bonilla}, {Bonilla}, {Bonino}, {Bonnand}, {Borchers}, {Borghi}, {Boschi}, {Bose}, {Bossilkov}, {Bothra}, {Boudon}, {Boybeyi}, {Boyle}, \& {Bozzi}}]{theligoscientificcollaboration2026gwtc50populationpropertiesmerging}
{The LIGO Scientific Collaboration}, {the Virgo Collaboration}, {the KAGRA Collaboration}, {et~al.} 2026, arXiv e-prints, arXiv:2605.27226

\bibitem[{{Toubiana} \& {Dvorkin}(2026)}]{2026A&A...711A.211T}
{Toubiana}, A. \& {Dvorkin}, I. 2026, \aap, 711, A211

\bibitem[{{Toubiana} {et~al.}(2026){Toubiana}, {Gerosa}, {Mould}, {Rinaldi}, {Arca Sedda}, {Bruel}, {Buscicchio}, {Gair}, {Paiella}, {Santoliquido}, {Tenorio}, \& {Ugolini}}]{2026PhRvD.113h3006T}
{Toubiana}, A., {Gerosa}, D., {Mould}, M., {et~al.} 2026, \prd, 113, 083006

\bibitem[{{van Son} {et~al.}(2025){van Son}, {Roy}, {Mandel}, {Farr}, {Lam}, {Merritt}, {Broekgaarden}, {Sander}, \& {Andrews}}]{2025ApJ...979..209V}
{van Son}, L.~A.~C., {Roy}, S.~K., {Mandel}, I., {et~al.} 2025, \apj, 979, 209

\bibitem[{{Verbunt} \& {Cator}(2017)}]{2017JApA...38...40V}
{Verbunt}, F. \& {Cator}, E. 2017, Journal of Astrophysics and Astronomy, 38, 40

\bibitem[{Vigna-Gómez(2025)}]{Vigna_G_mez_2025}
Vigna-Gómez, A. 2025, Astronomy \& Astrophysics, 701, L3

\bibitem[{{Vink} \& {de Koter}(2005)}]{2005A&A...442..587V}
{Vink}, J.~S. \& {de Koter}, A. 2005, \aap, 442, 587

\bibitem[{{Vink} {et~al.}(2001){Vink}, {de Koter}, \& {Lamers}}]{2001A&A...369..574V}
{Vink}, J.~S., {de Koter}, A., \& {Lamers}, H.~J.~G.~L.~M. 2001, \aap, 369, 574

\bibitem[{Wang(2016)}]{Wang_2016}
Wang, J. 2016, Advances in Astronomy, 2016, 1–15

\bibitem[{{Webbink}(2008)}]{2008ASSL..352..233W}
{Webbink}, R.~F. 2008, in Astrophysics and Space Science Library, Vol. 352, Astrophysics and Space Science Library, ed. E.~F. {Milone}, D.~A. {Leahy}, \& D.~W. {Hobill}, 233

\bibitem[{{Xin} {et~al.}(2022){Xin}, {Renzo}, \& {Metzger}}]{2022MNRAS.516.5816X}
{Xin}, C., {Renzo}, M., \& {Metzger}, B.~D. 2022, \mnras, 516, 5816

\end{thebibliography}

\appendix
\section{$\alpha-\lambda$ formalism}
\label{Appendix}

The common-envelope (CE) phase is described using the energy formalism implemented in \textit{BSE} \citep{2002MNRAS.329..897H}, which follows the prescription developed by \citet{1997A&A...327..620S}. In this formalism, an unstable mass-transfer episode leads to the formation of a common envelope surrounding both stellar components. A fraction of the orbital energy released during the inspiral phase is assumed to be transferred to the envelope and used to eject it from the system.

The outcome of the CE phase is governed by two main parameters. The first one, $\alpha_{\rm CE}$, is the common-envelope efficiency parameter and quantifies the fraction of the released orbital energy that contributes to the envelope ejection. In its simplest interpretation, $\alpha_{\rm CE}$ is expected to lie between 0 and 1. However, values larger than unity can be considered if additional energy sources, such as recombination energy or other internal energy contributions, participate in the envelope ejection process \citep{2013A&ARv..21...59I}.

The second parameter, $\lambda$, describes the binding energy of the stellar envelope and accounts for the internal structure of the donor star. Its value depends on the evolutionary state of the star, including its density profile and the location of the core-envelope boundary. The interaction between these two parameters is described by the following energy balance:
\begin{equation}
E_{\rm bind,i} = \alpha_{\rm CE}\left(E_{\rm orb,f} - E_{\rm orb,i}\right).
\end{equation}

where the initial binding energy of the envelope is given by

\begin{equation}
E_{\rm bind,i} =
-\frac{G}{\lambda}
\left(
\frac{M_1 M_{\rm env,1}}{R_1}
+
\frac{M_2 M_{\rm env,2}}{R_2}
\right)
\end{equation}

and the initial and final orbital energies are

\begin{equation}
E_{\rm orb,i} =
-\frac{G M_{\rm c,1} M_{\rm c,2}}{2a_{\rm i}},
\end{equation}

\begin{equation}
E_{\rm orb,f} =
-\frac{G M_{\rm c,1} M_{\rm c,2}}{2a_{\rm f}} .
\end{equation}

Here, $G$ is the gravitational constant, $M_1$ and $M_2$ are the total masses of the two stars at the onset of the CE phase, and $M_{\rm env,1}$ and $M_{\rm env,2}$ are their corresponding envelope masses. $R_1$ and $R_2$ denote the stellar radii at the beginning of the CE phase. The quantities $M_{\rm c,1}$ and $M_{\rm c,2}$ correspond to the masses of the stellar cores remaining after envelope removal. Finally, $a_{\rm i}$ and $a_{\rm f}$ are the orbital separations before and after the CE phase, respectively.

The CE phase is considered successful if the released orbital energy is sufficient to eject the envelope and if neither stellar core overfills its Roche lobe after the inspiral. Otherwise, the two cores are assumed to merge. In that case, systems experiencing an unsuccessful CE phase have no possibility of forming BNS systems. 

\section{Additional \textit{COSMIC} parameters}
\label{section:additionnal cosmic parameters}
\begin{table}[!ht]
\centering
\small
\setlength{\tabcolsep}{2.0pt}
\renewcommand{\arraystretch}{1.1}

\begin{tabular}{@{}p{5cm}p{\dimexpr\columnwidth-5cm\relax}@{}}
\hline\hline
Parameter &
Values / description \\
\hline

\multicolumn{2}{@{}l@{}}{\textbf{Sampling}} \\
\hline

sampling\_method &
Independent \\

\hline

\multicolumn{2}{@{}l@{}}{\textbf{Numerical time resolution}} \\
\hline

pts1 &
0.01 \\

pts2 &
0.01 \\

pts3 &
0.02 \\

\hline

\multicolumn{2}{@{}l@{}}{\textbf{Other fixed parameters}} \\
\hline

Remnant\_flag &
Model 4 (\citet{2012ApJ...749...91F}) \\

$m_{\rm NS,max}$ &
3 \\

wd\_mass\_lim &
1 \\

\hline\hline
\end{tabular}
\label{tab:model_parameters_annex}
\end{table}

\FloatBarrier

\section{\textit{KICKFLAG = 1} kicks distributions} \label{section:kickflag_1_distributions}
\begin{figure}[!ht]
    \centering
    \includegraphics[width=0.9\linewidth]{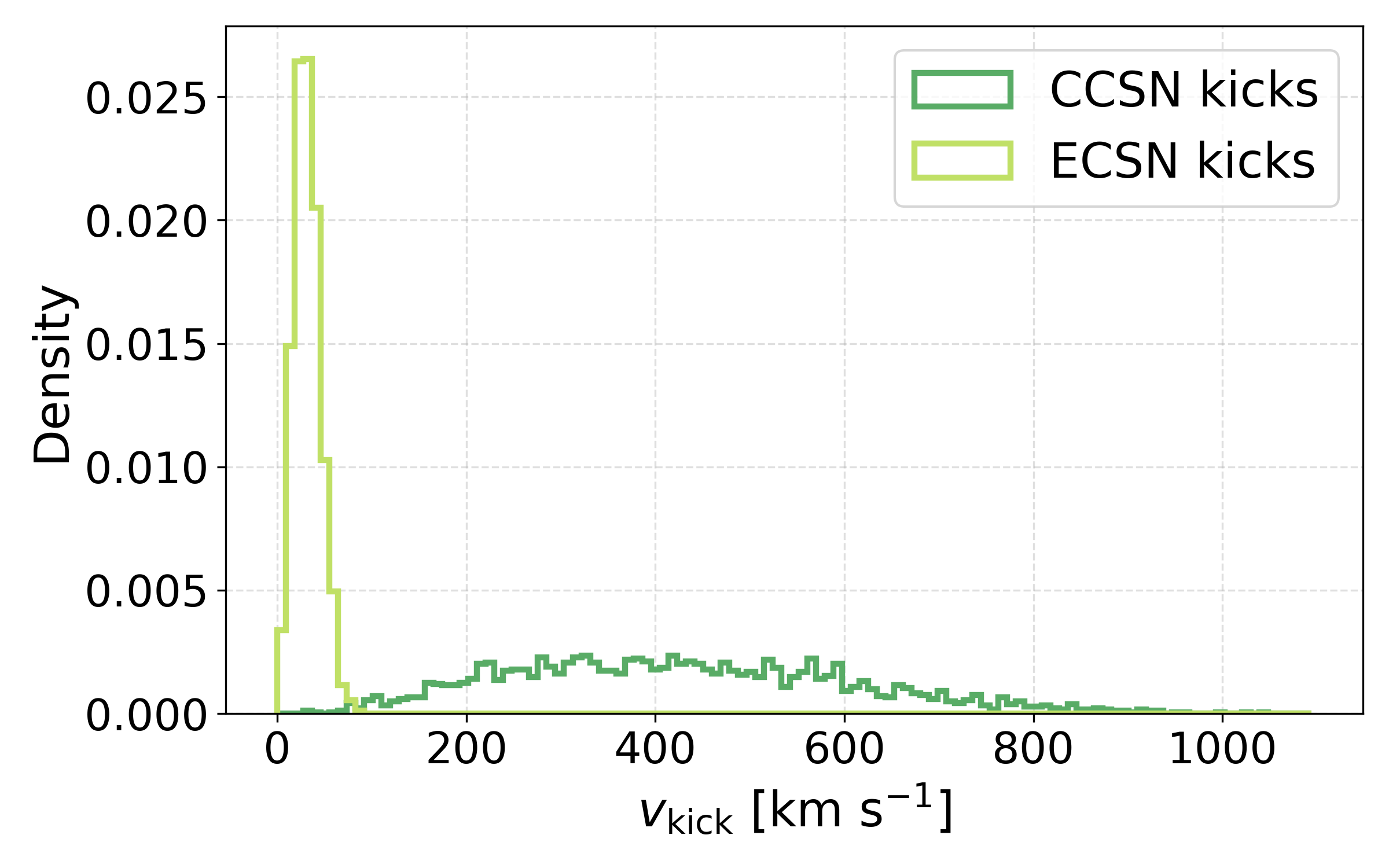}
    \caption{Distribution of natal kicks for CCSN and ECSN events obtained with the \textit{KICKFLAG = 1} prescription from 2000 random realizations. The distributions peak at $ 425 \ km.s^{-1}$ and $ 32  \ km.s^{-1}$ for CCSN and ECSN, respectively, as discussed it \citet{Disberg_2025}, whereas the most recent observational values are lower, with peak velocities at $\sim 200\ km.s^{-1}$ for CCSN \citep{2017JApA...38...40V,2021MNRAS.508.3345I}.}
    \label{fig:kicks distribution}
\end{figure}

\FloatBarrier

\section{First Giant Branch Skip}
\label{section:FGB skip}
\begin{figure}[!ht]
    \centering
    \includegraphics[width=0.9\linewidth]{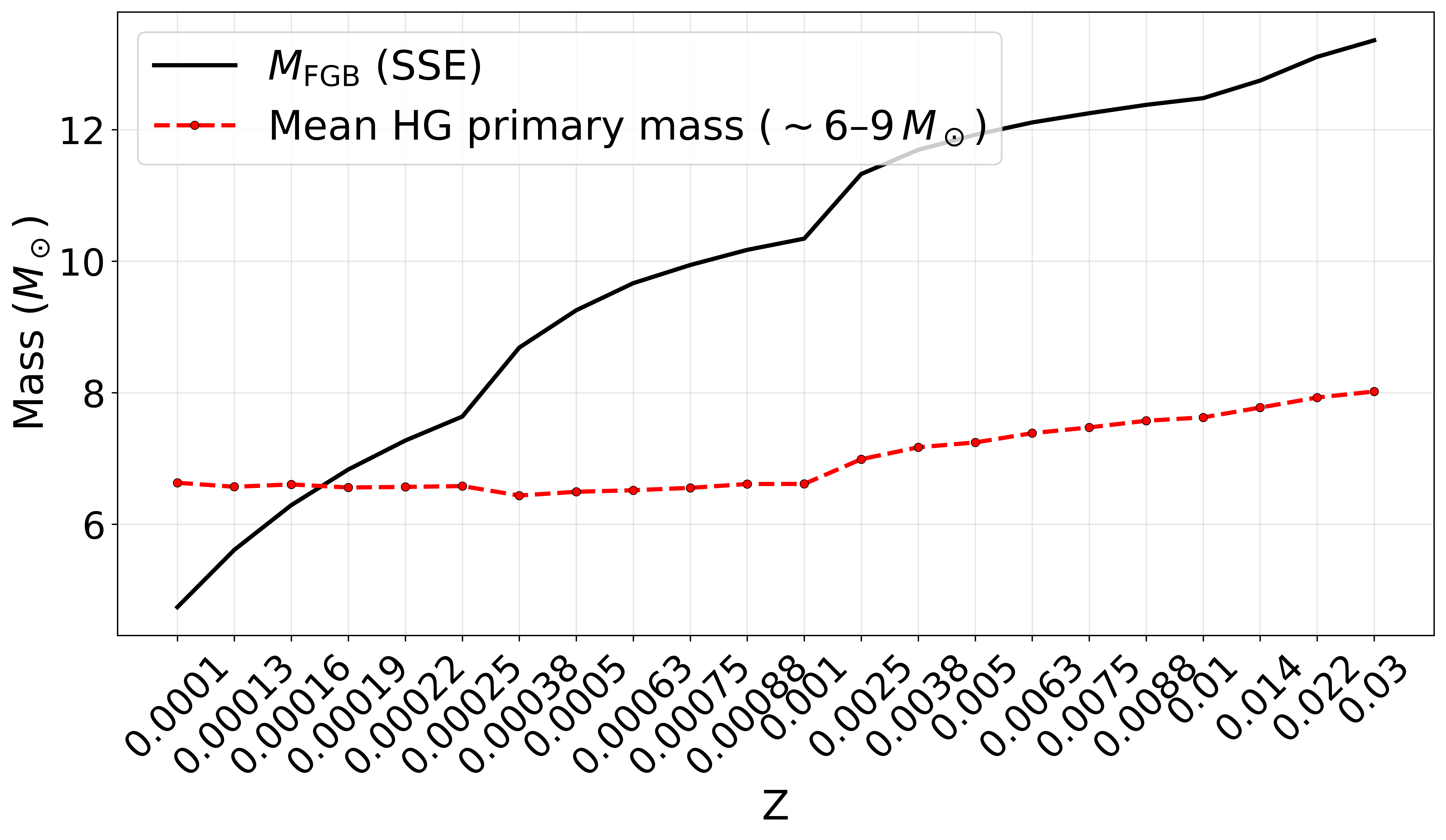}
    \caption{The black curve shows the evolution of $M_{\mathrm{FGB}}$ as a function of metallicity. The dashed red curve represents the mean stellar mass at the end of the Hertzsprung Gap phase for low-mass progenitor stars for the model \textit{QCFLAG = 5, CE\_KICKFLAG = 2, ALPHA = 7.0, KICKFLAG = 2} and a \citet{2003PASP..115..763C} IMF. For $Z < 0.0019$, the majority of systems are expected to bypass the FGB phase.}
    \label{fig:skip_FGB}
\end{figure}
\FloatBarrier
\onecolumn

\section{Complementary $dN/dM$ histogramms}
\label{section:Complementary histogramms}

\begin{figure}[!ht]
    \centering
    \includegraphics[width=0.85\textwidth]{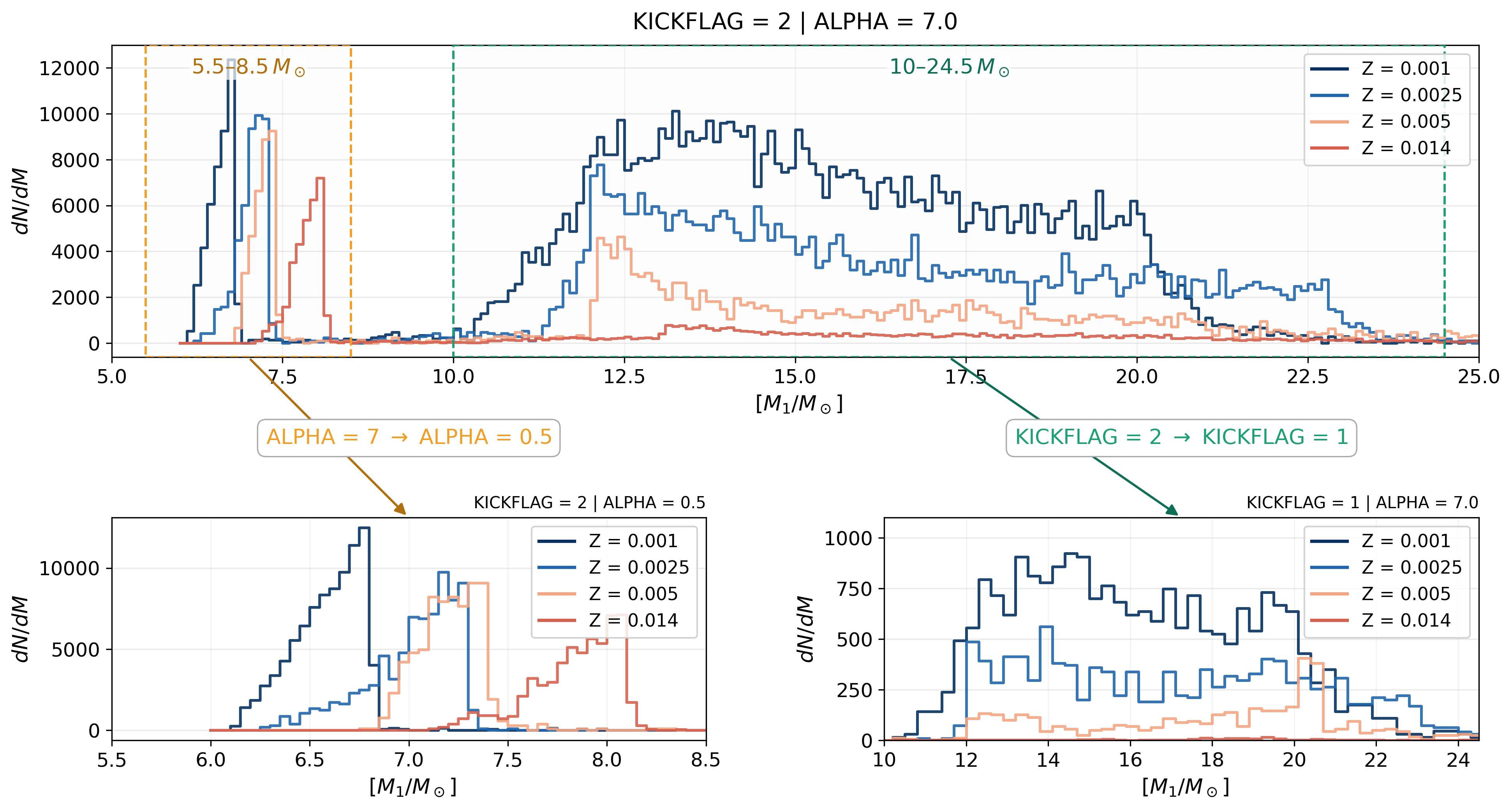}
    \caption{Histograms of the $dN/dM$ distribution as a function of the primary progenitor mass of BNS systems for $Z = 0.001$, $0.0025$, $0.005$, and $0.014$. All panels correspond to models computed with the \citet{2003PASP..115..763C} IMF, \textbf{QCFLAG = 5}, and \textbf{CE\_KICKFLAG = 2}. The top panel shows the complete distribution for a reference model with \textbf{KICKFLAG = 2} and \textbf{ALPHA = 7.0}. Two regions exhibiting distinct behaviours with respect to variations of these parameters are highlighted. The orange region corresponds to low-mass progenitors in the range $5.5$--$8.5\,M_\odot$. The bottom-right panel shows this same region for an otherwise identical model with \textbf{ALPHA = 0.5}, illustrating the weak impact of this parameter on the changes in formation efficiency evolution with $Z$ in this mass range. The green region corresponds to high-mass progenitors in the range $10$--$24.5\,M_\odot$. The bottom-left panel shows this same region for an otherwise identical model with \textbf{KICKFLAG = 2}, illustrating the weak impact of this parameter on the high-mass progenitor distribution evolution with $Z$.}
    \label{fig:histogramms_complementary}
\end{figure}

\vspace{1cm}

\noindent
\begin{minipage}[t]{0.48\textwidth}
    \centering
    
    \section{Metallicity-redshift relation}
    \label{sec:met-Z}

    \begin{figure}[H]
        \centering
        \includegraphics[width=1.0\linewidth]{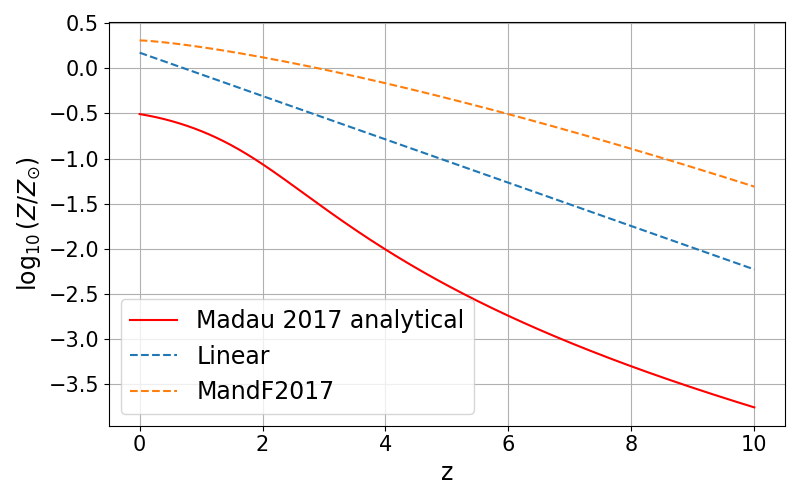}
        \caption{Mean metallicity as a function of redshift for our three models. The red curve represents our integration method, described in Section~\ref{section : merger rate calculations}, using the SFR from \citet{2017ApJ...840...39M}. The blue and yellow dashed curves represent observational fits from \citet{2008MNRAS.383.1439G}, \citet{2018A&A...611A..76D}, and \citet{2017ApJ...840...39M}, respectively.}
        \label{fig:metallicity}
    \end{figure}

\end{minipage}
\hfill
\begin{minipage}[t]{0.48\textwidth}
    \centering
    
    \section{Merger proportions}
    \label{section:additionnal-plots}

    \begin{figure}[H]
        \centering
        \includegraphics[width=1.0\linewidth]{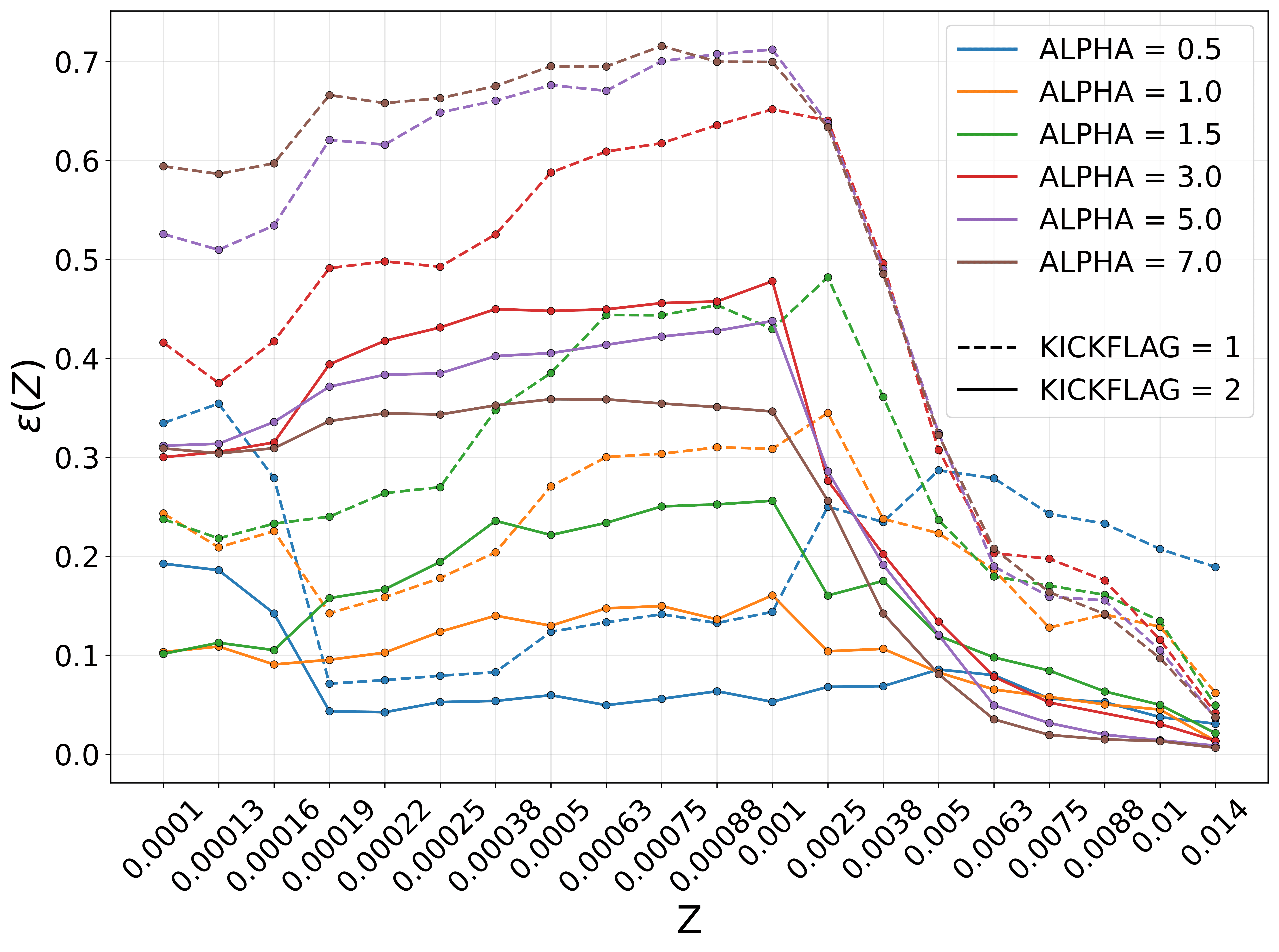}
        \caption{Merger proportion $\epsilon(Z)$ as a function of metallicity $Z$ for the selected models with a \textit{Chabrier} IMF, \textit{QCFLAG} = 5, and \textit{CE\_kickflag} = 2.}
        \label{fig:merger-proportion}
    \end{figure}

\end{minipage}

\newpage
\FloatBarrier
\section{Additional merger rates}
\label{section:additionnal merger rates}

\begin{figure*}[!ht]
    \centering
    \includegraphics[width=0.95\textwidth]{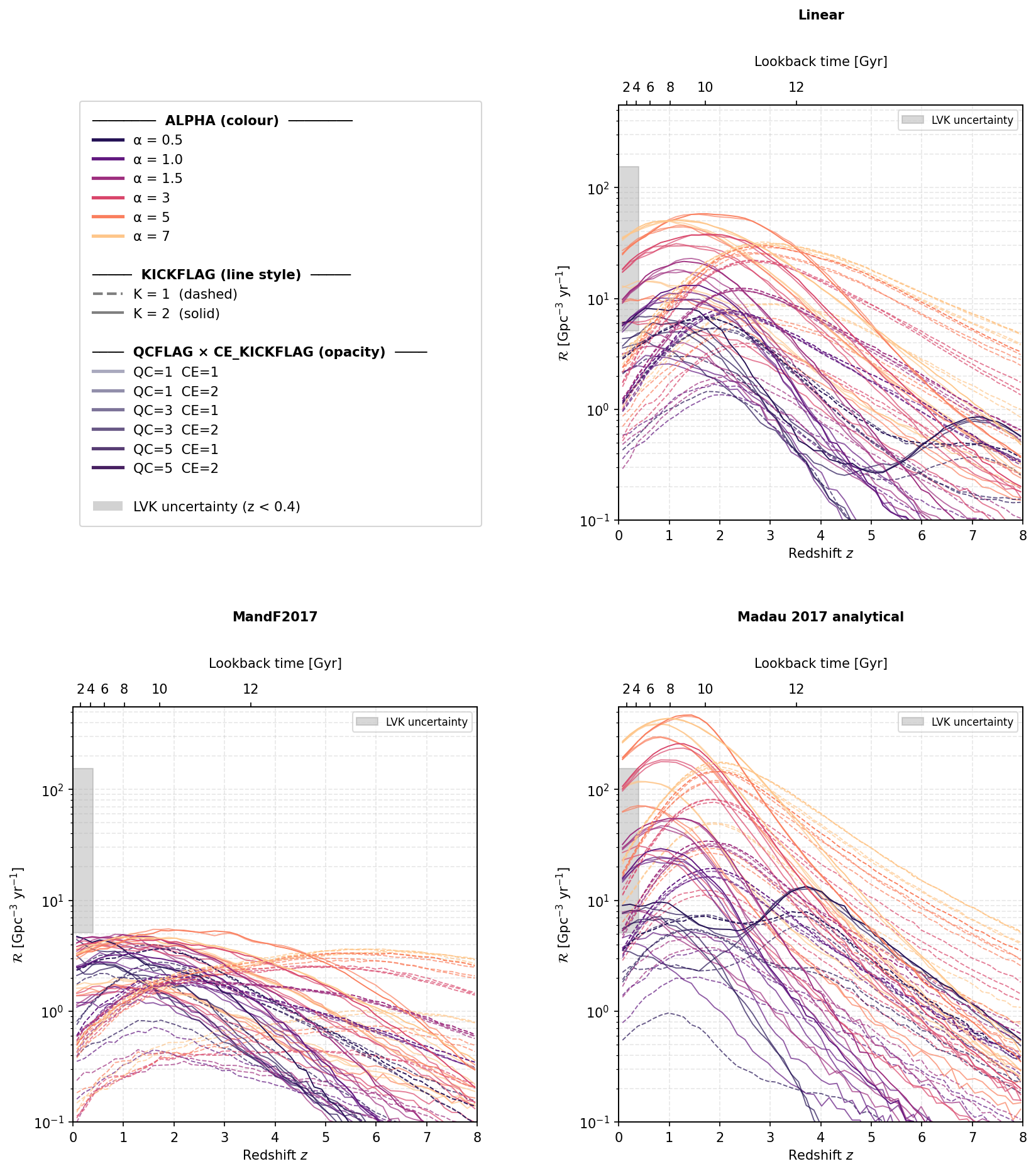}
    \caption{Cosmological merger rate as a function of redshift for three different methods described in Section~\ref{section : merger rate calculations}. For all three approaches, we adopt the star formation rate (SFR) from \citet{2017ApJ...840...39M} and implement different prescriptions for the mean metallicity evolution. The \textbf{linear} and \textbf{MandF2017} models are already implemented in \textit{cosmo\_Rate} and are based on observational fits from \citet{2008MNRAS.383.1439G}, \citet{2018A&A...611A..76D}, and \citet{2017ApJ...840...39M}, respectively. The \textbf{Madau 2017 analytical} model relies on an analytical expression for $\bar{Z}(z)$ derived in \citet{2016Natur.534..512B}. Different curves correspond to various combinations of model parameters: \textbf{ALPHA} is represented by a color gradient, \textbf{KICKFLAG} by different line styles, and combinations of \textbf{QCFLAG} and \textbf{CE\_KICKFLAG} by varying opacity levels. The grey shaded area represents the LVK uncertainty band for detected BNS mergers, as reported by \citet{theligoscientificcollaboration2026gwtc50populationpropertiesmerging}. Its extent along the redshift axis is shown for visual guidance only and does not represent the redshift range over which the constraint is measured.}
    \label{fig:merger rates}
\end{figure*}

\end{document}